\documentclass{CSML}

\def\dOi{13(1:13)2017}
\lmcsheading%
{\dOi}
{1--46}
{}
{}
{Feb.~\phantom05, 2016}
{Mar.~22, 2017}
{}

\usepackage{hyperref} 

\usepackage{stmaryrd}
\usepackage{amsmath}
\usepackage{amssymb}
\usepackage{latexsym}
\usepackage{mathrsfs}
\usepackage[arrow,matrix,frame,curve,cmtip]{xy}
\usepackage{xcolor}
\usepackage{rotating}
\usepackage{url}
\usepackage{epstopdf}
\usepackage{listings}
\usepackage{subfigure}
\usepackage{wrapfig}
\usepackage{xspace}
\usepackage{placeins}
\usepackage{graphicx}
\usepackage{tikz}
\usetikzlibrary{arrows}
\usepackage[draft]{todonotes}
\usepackage{bm}
\usepackage{wasysym}
\usepackage{mathtools}
\usepackage{soul}
\usepackage{thmtools, thm-restate}

\newcommand{\Nat}{{\mathbb N}}

\newcommand{\Real}{{\mathbb R}}

\newcommand{\smuc}{\textsc{SMuC}}

\usepackage{tikz}
\usetikzlibrary{calc}
\newcommand*\circled[1]{\tikz[baseline=(char.base)]{
        \node[shape=circle,draw,minimum size=4mm, inner sep=0pt] (char)
        {\rule[-3pt]{0pt}{\dimexpr2ex+2pt}#1};}}
\newcommand*\circledBlack[1]{\tikz[baseline=(char.base)]{
        \node[shape=circle,fill=black,draw,minimum size=4mm, inner sep=0pt] (char)
        {\rule[-3pt]{0pt}{\dimexpr2ex+2pt}\color{white}#1};}}

\newcommand{\aggregateOut}[3]{#2 \circled{#1} #3}
\newcommand{\aggregateIn}[3]{#2 \circledBlack{#1} #3}

\usepackage[framemethod=TikZ]{mdframed}
\mdfdefinestyle{change}{%
    linecolor=white,
    innerrightmargin=0pt,
    innerleftmargin=0pt,
    backgroundcolor=white}
    
\newcommand{\change}[1]{#1}
\newcommand{\mchange}[1]{#1}
\newcommand{\changeR}[1]{#1}
\newcommand{\mchangeR}[1]{\colorbox{white}{$\displaystyle #1$}}

\usepackage{hyperref}

\begin{document}

\title[Asynchronous Distributed Execution Of 
Fixpoint-Based Computational Fields]{Asynchronous Distributed Execution Of\\ 
Fixpoint-Based Computational Fields
}

\author[A.~{Lluch Lafuente}]{Alberto {Lluch Lafuente}\rsuper a}	
\address{{\lsuper a}DTU Compute, Technical University of Denmark, Denmark}	
\email{albl@dtu.dk}  

\author[M.~Loreti]{Michele Loreti\rsuper b}	
\address{{\lsuper b}University of Florence, Italy}	
\email{michele.loreti@unifi.it}  

\author[Ugo Montanari]{Ugo Montanari\rsuper c}	
\address{{\lsuper c}Computer Science Department, University of Pisa, Italy}	
\email{ugo@di.unipi.it}  
\thanks{
Research supported by the European projects IP 257414 ASCENS and STReP 600708 QUANTICOL, and the Italian PRIN 2010LHT4KM CINA}




\begin{abstract}

  Coordination is essential for dynamic distributed systems whose
  components exhibit interactive and autonomous behaviors. Spatially
  distributed, locally interacting, propagating computational fields
  are particularly appealing for allowing components to join and leave
  with little or no overhead. Computational fields are a key
  ingredient of {\em aggregate programming}, a promising software
  engineering methodology particularly relevant for the Internet of
  Things.  In our approach, space topology is represented by a
  \changeR{fixed graph-shaped} field, namely a network with attributes
  on both nodes and arcs, where arcs represent interaction
  capabilities between nodes. We propose a \smuc{} calculus where
  $\mu$-calculus-like modal formulas represent how the values stored
  in neighbor nodes should be combined to update the present
  node. Fixpoint operations can be understood globally as recursive
  definitions, or locally as asynchronous converging propagation
  processes.  We present a distributed implementation of our calculus.
  The translation is first done mapping \smuc{} programs into normal
  form, purely iterative programs and then into distributed programs.
%
\changeR{%
Some key results are presented that show convergence of fixpoint computations under fair asynchrony and under reinitialization of nodes. The first result allows nodes to proceed at different speeds, while the second one provides robustness against certain kinds of failure. %
}%
We illustrate our approach with a case study based on a disaster
recovery scenario, implemented in a prototype simulator that we use to
evaluate the performance of a recovery strategy.
\end{abstract}

\maketitle


\section{Introduction}\label{section:introduction}
Coordination is essential in all the activities where an ensemble of agents interacts within a distributed system. Particularly interesting is the situation where the ensemble is dynamic, with agents entering and exiting, and when the ensemble must adapt to new situations and must have in general an autonomic behavior. Several models of coordination have been proposed and developed in the past years. Following the classification of~\cite{DBLP:conf/biowire/MameiZ07}, we mention: (i) point-to-point direct coordination, (ii) connector-based coordination, (iii) shared data spaces, (iv) shared deductive knowledge bases, and (v) spatially distributed, locally interacting, propagating computational fields. 

\subsection*{Computational Fields.} 
Among them, \change{computational fields} are particularly appealing for their ability of allowing new interactions with little or no need of communication protocols for initialization. Computational fields are analogous to fields in physics: classical fields are scalars, vectors or tensors, which are functions defined by partial differential equations with initial and/or boundary conditions. Analogously, computational fields consist of suitable space dependent data structures, where interaction is possible only between neighbors.

Computational fields have been proposed as models for several coordination applications, like amorphous computing, routing in mobile ad hoc and sensor networks, situated multi agent ecologies, like swarms, and finally for robotics applications, like coordination of teams of modular robots. Physical fields, though, have the advantage of a regular structure of space, e.g. the one defined by Euclidean geometry, while computational fields are sometimes based on specific (logical) networks of connections. The topology of such a network may have little to do with Euclidean distance, in the sense that a node can be directly connected with nodes which are far away, e.g. for achieving a logarithmic number of hops in distributed hash tables. However, for several robotics applications, and also for swarms and ad hoc networking, one can reasonably assume that an agent can directly interact only with peers located within a limited radius. Thus locality of interaction and propagation of effects become reasonable assumptions.

The computational fields approach has a main conceptual advantage: it offers to the analyst/programmer a high level interface for the collection of possibly inhomogeneous components and connectors which constitute the distributed system under consideration. This view is not concerned with local communication and computation, but only with the so called {\em emergent behavior} of the system. Coordination mechanisms should thus be resilient and self-stabilizing, they should adjust to network  structure and should scale to large networks. As for physical fields, this approach distinguishes clearly between local parameters, which express (relatively) static initial/boundary/inhomogeneity conditions, and field values, which are computed in a systematic way as a result of the interaction with neighbour components. Analogously to the very successful {\em map-reduce} strategy~\cite{DBLP:journals/cacm/DeanG10}, typical operations are propagation and accumulation of values, with selection primitives based on distance and gradient. However, actual computation may require specific control primitives, guaranteeing proper sequentialization of possibly different fields. In particular, if coordination primitives are executed in asynchronous form, suitable termination/commit events should be available in the global view.

\subsection*{Aggregate Programming.}
Recently, computational fields have been integrated in a software engineering methodology called {\em aggregate programming}~\cite{DBLP:journals/computer/BealPV15}. Several abstraction layers are conceptualized, from component/connector capabilities, to coordination primitives, to API operations for global programming. As a main application, the {\em Internet of Things} (IoT) has been suggested, a context where the ability to offer a global view is particularly appreciated.

Another programming style where the advantages of aggregate programming could be felt significantly is {\em fog} or {\em edge computing}~\cite{DBLP:conf/mobihoc/YiLL15}. In this approach, a conceptual level below (or at the edge) of the cloud level is envisaged, where substantial computing and storage resources are available in a located form. Organizing such resources in an aggregated, possibly hierarchical, style might combine the positive qualities of pattern-based distributed programming with the abstract view of cloud virtualization.

\change{
A third potential application domain for aggregate programming is Big Data analytics, where many analyses are essentially based on the computation of fixpoints over a graph or network. Typical examples are centrality measures like PageRank or reachability properties like shortest paths. Entire parallel graph analysis frameworks like Google's Pregel~{\cite{DBLP:conf/sigmod/MalewiczABDHLC10}} and Apache's Giraph~{\cite{DBLP:journals/pvldb/ChingEKLM15}} are built precisely around this idea, originally stemming from the bulk synchronous parallel model of computation~{\cite{DBLP:journals/cacm/Valiant90}}. Those frameworks include features for propagating and aggregating values from and among neighbour nodes, as well as termination detection mechanisms. 
}

\subsection*{Contributions.} 
\change{This paper introduces \changeR{\smuc{}, the \emph{Soft Mu-calculus for Computational fields},} and presents some fundamental results}. \change{In particular the main contributions of the paper are (i) a detailed presentation of the \smuc{} calculus, (ii) results on robustness against node unavailability, (iii) results on robustness against node failures, (iv) a distributed implementation, (v) a case study.}

\paragraph*{(i) \smuc{}} 
\change{In \smuc{}} execution corresponds to sequential  
\changeR{computation of fixpoints in a computational field that represents a fixed graph-shaped topology.}
Fields are essentially networks with attributes on both nodes and arcs, where arcs represent interaction capabilities between nodes.
We \change{originally} introduced \smuc{} in~\cite{DBLP:conf/coordination/Lluch-LafuenteL15} as based on the {\em semiring $\mu$-calculus}~\cite{DBLP:journals/tcs/Lluch-LafuenteM05}, a constraint semiring-valued
generalisation of the modal $\mu$-calculus, which provides a flexible mechanism to specify
the neighbor range (according to path formulae) and the way attributes should be combined
(through semiring operators). Constraint semirings are semirings where the additive operation
is idempotent and the multiplicative operation is commutative. The former allows
one to define a partial ordering as $a \sqsubseteq b$ iff $a+b = a$, under which both additive and multiplicative
operations are monotone. The diamond modality corresponds, in the ordinary
$\mu$-calculus, to disjunction of the logical values on the nodes reached by all the outgoing
arcs. In soft $\mu$-calculus the values are semiring values and \changeR{the diamond modality corresponds to} the additive
operation of the semiring. Similarly for the box modality and the multiplicative operation
of the semiring. In the present version of \smuc{} there is no distinction between the two
modalities: we have only a parametric modality labeled by monotone associative and commutative
operations. More precisely, we have a forward and a backward modality, referring
to outgoing and ingoing arcs. \change{This generalisation allows us to cover more cases of domains and operations.

We believe that our approach based on $\mu$-calculus-like modalities can be particularly
convenient for aggregate programming scenarios. In fact, the $\mu$-calculus, both in its original and
in its soft version, offers a high level, global meaning expressed by its recursive formulas,
while their interpretation in the evaluation semantics computes the fixpoints via iterative approximations which can be interpreted as propagation processes. Thus the \smuc{} calculus provides a well-defined, general, expressive link bridging the gap between the component/connector view and the emergent behavior view.}

\paragraph*{(ii) Robustness against node unavailability} 
\change{Under} reasonable conditions, fixpoints can be computed by asynchronous iterations, where at each iteration certain node attributes are updated based on the attributes of the neighbors in the previous iteration. Not necessarily all nodes must be updated at every iteration: to guarantee convergence it is enough that every node is updated infinitely often. Furthermore, the fixpoint does not depend on the particular sequence of updates. If the partial ordering has only  finite chains, the unique (minimal or maximal) fixpoint is reached in a finite number of iterations. \change{In order to guarantee convergence,  basic constructs must be monotone.  Theorem~{\protect{\ref{th:asynchrony-theorem2}}} formalises this key result.
}

\paragraph*{(iii) Robustness against node failures} 
\changeR{Another concern is about dependability in the presence of failure. 
In our model, only a limited kind of failure is taken care of: 
nodes and links may fail, but then they regularly become active, and the underlying mechanism guarantees that they start from some previous back up state or are restarted.
Robustness against such failures is precisely provided by 
Theorem~{\protect{\ref{theorem:failure}}}, which guarantees
that if at any step some nodes are updated with the values they had in previous steps, possibly the initialization value, but then from some time on they work correctly, the limit value is still the fixpoint. In fact, from a semantical point of view the equivalence class of possible computations for a given formula is characterized as having the same set of upper bounds (and thus the same least fixpoint).

A more general concern is about possible changes in the structure of the network. The meaning of a $\mu$-calculus formula is supposed to be independent of the network itself: for instance a formula expressing node assignment as the minimal distance of every node from some set of final nodes is meaningful even when the network is modified: if the previous network is not changed, and an additional part, just initialised, is connected to it, the fixpoint computation can proceed without problems: it just corresponds to the situation where the additional part is added at the very beginning, but its nodes have never been activated according to the chosen asynchronous computation policy. In general, however, specific recovery actions must be foreseen for maintaining networks with failures, which apply to our approach just as they concern similar coordination styles. Some remedies to this can be found for example in~{\cite{DBLP:conf/coordination/PianiniBV16}} where overlapping fields are used to adapt to network changes.}

\paragraph*{(iv) Distributed implementation}
We present a possible distributed implementation of our calculus. 
The translation is done in two phases, from \smuc{} programs into normal form \smuc{} programs (a step which explicits communication and synchronisation constraints) and then into distributed programs. The correctness of translations exploits the above mentioned results on asynchronous computations.

\change{A delicate issue in the distributed implementation is how to detect termination of a fixpoint computation. 
Several approaches are possible. We considered the Dijkstra-Scholten algorithm~{\protect{\cite{DS80}}}, based on the choice of a 
fixed spanning tree. We do not discuss how to construct and deploy such a tree, 
or how to maintain it in the presence of certain classes of failures and of attackers.
However, most critical aspects are common to all the models based \changeR{on} computational fields. On a related issue, spanning trees can be computed in \smuc{} as illustrated in one of the several examples we provide in this paper.
}

\paragraph*{(v) Case study} 
As a meaningful case study, we present a novel disaster recovery coordination strategy. The goal of the coordination strategy is to direct several rescuers present in the network to help a number of victims, where each victim may need more than one rescuer. While an optimal solution is not required, each victim should be reached by its closest rescuers, so to minimise intervention time.  Our proposed approach may need several iterations of a sequence of three propagations: the first to determine the distance of each rescuer from his/her closest victim, the second to associate to every victim $v$ the list of rescuers having $v$ as their closest victim, so to select the best $k$ of them, if $k$ helpers are needed for $v$; finally, the third propagation is required for notifying each selected rescuer to reach its specific victim.

We have also developed a prototype tool for our language, equipped with a graphical interface that provides useful visual feedback. Indeed we employ those visual features to illustrate the application of our approach to the aforementioned case study.


\subsection*{Previous work}
A first, initial version of \smuc{} was presented in~\cite{DBLP:conf/coordination/Lluch-LafuenteL15}. However the present version offers important improvements in many aspects.

\begin{itemize}
\item Graph-based fields and \smuc{} formulas are generalised here to $\omega$-chain-complete partial orders, with constraint semirings (and their underlying partial orders) being a particularly interesting instance. The main motivation behind such extension is that some of the values transmitted and updated, as data and possibly \smuc{} programs themselves, can be given a partial ordering structure relatively easily, while semirings require lots of additional structure, which sometimes is not available and not fully needed.
\item We have formalised the notion of asynchronous computation of fixpoints in our fields and have provided results ensuring that, under reasonable conditions, nodes can proceed at different speeds without synchronising at each iteration, while still computing the same, desired fixpoint. 
\item \change{We have formalised a notion of safe computation, that can handle certain kinds of failures and have shown that fixpoint computations are robust against such failures.} 
\item The simple imperative language on which \smuc{} is embedded has been simplified. In particular it is now closer to standard \textsf{while}~\cite{DBLP:series/utcs/NielsonN07}. The motivations, besides simplicity and adherence to a well-known language, is that it becomes easier to define control flow based on particular agreements and not just any agreement (as it was in~\cite{DBLP:conf/coordination/Lluch-LafuenteL15}). Of course, control flow based on any agreement can still be achieved, as explained in the paper.
\item The distributed realisation of \smuc{} programs has been fully re-defined, refined and improved. Formal proofs of correctness have been added. Moreover, the global agreement mechanism is now related to the \emph{Dijkstra-Scholten} algorithm~\cite{DS80} for termination detection.    
%
\end{itemize}

\subsection*{Structure of the paper.}
The rest of the paper is structured as follows. 
Sect.~\ref{section:background} recalls some basic definitions related to partial orders and semirings.  
Sect.~\ref{section:calculus-global} presents the \smuc{} calculus \change{and the results related to robustness against unavailability and failures}.  
Sect.~\ref{section:case-study} presents the \smuc{} specification of our disaster recovery case study, which is illustrated with figures obtained with our prototypical tool.
Sect.~\ref{section:calculus-local} discusses several performance and synchronisation issues related to distributed implementations of the calculus.  
Sect.~\ref{section:related} discusses related works. 
Sect.~\ref{section:conclusion} concludes the paper, describes our current work and identifies opportunities for future research. 
\change{Formal proofs can be found in the appendix, together with a table of symbols.}

\section{Background} \label{section:background}

Our computational fields are essentially networks of inter-connected agents, where both agents and their connections have attributes. One key point in our proposal is that the domains of attributes are partially ordered and possibly satisfy other properties. Attributes, indeed, can be natural numbers (e.g. ordered by $\leq$ or $\geq$), sets of nodes (e.g. ordered by containement), paths in the graph (e.g. lexicographically ordered), etc. 
%
We call here such domains \emph{field domains} \change{since node-distributed attributes form, in a certain sense, a computational field of values}. The basic formal underlying structure we will consider for attributes is that of complete partial orders (CPOs) with top and bottom elements, but throughout the paper we will see that requiring some additional conditions is fundamental for some results. 

\begin{defi}[field domain]
Our \emph{field domains} are tuples $\langle A, \sqsubseteq , \bot, \top \rangle$ such that  $\langle A, \sqsubseteq \rangle$ is an $\omega$-chain complete partially $\sqsubseteq$-ordered set $A$ with bottom element $\bot \in A$ and top element $\top \in A$, and $\langle A, \sqsupseteq \rangle$ is an $\omega$-chain complete partially $\sqsupseteq$-ordered set $A$ with bottom element $\top \in A$ and top element $\bot \in A$ . 
\end{defi}

\change{
Recall that an $\omega$-chain in a complete partially $\sqsubseteq$-ordered (resp. $\sqsupseteq$-ordered) set $A$ is an infinite sequence $a_0 \sqsubseteq a_1 \sqsubseteq a_2  \sqsubseteq \dots$ (resp. $a_0 \sqsupseteq a_1 \sqsupseteq a_2  \sqsupseteq \dots$) and that in such domains all $\omega$-chains have a least upper (resp. greatest lower) bound.
}

With an abuse of notation we sometimes refer to a field domain $\langle A, \sqsubseteq , \bot, \top  \rangle$ with the carrier $A$ and to its components by subscripting them with the carrier, i.e. $\sqsubseteq_A$, $\bot_A$ and $\top_A$. For the sake of a lighter notation we drop the subscripts if clear from the context.

\newcounter{example}
\newenvironment{example}[1][]{\refstepcounter{example}\par\medskip
   \noindent \textbf{Example~\theexample. #1} \rmfamily}{\medskip}
 
\begin{exa}\label{ex:domains}
Some typical examples of field domains are:
\begin{itemize}
\item Boolean and quasi-boolean partially ordered domains such as the classical Boolean domain $\langle \{\mathit{true},\mathit{false}\} , \rightarrow , \mathit{false} , \mathit{true} \rangle$,  Belnap's 4-valued domains, etc.
\item Totally ordered numerical domains such as $\langle A , \leq , 0 , +\infty  \rangle$, 
with $A$ being $\Nat \cup \{ +\infty \}$, $\Real^{+} \cup \{ +\infty \}$, or $\langle [a..b] , \leq , a , b \rangle$ with $a,b \in \Real$ and $a \leq b$,  etc.;
\item Sets with containment relations such as $\langle 2^A,  \subseteq, \emptyset, A\rangle$; 
\item Words with lexicographical relations such as $\langle A^* \cup \{\bullet\} , \sqsubseteq , \epsilon , \bullet  \rangle$, with $A$ being a partially ordered alphabet of symbols, \change{$A^*$ denoting possibly empty sequences of symbols of $A$} and $\sqsubseteq$ being a lexicographical order with the empty word $\epsilon$ as bottom and $\bullet$ as top element (an auxiliary element that dominates all words).
\end{itemize}
\end{exa}

\noindent Many other domains can be constructed by \emph{reversing} the domains of the above example. For example,  $\langle \{\mathit{true},\mathit{false}\} , \leftarrow , \mathit{true} , \mathit{false} \rangle$, $\langle A , \geq , +\infty, 0  \rangle$, $\langle 2^A, \supseteq, A, \emptyset  \rangle$, $\langle (A^* \cup \{\bullet\} , \sqsupseteq , \bullet , \epsilon \rangle$... can be considered as domains as well.
Moreover, additional domains can be constructed by composition of domains, e.g. based on Cartesian products and power domains. \change{The Cartesian product is indeed useful whenever one needs to combine two different domains.}

\begin{mdframed}[style=change]

\begin{defi}[Cartesian product]
Let $\langle A_1, \sqsubseteq_1 , \bot_1, \top_1 \rangle$ and $\langle A_2, \sqsubseteq_2 , \bot_2, \top_2 \rangle$ be two field domains. Their \emph{Cartesian product} $\langle A_1, \sqsubseteq_1 , \bot_1, \top_1 \rangle \times  \langle A_2, \sqsubseteq_2 , \bot_2, \top_2 \rangle$ is the field domain $\langle A_1 \times A_2, \sqsubseteq , (\bot_1,\bot_2 ), (\top_1,\top_2) \rangle$ where $\sqsubseteq$ is defined as $(a_1,a_2) \sqsubseteq (a'_1,a'_2)$ iff $a_1 \sqsubseteq_1 a'_1$ and $a_2 \sqsubseteq_1 a'_2$.
\end{defi}

In some of the examples we shall use a variant of the Cartesian product where pairs of values are ordered lexicographically, \changeR{corresponding to the case in which there is a priority between the two dimensions being combined.}

\begin{defi}[Lexicographical Cartesian product]
Let $\langle A_1, \sqsubseteq_1 , \bot_1, \top_1 \rangle$ and $\langle A_2, \sqsubseteq_2 , \bot_2, \top_2 \rangle$ be two field domains. Their \emph{lexicographical cartesian product} $\langle A_1, \sqsubseteq_1 , \bot_1, \top_1 \rangle \times_1  \langle A_2, \sqsubseteq_2 , \bot_2, \top_2 \rangle$ is the field domain $\langle A_1 \times A_2, \sqsubseteq , (\bot_1,\bot_2 ), (\top_1,\top_2) \rangle$ where $\sqsubseteq$ is defined as $(a_1,a_2) \sqsubseteq (a'_1,a'_2)$ iff $a_1 \sqsubset_1 a'_1$ or ($a_1 = a'_1$ and $a_2 \sqsubseteq_1 a'_2$).
\end{defi}

Sometimes one needs to deal with sets of non-dominated values, \changeR{for example when considering multi-criteria optimisation problems}. A suitable construction in this case is to consider the Hoare Power Domain~{\cite{DBLP:journals/jcss/Smyth78}}.

\begin{defi}[Hoare Power Domain]
Let $\langle A, \sqsubseteq , \bot, \top \rangle$ be a field domain. \changeR{The} \emph{Hoare Power Domain} $P^H(\langle A, \sqsubseteq , \bot, \top \rangle)$ is the field domain $\langle \{ B \subseteq A \mid a \in B \wedge b \sqsubseteq a \Rightarrow b \in B \} , \subseteq, \emptyset, A \rangle$.
\end{defi}

In words, the obtained domain contains downward closed sets of values, ordered by set inclusion.

\end{mdframed}

\medskip 
Our agents will use arbitrary functions to operate on attributes and to coordinate. 
Among other things, agents will compute (least or greatest) fixpoints of functions on field domains.  
 Of course, for fixpoints to be well-defined some restrictions may need to be imposed, in particular regarding monotonicity and continuity properties.

A sufficient condition for the least and greatest fixpoints of a function $f : A \rightarrow A$ on an $\omega$-chain complete field domain $A$ to be well-defined is for $f$ to be continuous and monotone. 
\change{Recall that} our domains \change{are} such that all infinite chains of partially ordered (respectively reverse-ordered) elements have a \change{least} upper bound (respectively a greatest lower bound). 
Indeed, under such conditions the least upper bound of the chain $\bot \sqsubseteq f \, \bot \sqsubseteq f^2\, \bot \sqsubseteq \dots$ is the least fixpoint of $f$.
Similarly, the greatest lower bound of chain \change{$\top \sqsupseteq  f \,\top \sqsupseteq f^2\, \top \sqsupseteq \dots$} is the greatest fixpoint of $f$.

Another desirable property is for fixpoints to be computable by iteration. This means that the least and greatest fixpoints of $f$ are equal to $f^n\, \top$ and $f^m\, \bot$, respectively, for some $n,m \in \Nat$.
In some cases, we will indeed require that all chains of partially ordered elements are finite. 
In that case we say that the chains \emph{stabilize}, which refers to the fact that, for example, $f^n\, \bot = f^{n+k}\, \bot$, for all $k \in \Nat$. This guarantees that the computation of a fixpoint by successive approximations eventually terminates \change{since every iteration corresponds to an element in the chain}. If this is not the case the fixpoint can only be approximated or solved with some alternative method that may depend on the concrete field domain and the class of functions under consideration. 
\medskip

To guarantee some of those properties, we will often instantiate our approach on algebraic structures based on a class of semirings called \emph{constraint semirings} (just \emph{semirings} in the following). 
Such class of semirings has been shown to be very flexible, expressive and convenient for a wide range of problems, in particular for optimisation and solving in problems with soft constraints and multiple criteria~\cite{DBLP:journals/jacm/BistarelliMR97}. 

\begin{defi}[semiring]
A \emph{semiring} is a tuple $\langle A, \boldsymbol{+}, \times, \bot, \top \rangle$ composed by a set $A$, two operators $\boldsymbol{+}$, $\times$ and two constants $\bot$, $\top$ such that: 
\begin{itemize}
\item $\boldsymbol{+} : 2^A \rightarrow A$ is an associative, commutative, idempotent operator to ``choose'' among values;
\item $\times : A \times A \rightarrow A$ is an associative, commutative operator  to ``combine'' values;
\item $\times$ distributes over $\boldsymbol{+}$;
\item $\bot \boldsymbol{+} a = a$, $\top \boldsymbol{+} a = \top$, $\top \times a = a$, $\bot \times a = \bot$ for all $a \in A$;
\item $\sqsubseteq$, which is defined as $a\sqsubseteq b$ iff $a+b=b$, provides a field domain of preferences $\langle A , \sqsubseteq , \bot, \top \rangle$ (which is actually a complete lattice~\cite{DBLP:journals/jacm/BistarelliMR97}).  
\end{itemize}
\end{defi}

\change{
\noindent Recall that \emph{classical} semirings are algebraic structures that are more general than the (constraint) semirings we consider here. In fact, classical semirings do not require the additive operation $\boldsymbol{+}$ to be idempotent or the multiplicative operation $\times$ to be commutative. Such axiomatic properties, however, turn out to yield many useful and interesting features (e.g. in constraint solving~{\protect{\cite{DBLP:journals/jacm/BistarelliMR97}}} and model checking~{\protect{\cite{DBLP:journals/tcs/Lluch-LafuenteM05}}}) and are actually provided by many semirings, such as the ones in Example~{\protect{\ref{ex:semirings}}}.
}

Again, we shall use the notational convention for semirings that we mentioned for field domains, i.e. we sometimes denote a semiring by its carrier $A$ and the rest of the components by subscripting them with $A$. 
Note also that since the underlying field domain of \changeR{a semiring} is a complete lattice, all partially ordered chains have \change{least upper \changeR{and greatest lower} bounds}.

\begin{exa}\label{ex:semirings}
Typical examples of semirings are:
\begin{itemize}
%
\item the Boolean semiring $\langle  \{\mathit{true},\mathit{false}\}, \vee, \wedge, \mathit{false}, \mathit{true} \rangle$; 
\item the tropical semiring $\langle \Real^+ \cup \{+\infty\},\emph{min},+,+\infty,0 \rangle$;
%
%
\item the possibilistic semiring:  $\langle [0..1], \emph{max}, \cdot, 0, 1 \rangle$;
\item the fuzzy semiring $\langle [0..1],  \emph{max}, \emph{min}, 0, 1 \rangle$;
\item and the set semiring $\langle 2^A,  \cup, \cap, \emptyset, A \rangle$\changeR{.}
\end{itemize}
All these examples have an underlying domain that can be found among the examples of field domains in Example~\ref{ex:domains}. As for domains, additional semirings can be obtained in some cases by reversing the underlying order. For instance, $\langle  \{\mathit{true},\mathit{false}\}, \wedge, \vee, \mathit{true} , \mathit{false} \rangle$, $\langle [0..1],  \emph{min}, \emph{max}, 1, 0 \rangle$, ... are semirings as well. A useful property of semirings is that Cartesian products and power constructions yield semirings, which allows one for instance to lift techniques for single criteria to multiple criteria. 
\end{exa}

%
%



\section{\smuc: A Soft $\mu$-calculus for Computations fields}\label{section:calculus-global}


\subsection{Graph-based Fields}

We are now ready to provide our notion of field, which is essentially a \changeR{fixed} graph equipped with field-domain-valued node and edge labels. %
The idea is that nodes play the role of agents, and (directed) edges play the role of (directional) connections. 
\change{Labels in the graph are of two different natures.} Node labels are used as \change{the names of} attributes of the agents. \change{On the other hand,} edge labels correspond to functions associated to the connections, e.g. representing how attribute values are transformed when traversing a connection. 

\begin{defi}[field]
A \emph{field} is a tuple $\langle N, E, A, L = L_N \uplus L_E, I = I_N \uplus I_E\rangle$ formed by 
\begin{itemize}
\item a set $N$ of nodes;
\item a relation $E \subseteq N \times N$ of edges;
\item a set $L$ of node labels $L_N$ and edge labels $L_E$; 
\item a field domain $\langle A , \sqsubseteq , \bot , \top \rangle$; 
\item an interpretation function $I_N: L_N \rightarrow N \rightarrow A$ associating a function from nodes to values to every node label in $L_N$; 
\item an interpretation function $I_E: L_E \rightarrow E \rightarrow A \rightarrow A$ 
associating a function from edges to functions from values to values to every edge label in $P$;
\end{itemize}
\noindent
where node, edge, and label sets are drawn from a corresponding universe, i.e. $N \subseteq \mathcal{N}$, $E \subseteq \mathcal{E}$, $L_N \subseteq \mathcal{L}$, $L_E \subseteq \mathcal{L}'$. 
\end{defi}

As usual, we may refer to the components of a field $F$ using subscripted symbols (i.e. $N_F$, $E_F$, \dots). We will denote the set of all fields by $\mathcal{F}$. 

%
It is worth to remark that while standard notions of computational fields tend to be restricted to nodes (labels) and their mapping to values, our notion of field includes the topology of the network and the mapping of edge (labels) to functions. As a matter of fact, the topology plays a fundamental role in our field computations as it defines how agents are connected and how their attributes are combined when communicated. \change{Note that we consider directed edges since there are many cases in which the direction of the connection matters as we shall see in applications based on spanning trees or shortest paths.} 
On the other hand,  the role of node and edge labels is different \changeR{in our approach}. In fact, some node labels are computed as the result of a fixpoint approximation which corresponds to a propagation procedure. They thus represent the genuine computational fields. Edge labels, instead, are assigned directly in terms of the data of the problem (e.g. distances) or in terms of the results of previous propagations. They thus represent more properly equation coefficients and boundary conditions as one can have in \emph{partial differential equations} in physical fields.


\subsection{\smuc{} Formulas}

\smuc{} (\emph{Soft $\mu$-calculus for Computations fields}) is meant to specify global computations on fields. One key aspect of our calculus are atomic computations denoted with expressions reminiscent of the semiring modal $\mu$-calculus proposed in~\cite{DBLP:journals/tcs/Lluch-LafuenteM05}.
\change{
The semiring $\mu$-calculus departed from the modal $\mu$-calculus, a very flexible and expressive calculus that subsumes other modal temporal logics such as CTL* (and hence also CTL and LTL). The semiring $\mu$-calculus inherits essentially the same syntax as the modal $\mu$-calculus (i.e. predicate logic enriched with temporal operators and fixpoint operators) but changes the domain of interpretation from Booleans (i.e. set of states that satisfy a formula) to semiring valuations (i.e. mappings of states to semiring values), and the semantic interpretation of operators, namely disjunction and existential quantification are interpreted as the semiring addition, while conjunction and universal quantification are interpreted as semiring multiplication. In that manner, the semiring $\mu$-calculus captures the ordinary  $\mu$-calculus for the Boolean semiring but, in addition, allows one to reason about quantitative properties of graph-based structures like transition systems (i.e. quantitative model checking) and network topologies (e.g. shortest paths and similar properties).  
}

In \smuc{} similar expressions will be used to specify the functions being calculated by global computations, to be recorded by updating the interpretation functions of the nodes.  

Given a field domain $A$, we shall call functions $f,g,\dots$ on $A$  \emph{attribute} operations. Functions $f : A^* \rightarrow A$ will be used to combine values, while functions $g: \mathit{mset}(A) \rightarrow A$ will be used to aggregate values, where $\mathit{mset}(A)$ denotes the domain of finite multisets on $A$.  The latter hence have finite multisets of $A$-elements as domain. The idea is that they are going to be used to aggregate values from neighbour nodes using associative and commutative functions, so that the order of the arguments does not matter. 
A function $N \rightarrow A$ is called a \emph{node valuation} that we typically range over by $\mathsf{f},\mathsf{g},\dots$. 
Note that we use the same symbols but a different font since we sometimes lift an attribute operation to a set of nodes. For instance a zero-adic attribute operation $f :\ \rightarrow A$ can be lifted to a node valuation $\mathsf{f} : N \rightarrow A$ in the obvious way, i.e. $\mathsf{f} = \lambda n . f$. 
A function $\psi : (N \rightarrow A) \rightarrow N \rightarrow A$ is called an \emph{update function} and is typically ranged over by $\psi, \psi_1, \psi_2, \dots$. 
As we shall see, the computation of fixpoints of such update functions is at the core of our approach.
Such fixpoints do not refer to functions on the field domain of attribute values $\langle A , \sqsubseteq_A, \bot_A, \top_A \rangle$ but to the field domain of node valuations $\langle (N \rightarrow A) , \sqsubseteq_{N \rightarrow A} , \bot_{N \rightarrow A}, \top_{N \rightarrow A} \rangle$. That field domain is obtained by lifting  $\langle A , \sqsubseteq_A, \bot_A, \top_A \rangle$ to set $N$, i.e. the carrier of the new field domain is the set of node valuations $N \rightarrow A$, the partial ordering relation $\sqsubseteq_{N \rightarrow A}$ is such that $\mathsf{f}_1 \sqsubseteq_{N \rightarrow A} \mathsf{f}_2$ iff $\forall n . \mathsf{f}_1\, n \sqsubseteq_A \mathsf{f}_2\, n$ and the bottom and top elements $\bot_{N \rightarrow A}$, $\top_{N \rightarrow A}$ are such that  $\bot_{N \rightarrow A}\, n=\bot_A$ and $\top_{N \rightarrow A}\, n=\top_A$. 
%

\medskip 
Given a set $\mathcal{Z}$ of formula variables, an environment is a partial function $\rho : \mathcal{Z} \rightarrow \mathcal{N} \rightarrow A$. We shall also use a set $\mathcal{M}$ of function symbols for attribute operations, of functional types $f: A^* \rightarrow A$ for combining values or \change{$g: \mathit{mset}(A) \rightarrow A$} for aggregating values.  

\begin{defi}[syntax of \smuc{} formulas] \label{def:formulas}
The syntax of \smuc{} formulas is as follows:
\[
\Psi ::= i \mid z \mid 
%
f(\Psi,\dots,\Psi) 
\mid \aggregateOut{$\alpha$}{g}{\Psi}
\mid \aggregateIn{$\alpha$}{g}{\Psi}
\mid \mu z.\Psi \mid \nu z.\Psi
\]
\noindent 
with $i \in \mathcal{L}$, $\alpha \in \mathcal{L}'$, $f , g \in \mathcal{M}$ and $z \in \mathcal{Z}$. 
\end{defi}

The formulas allow one to combine atomic node labels $i$, functions $f$, \change{classical least ($\mu$) and greatest ($\nu$) fixpoint operators} and the modal operators $\aggregateOut{}{}{}$ and $\aggregateIn{}{}$.
\changeR{Including both least and greatest fixpoints is needed since we consider cases in which it is not always possible to express one in terms of the other. It is also useful to consider both since they provide two different ways of describing computations: recursive (in the case of least fixpoints) and co-recursive (in the case of greatest fixpoints). The operational view of formulas can provide a useful intuition of when to use least or greatest fixpoints. Informally, least fixpoints are useful when we conceive the computation being described as starting from none or few information that keeps being accumulated until enough (a fixpoint) is reached. The typical such property in a graph is the reachability of a node satisfying some property. Conversely, the computation corresponding to a greatest fixpoint starts with a lot (possibly irrelevant) amount of information that keeps being refined until no irrelevant information is present. The typical such property in a graph is the presence of an infinite path where all nodes have some feature.
Usually infinite paths can be easily represented when they traverse finite cycles in the graph, but in some practical cases the greatest fixpoint approach may be difficult to implement when it requires a form of global information to be available to all nodes. 
} 
The modal operators are used to \change{aggregate (with function $g$) values  obtained from neighbours} following outgoing ($\aggregateOut{}{}{}$) or incoming ($\aggregateIn{}{}{}$) edges and using the edge capability $\alpha$ \changeR{(i.e. the function transforming values associated to label $\alpha$)}. We sometimes use $\mathit{id}$ as the identity edge capability and abbreviate $\aggregateOut{\it{id}}{}{}$ and $\aggregateIn{\it{id}}{}{}$ with $\aggregateOut{}{}{}$ and $\aggregateIn{}{}{}$, respectively. 
 The choice of the symbol is reminiscent of modal temporal logics with past operators.

If we choose a semiring as our field domain, the set of function symbols may include, among others, the semiring operator symbols $\boldsymbol{+}$ and $\times$ and possibly some additional ones, for which an interpretation on the semiring of interest can be given. 
In that case, for instance we can instantiate modal operators $\aggregateOut{}{}{}$ and $\aggregateIn{}{}{}$ to ``choose'' or ``combine'' values from neighbour nodes as \change{we did} in~\cite{DBLP:conf/coordination/Lluch-LafuenteL15}, i.e. \change{by using box and diamond operators} $[ \alpha ] \Psi \equiv \aggregateOut{$\alpha$}{\mathbf{+}}{\Psi}$, $\langle \alpha \rangle \Psi \equiv \aggregateOut{$\alpha$}{\times}{\Psi}$, $[[ \alpha ]] \Psi \equiv \aggregateIn{$\alpha$}{\mathbf{+}}{\Psi}$ and $\langle\langle \alpha \rangle\rangle \Psi \equiv \aggregateIn{$\alpha$}{\times}{\Psi}$. 
\change{
Those operators were inspired by classical operators of modal logics: in modal logics the box ($\Box$) and diamond ($\Diamond$) modalities are used to universally or existentially quantify among all possible next worlds, which amounts to aggregate values with logical conjunction and disjunction, respectively. Generalising conjunction and disjunction to the multiplicative and additive operations of a semiring yields the modalities in~{\protect{\cite{DBLP:conf/coordination/Lluch-LafuenteL15}}}.
} 

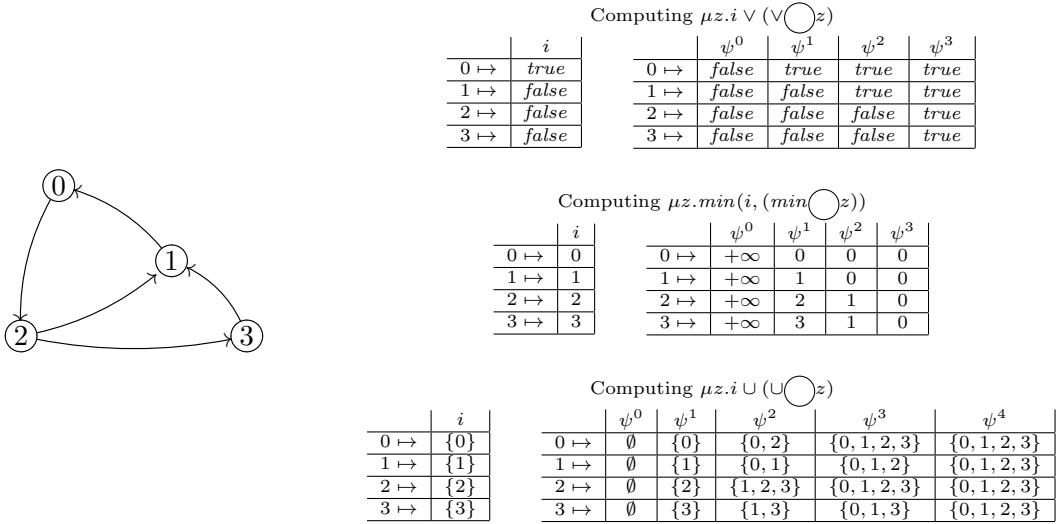
\begin{figure}
\begin{center}
\begin{minipage}{0.29\textwidth}
\begin{mdframed}[style=change]
\[
\begin{matrix}
\begin{xy}
(5,40)*{0}*\cir<6pt>{}="0";
(20,30)*{1}*\cir<6pt>{}="1";
(0,20)*{2}*\cir<6pt>{}="2";
(30,20)*{3}*\cir<6pt>{}="3";
\ar@{->}@/_0.4pc/ "0"; "2"
\ar@{->}@/_0.4pc/ "2"; "1"
\ar@{->}@/_0.4pc/ "2"; "3"
\ar@{->}@/_0.4pc/ "3"; "1"
\ar@{->}@/_0.4pc/ "1"; "0"
\end{xy}
\end{matrix}
\]
\end{mdframed}
\end{minipage}
\begin{minipage}{0.7\textwidth}
\begin{mdframed}[style=change]
{\tiny \centering

Computing $\mu z . i \vee (\aggregateOut{}{\mathbf{\vee}}{z})$

\begin{tabular}{c|c|}
                     & $i$ \\\hline
$0 \mapsto$ & $\mathit{true}$  \\\hline
$1 \mapsto$ & $\mathit{false}$  \\\hline
$2 \mapsto$ & $\mathit{false}$  \\\hline
$3 \mapsto$ & $\mathit{false}$  \\\hline
\end{tabular}
\hspace{0.5cm}
\begin{tabular}{c|c|c|c|c|c|c}
                     & $\psi^0$ & $\psi^1$  & $\psi^2$  & $\psi^3$   \\\hline
$0 \mapsto$ & $\mathit{false}$ & $\mathit{true}$ & $\mathit{true}$ & $\mathit{true}$  \\\hline
$1 \mapsto$ & $\mathit{false}$ & $\mathit{false}$ & $\mathit{true}$ & $\mathit{true}$ \\\hline
$2 \mapsto$ & $\mathit{false}$ & $\mathit{false}$ & $\mathit{false}$ & $\mathit{true}$ \\\hline
$3 \mapsto$ & $\mathit{false}$ & $\mathit{false}$ & $\mathit{false}$ & $\mathit{true}$ \\\hline
\end{tabular}

\vspace{0.5cm}
Computing $\mu z . \mathit{min}(i, (\aggregateOut{}{\mathit{min}}{z}))$

\begin{tabular}{c|c|}
                     & $i$ \\\hline
$0 \mapsto$ & $0$  \\\hline
$1 \mapsto$ & $1$  \\\hline
$2 \mapsto$ & $2$  \\\hline
$3 \mapsto$ & $3$  \\\hline
\end{tabular}
\hspace{0.5cm}
\begin{tabular}{c|c|c|c|c|c|c}
                     & $\psi^0$ & $\psi^1$  & $\psi^2$  & $\psi^3$   \\\hline
$0 \mapsto$ & $\mathit{+\infty}$ & $0$ & $0$ & $0$  \\\hline
$1 \mapsto$ & $\mathit{+\infty}$ & $1$ & $0$ & $0$  \\\hline
$2 \mapsto$ & $\mathit{+\infty}$ & $2$ & $1$ & $0$  \\\hline
$3 \mapsto$ & $\mathit{+\infty}$ & $3$ & $1$ & $0$  \\\hline
\end{tabular}

\vspace{0.5cm}
Computing $\mu z . i \cup (\aggregateOut{}{\mathbf{\cup}}{z})$

\begin{tabular}{c|c|}
                     & $i$ \\\hline
$0 \mapsto$ & $\{0\}$  \\\hline
$1 \mapsto$ & $\{1\}$  \\\hline
$2 \mapsto$ & $\{2\}$  \\\hline
$3 \mapsto$ & $\{3\}$  \\\hline
\end{tabular}
\hspace{0.5cm}
\begin{tabular}{c|c|c|c|c|c|c}
                     & $\psi^0$ & $\psi^1$  & $\psi^2$  & $\psi^3$ & $\psi^4$  \\\hline
$0 \mapsto$ & $\emptyset$ & $\{0\}$ & $\{0,2\}$ & $\{0,1,2,3\}$ & $\{0,1,2,3\}$  \\\hline
$1 \mapsto$ & $\emptyset$ & $\{1\}$ & $\{0,1\}$ & $\{0,1,2\}$  & $\{0,1,2,3\}$ \\\hline
$2 \mapsto$ & $\emptyset$ & $\{2\}$ & $\{1,2,3\}$ & $\{0,1,2,3\}$ & $\{0,1,2,3\}$  \\\hline
$3 \mapsto$ & $\emptyset$ & $\{3\}$ & $\{1,3\}$ & $\{0,1,3\}$ & $\{0,1,2,3\}$ \\\hline
\end{tabular}

}
\end{mdframed}
\end{minipage}
\end{center}
\caption{\change{The underlying graph of a field (left) and the computation of three \smuc{} formulas in detail (right).}}
\label{fig:formulas1}
\end{figure}

A set of typical example formulas can be obtained by instantiating the simple pattern formula $\mu z . i \sqcup (\aggregateOut{}{\mathbf{\sqcup}}{z})$ in different domains that happen to be complete lattices, such as the ones in Example~\ref{ex:domains} and~\ref{ex:semirings}, where $\sqcup$ is a well-defined join operation. %

\begin{exa} \label{ex:formulas1}
Formula $\mchange{\mu z . i \vee (\aggregateOut{}{\mathbf{\vee}}{z})}$ is equivalent to the temporal property \emph{``eventually $i$''} if we use the Boolean domain. Indeed \change{the formula} amounts to the fixpoint characterization of the \change{Computation Tree Logic (CTL)} formula $\mathbf{EF} i$ \change{stating that, starting from the current state (represented as a node of the graph), there is an execution path in the transition system (represented as a graph) and some state on that path that has property $i$}.  Other well-known temporal properties can be similarly obtained and used for agents to check, for example, complex reachability properties.
\changeR{Recall that the entire CTL and CTL* temporal logic languages can be encoded in the $\mu$-calculus. 
For example, the CTL formula $\mathbf{AG} i$ stating that 
\emph{``starting from the current state (represented as a node of the graph), all execution paths in the transition system (represented as a graph) are such that all states along those
paths have property $i$''} can be easily expressed as} 
$\mchangeR{\nu z . i \wedge (\aggregateOut{}{\mathbf{\wedge}}{z})},$
\changeR{ and similarly for other properties.
}
\end{exa}

\begin{exa} \label{ex:formulas2}
Formula $\mchange{\mu z . \mathit{min}(i, (\aggregateOut{}{\mathit{min}}{z}))}$ yields the minimal value in a totally ordered numerical domain, like the tropical semiring. This can be used by agents to discover the best value for some attribute. A typical example could be the discovery of a leader agent, in case totally ordered agent identifiers are used. 
\changeR{In the same setting the maximal value could be obtained with} $\mchangeR{\nu z . \mathit{max}(i, (\aggregateOut{}{\mathit{max}}{z}))}$.

\end{exa}

\begin{exa} \label{ex:formulas3}
In a set-based domain, $\mchange{\mu z . i \cup (\aggregateOut{}{\mathbf{\cup}}{z})}$ can provide the union of all elements in a graph. For example, if the set $A$ \change{coincides with the set $N$ of nodes} and $i$ records \change{each node's identifier as a singleton}, then \change{the formula} can be used for \change{each node to compute the set of nodes that it can reach.} 
%
\changeR{Assume now that $i$ records some arbitrary set, say the set of nodes that every node happens to know. Then the formula} $\mchangeR{\nu z . i \cap (\aggregateOut{}{\mathbf{\cap}}{z})}$ \changeR{can be used to compute the set of nodes that every node knows.} 
\end{exa}

\change{The computation of some of the above formulas can be found in 
Fig.~{\ref{fig:formulas1}}. The figure includes a simple graph (underlying a field), the instance of each formula $\psi$ on the considered domain of interpretation, and the details of the computation of each of the formulas. For each computation a table is used to represent the value of $i$ on each node and the evaluation of the (fixpoint) formula by iteration, as a sequence $\psi^0, \psi^1, \psi^2,\dots$}

\begin{figure}
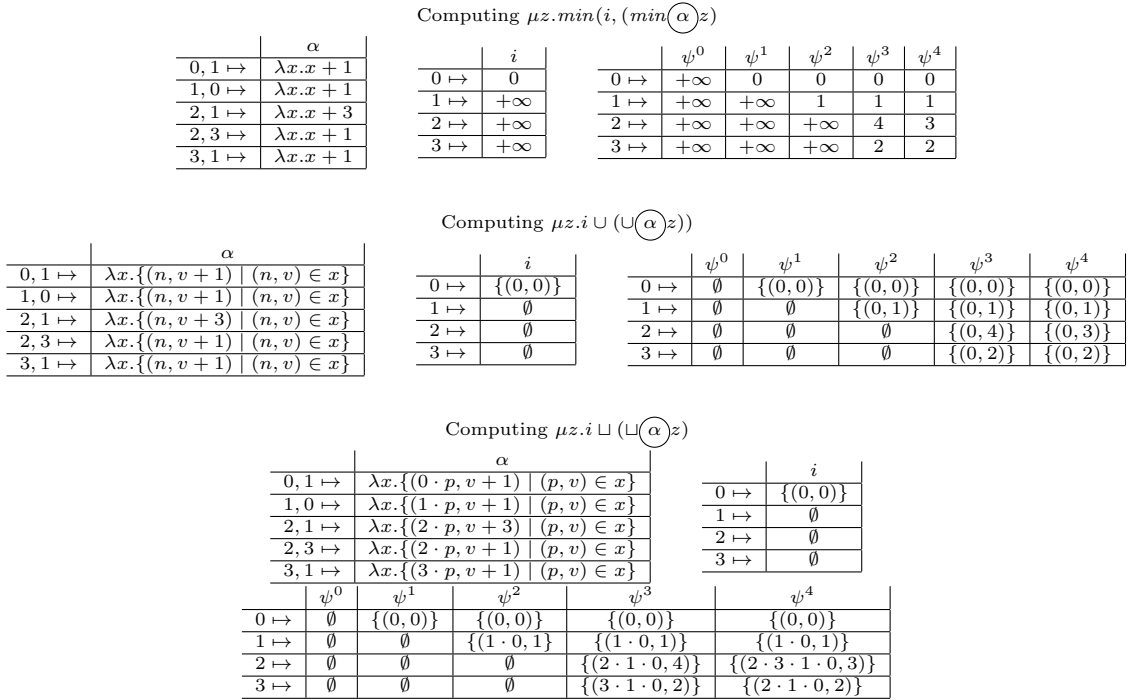

\begin{center}
\begin{mdframed}[style=change]
{\tiny \centering

Computing $\mu z . \mathit{min} (i ,  (\aggregateOut{$\alpha$}{\mathit{min}}{z})$

\begin{tabular}{c|c|}
                     & $\alpha$ \\\hline
$0,1 \mapsto$ & $\lambda x . x + 1$  \\\hline
$1,0 \mapsto$ & $\lambda x . x + 1$  \\\hline
$2,1 \mapsto$ & $\lambda x . x + 3$  \\\hline
$2,3 \mapsto$ & $\lambda x . x + 1$  \\\hline
$3,1 \mapsto$ & $\lambda x . x + 1$  \\\hline
\end{tabular}
\hspace{0.5cm}
\begin{tabular}{c|c|}
                     & $i$ \\\hline
$0 \mapsto$ & $0$  \\\hline
$1 \mapsto$ & $+\infty$  \\\hline
$2 \mapsto$ & $+\infty$  \\\hline
$3 \mapsto$ & $+\infty$  \\\hline
\end{tabular}
\hspace{0.5cm}
\begin{tabular}{c|c|c|c|c|c|c}
                     & $\psi^0$ & $\psi^1$  & $\psi^2$  & $\psi^3$  & $\psi^4$   \\\hline
$0 \mapsto$ & $+\infty$ & $0$ & $0$ & $0$  & $0$ \\\hline
$1 \mapsto$ & $+\infty$ & $+\infty$ & $1$ & $1$  & $1$ \\\hline
$2 \mapsto$ & $+\infty$ & $+\infty$ & $+\infty$ & $4$  & $3$ \\\hline
$3 \mapsto$ & $+\infty$ & $+\infty$ & $+\infty$ & $2$  & $2$ \\\hline
\end{tabular}

\vspace{0.5cm}
Computing $\mu z . i \cup (\aggregateOut{$\alpha$}{\cup}{z}))$

\begin{tabular}{c|c|}
                     & $\alpha$ \\\hline
$0,1 \mapsto$ & $\lambda x . \{ (n,v+1) \mid (n,v) \in x \}$  \\\hline
$1,0 \mapsto$ & $\lambda x . \{ (n,v+1) \mid (n,v) \in x \}$  \\\hline
$2,1 \mapsto$ & $\lambda x . \{ (n,v+3) \mid (n,v) \in x \}$  \\\hline
$2,3 \mapsto$ & $\lambda x . \{ (n,v+1) \mid (n,v) \in x \}$  \\\hline
$3,1 \mapsto$ & $\lambda x . \{ (n,v+1) \mid (n,v) \in x \}$  \\\hline
\end{tabular}
\hspace{0.5cm}
\begin{tabular}{c|c|}
                     & $i$ \\\hline
$0 \mapsto$ & $\{(0,0)\}$  \\\hline
$1 \mapsto$ & $\emptyset$  \\\hline
$2 \mapsto$ & $\emptyset$  \\\hline
$3 \mapsto$ & $\emptyset$  \\\hline
\end{tabular}
\hspace{0.5cm}
\begin{tabular}{c|c|c|c|c|c|c}
                     & $\psi^0$ & $\psi^1$  & $\psi^2$  & $\psi^3$  & $\psi^4$  \\\hline
$0 \mapsto$ & $\emptyset$ & $\{(0,0)\}$ & $\{(0,0)\}$ & $\{(0,0)\}$  & $\{(0,0)\}$ \\\hline
$1 \mapsto$ & $\emptyset$ & $\emptyset$ & $\{(0,1)\}$ & $\{(0,1)\}$  & $\{(0,1)\}$ \\\hline
$2 \mapsto$ & $\emptyset$ & $\emptyset$ & $\emptyset$ & $\{(0,4)\}$  & $\{(0,3)\}$ \\\hline
$3 \mapsto$ & $\emptyset$ & $\emptyset$ & $\emptyset$ & $\{(0,2)\}$  & $\{(0,2)\}$ \\\hline
\end{tabular}

\vspace{0.5cm}
Computing $\mu z . i \sqcup (\aggregateOut{$\alpha$}{\mathbf{\sqcup}}{z}) $

\begin{tabular}{c|c|}
                     & $\alpha$ \\\hline
$0,1 \mapsto$ & $\lambda x . \{ (0 \cdot p,v+1) \mid (p,v) \in x \}$  \\\hline
$1,0 \mapsto$ & $\lambda x . \{ (1 \cdot p,v+1) \mid (p,v) \in x \}$  \\\hline
$2,1 \mapsto$ & $\lambda x . \{ (2 \cdot p,v+3) \mid (p,v) \in x \}$  \\\hline
$2,3 \mapsto$ & $\lambda x . \{ (2 \cdot p,v+1) \mid (p,v) \in x \}$  \\\hline
$3,1 \mapsto$ & $\lambda x . \{ (3 \cdot p,v+1) \mid (p,v) \in x \}$  \\\hline
\end{tabular}
\hspace{0.5cm}
\begin{tabular}{c|c|}
                     & $i$ \\\hline
$0 \mapsto$ & $\{(0,0)\}$  \\\hline
$1 \mapsto$ & $\emptyset$  \\\hline
$2 \mapsto$ & $\emptyset$  \\\hline
$3 \mapsto$ & $\emptyset$  \\\hline
\end{tabular}

\begin{tabular}{c|c|c|c|c|c|c}
                     & $\psi^0$ & $\psi^1$  & $\psi^2$  & $\psi^3$ & $\psi^4$  \\\hline
$0 \mapsto$ & $\emptyset$ & $\{(0,0)\}$ & $\{(0,0)\}$ & $\{(0,0)\}$  & $\{(0,0)\}$ \\\hline
$1 \mapsto$ & $\emptyset$ & $\emptyset$ & $\{(1 \cdot 0,1\}$ & $\{(1 \cdot 0,1)\}$  & $\{(1 \cdot 0,1)\}$ \\\hline
$2 \mapsto$ & $\emptyset$ & $\emptyset$ & $\emptyset$ & $\{(2 \cdot 1 \cdot 0,4)\}$  & $\{(2 \cdot 3 \cdot 1 \cdot 0,3)\}$ \\\hline
$3 \mapsto$ & $\emptyset$ & $\emptyset$ & $\emptyset$ & $\{(3 \cdot 1 \cdot 0,2)\}$  & $\{(2 \cdot 1 \cdot 0,2)\}$ \\\hline
\end{tabular}

}
\end{mdframed}
\end{center}
\caption{\change{Computation of \smuc{} optimal path formulas in detail: cost of the optimal path (above), optimal cost and actual goal node (center), optimal path and its cost (bottom).}}
\label{fig:formulas2}
\end{figure}

Another interesting family of properties can be obtained with the slightly extended pattern formula $\mu z . i \sqcup (\aggregateOut{$\alpha$}{\mathbf{\sqcup}}{z})$. This pattern formula can be used for several shortest-path related properties, considering $i$ to be a label providing information related to each node being or not a goal node and $\alpha$ providing a function that takes care of composing the cost of traversing an edge. 

\begin{exa} \label{ex:shortestpathformulas1}
Considering $\langle \Real^{+} \cup \{ +\infty \}, \geq , +\infty, 0  \rangle$ as domain, $i$ to yield $0$ for goal nodes and $+\infty$ for the rest of the nodes, and $\alpha$ being a function that adds the cost of traversing an edge, \change{formula} $\mchange{\mu z . \mathit{min} (i ,  (\aggregateOut{$\alpha$}{\mathit{min}}{z})}$ yields the shortest distance to a goal node.
\end{exa}

\begin{exa} \label{ex:shortestpathformulas2}
The actual sets of reachable goal nodes with their distances can be obtained if we consider \change{the evaluation of } $\mchange{\mu z . i \cup (\aggregateOut{$\alpha$}{\cup}{z}))}$ \change{under the Hoare power domain of the Cartesian product of nodes and costs, i.e. $P^H(N \times (\Real^{+} \cup \{ +\infty \}))$.} 
\change{In words, the domain consists of non-dominated sets of pairs (node,cost)}. 
In this case $i$ should be $\{(n,0)\}$ for every goal node $n$ and $\emptyset$ for the rest of the nodes, and $\alpha$ should be similar as before (pointwise applied to every pair, only on the second component of each pair).
\end{exa}

\begin{exa} \label{ex:shortestpathformulas3}
The actual set of paths to the goal nodes can be obtained in a similar way, by considering \change{formula } $\mchange{\mu z . i \sqcup (\aggregateOut{$\alpha$}{\mathbf{\sqcup}}{z})}$ \change{ under the Hoare power domain of the Cartesian product of paths and costs, i.e. $P^H((N^* \cup \{ \bullet \}) \times (\Real^{+} \cup \{ +\infty \}))$}. 
\change{In words, the domain consists of non-dominated sets of pairs (path,cost)}. In this case $i$ should be $\{(n,0)\}$ for every goal node $n$ and $\emptyset$ for the rest of the nodes, and $\alpha$ should be such that $\alpha(n,n')$ is a function that prefixes $n$ to a path. \change{Since the Hoare power domain deals with non-dominated paths, loops (that would require special treatment with an ordinary power construct) are implicitly dealt (i.e. extending a set of paths can only consist of adding non-dominated paths and loops can only worsen the cost of existing ones).}
\end{exa}

\change{The computation of the above formulas can be found in Fig.~{\ref{fig:formulas2}}, which follows a similar schema as  Fig.~{\ref{fig:formulas1}} but includes in addition the interpretation of edge label $\alpha$.}




\begin{figure}
\begin{center}
\begin{minipage}{0.29\textwidth}
\begin{mdframed}[style=change]
\[
\begin{matrix}
\begin{xy}
(5,40)*{0}*\cir<6pt>{}="0";
(20,30)*{1}*\cir<6pt>{}="1";
(0,20)*{2}*\cir<6pt>{}="2";
(30,20)*{3}*\cir<6pt>{}="3";
\ar@{->}@/_0.4pc/ "0"; "2"
\ar@{->}@/_0.4pc/ "2"; "1"
\ar@{->}@/_0.4pc/ "2"; "3"
\ar@{->}@/_0.4pc/ "3"; "1"
\ar@{->}@/_0.4pc/ "1"; "0"
\end{xy}
\end{matrix}
\]
\end{mdframed}
\end{minipage}
\begin{minipage}{0.7\textwidth}
\begin{mdframed}[style=change]
{\tiny \centering
Computing $\mu Z . \mathsf{min_1}(\mathsf{i}, \aggregateOut{$\alpha$}{\tiny\mathsf{min_1}}{Z})$

\begin{tabular}{c|c|}
                     & $\alpha$ \\\hline
$0,1 \mapsto$ & $\lambda x . \{ (0 \cdot p,v+1) \mid (p,v) \in x \}$  \\\hline
$1,0 \mapsto$ & $\lambda x . \{ (1 \cdot p,v+1) \mid (p,v) \in x \}$  \\\hline
$2,1 \mapsto$ & $\lambda x . \{ (2 \cdot p,v+3) \mid (p,v) \in x \}$  \\\hline
$2,3 \mapsto$ & $\lambda x . \{ (2 \cdot p,v+1) \mid (p,v) \in x \}$  \\\hline
$3,1 \mapsto$ & $\lambda x . \{ (3 \cdot p,v+1) \mid (p,v) \in x \}$  \\\hline
\end{tabular}
\hspace{0.5cm}
\begin{tabular}{c|c|}
                     & $\mathit{i}$ \\\hline
$0 \mapsto$ & $(0,0)$  \\\hline
$1 \mapsto$ & $(1,+\infty)$  \\\hline
$2 \mapsto$ & $(2,+\infty)$  \\\hline
$3 \mapsto$ & $(3,+\infty)$  \\\hline
\end{tabular}
\hspace{0.5cm}
\begin{tabular}{c|c|c|c|c|c|c}
                     & $\psi^0$ & $\psi^1$  & $\psi^2$  & $\psi^3$ & $\psi^4$  \\\hline
$0 \mapsto$ & $(+\infty,+\infty)$ & $(0,0)$ & $(0,0)$ & $(0,0)$  & $(0,0)$ \\\hline
$1 \mapsto$ & $(+\infty,+\infty)$ & $(1,+\infty)$ & $(0,1)$ & $(1 \cdot 0,1)$  & $(0,1)$ \\\hline
$2 \mapsto$ & $(+\infty,+\infty)$ & $(2,+\infty)$ & $(2,+\infty)$ & $(1,4)$  & $(3,3)$ \\\hline
$3 \mapsto$ & $(+\infty,+\infty)$ & $(3,+\infty)$ & $(3,+\infty)$ & $(1,2)$  & $(1,2)$ \\\hline
\end{tabular}

}
\end{mdframed}
\end{minipage}
\end{center}
\caption{\change{Computation of shortest path spanning tree.}}
\label{fig:mst}
\end{figure}
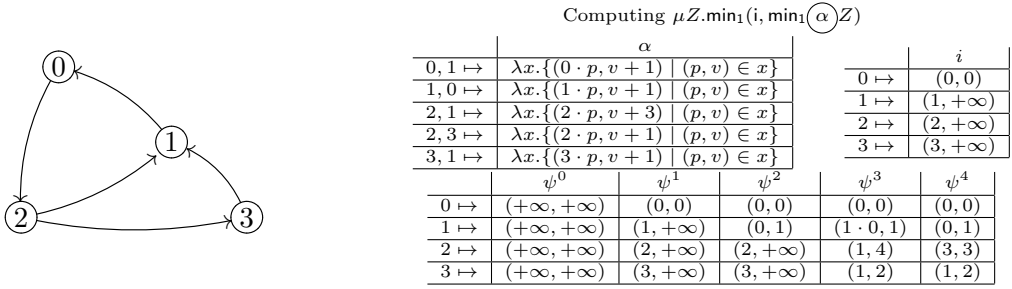

%

\begin{mdframed}[style=change]
%
The set of shortest path formulas we have discussed above \changeR{is} very flexible and can be used indeed to build useful field structures. A typical example are spanning trees. 
Indeed, a spanning tree can be computed as follows. 

\begin{exa}\label{ex:mst}
Consider as domain the lexicographical Cartesian product of domains $\langle \mathbb{N} \cup \{+\infty\}, \leq, +\infty, 0 \rangle$ and $N_{\sqsubseteq} = \langle N , \sqsubseteq_{N}, n_{|N|}, n_1\rangle $ given by some total ordering $n_1 \sqsubseteq_{N}  n_2 \sqsubseteq_{N} ..  \sqsubseteq_{N} n_{|N|}$ on nodes. Then, the formula for computing a (shortest path-based) spanning tree is $\mu Z . \mathsf{min_1}(\mathsf{i}, \aggregateOut{$\alpha$}{\tiny\mathsf{min_1}}{Z})$, where label $i$ is $(n,0)$ for the root $n$ and $(n'+\infty)$ for any other node $n'$ (i.e. the root points at itself with cost 0, while all other nodes point at themselves with infinite cost) and edge label $\alpha$ is used to append the source of an edge to the path component of a tuple $(p,v)$. The computation of the formula on a simple example is illustrated in Fig.~\ref{fig:mst}. 
\end{exa}

The exhaustive presentation of our case study in Section~{\ref{section:case-study}} exploits some of the above examples to solve a complex task.
\end{mdframed}

\medskip
\change{Now that we have provided a\changeR{n illustrative} set of examples, we are ready to formalise the meaning of formulas.} 
Given a formula $\Psi$ and an environment $\rho$ we say that $\Psi$ is $\rho$ \emph{closed} if $\rho$ is defined for the free formula variables of $\Psi$. 

\begin{defi}[semantics of \smuc{} formulas]
Let $F$ be a field and $\rho$ be an enviroment.  
The semantics of $\rho$-closed \smuc{} formulas is given by the interpretation function $\llbracket \cdot \rrbracket^F_\rho : \Psi \rightarrow N_F \rightarrow A_F$ defined by 
\[\def\arraystretch{1.3}
\begin{array}{rcl}
\llbracket i\rrbracket^F_{\rho} &=& I_F (i) \\
\llbracket z\rrbracket^F_{\rho} &=& \rho (z) \\ 
\llbracket f(\Psi_1,\dots,\Psi_k)\rrbracket^F_{\rho} &=&
\lambda n . \llbracket f \rrbracket_{A_F}  ( \llbracket \Psi_1\rrbracket^F_{\rho}\, n , .. , \llbracket \Psi_k\rrbracket^F_{\rho}\, n)\\ 
\llbracket \aggregateOut{$\alpha$}{g}{\Psi}\rrbracket^F_{\rho} &=& \lambda n . \llbracket g \rrbracket_{A_F} ( \{ I_F (\alpha)(n,n')(\llbracket\Psi\rrbracket^F_{\rho}(n')) \mid (n,n') \in  E_F\} ) \\ 
\llbracket \aggregateIn{$\alpha$}{g}{\Psi} \rrbracket^F_{\rho} &=& \lambda n . \llbracket g \rrbracket_{A_F} ( \{ I_F (\alpha)(n',n)(\llbracket\Psi\rrbracket^F_{\rho}(n')) \mid (n',n) \in  E_F\} )  \\ 
%
\llbracket \mu z.\Psi\rrbracket^F_{\rho} &=& \mathit{lfp}\ \lambda \mathsf{f} . \llbracket \Psi \rrbracket^F_{\rho[^\mathsf{f}/_z]} \\
\llbracket \nu z.\Psi\rrbracket^F_{\rho} &=& \mathit{gfp}\ \lambda \mathsf{f} . \llbracket \Psi \rrbracket^F_{\rho[^\mathsf{f}/_z]} \\ 
\end{array}
\]
where $\mathit{lfp}$ and $\mathit{gfp}$ stand for the least and greatest fixpoint, respectively\footnote{Notice that the value $I_F (\alpha)(n,n')(\llbracket\Psi\rrbracket^F_{\rho}(n'))$ will be passed to function $\llbracket g \rrbracket_{A_F}$ with a multiplicity which depends on the number of nodes $n'$.}. 
\end{defi}

As usual for such fixpoint formulas, the semantics is well defined if so are all fixpoints. As we mentioned in the previous section we require that all functions $\lambda \mathsf{f}. \llbracket \Psi \rrbracket^F_{\rho[^\mathsf{f}/_z]}$ are monotone and continuous.  
%
%

It is worth to remark that if we restrict ourselves to a semiring-valued field, then all \smuc{} formulas provide such guarantees.

\begin{restatable}[semiring monotony]{lemm}{semiringmonotony}
Let $F$ be a field, where $F_A$ is a semiring, $I_F$ is such that $I_F(\alpha)(e)$ is monotone for all $\alpha \in L_A, e \in E_A$,  $\mathcal{M}$ contains only function symbols that 
are obtained by composing additive and multiplicative operations of the semiring,
%
$\rho$ be an environment and $\Psi$ be a $\rho$-closed formula. Then, every function  
$\lambda \mathsf{f}. \llbracket \Psi \rrbracket^F_{\rho[^\mathsf{f}/_z]}$ is monotone and continuous.
\end{restatable}

%


\subsection{Robustness against unavailabilty}

\change{
This section provides a formal \changeR{characterisation} of unavailability and robustness results against  \changeR{situations where nodes are allowed} to proceed at different speeds.
For this purpose we introduce the notions of a \emph{pattern} and a \emph{strategy} which formalise the ability of nodes to participate to an iteration in the computation of a fixpoint. A pattern and the corresponding pattern-restricted application formalise which nodes will participate in an iteration. Note that the unavailability of a node $n$ does not mean that other nodes will ignore $n$ when aggregating values. We assume that the underlying system will ensure that the last known attributes of $n$ will be available (e.g. through a cache-based mechanism). 
}

\begin{defi}[pattern]
Let $F$ be a field. A  \emph{pattern} is a subset $\pi \subseteq N_F$ of the nodes of $F$.
\end{defi}

\begin{defi}[pattern-restricted application]\label{def:pattern-restricted-application}
Let $F$ be a field, $\pi \subseteq N_F$ be a pattern and $\psi : (N \rightarrow A) \rightarrow N \rightarrow A$ be an update function. The $\pi$-restricted application of $\psi$, denoted $\psi_\pi$ is a function $\psi_{\_}: 2^{N_F} \rightarrow (N_F \rightarrow A_F) \rightarrow N_F \rightarrow A_F$ such that:
\[
\psi_\pi\, \mathsf{f}\, n = \left\{
\begin{array}{ll}
\psi\, \mathsf{f}\, n & \text{ if } n \in \pi \\
\mathsf{f}\, n & \text{ otherwise } 
\end{array}
\right.
\]
\end{defi}

The intuition is that $\psi_\pi$ applies a node valuation $\mathsf{f}$ on the nodes in $\pi$ and ignores the rest. Note that $\psi_{N_F} = \psi$.

\change{
The concepts of pattern and pattern-restricted application are extended to sequences of executions that we call here \emph{strategies}. They can be seen as schedules determining which processes will be able to update their values in each step of an execution.   
}

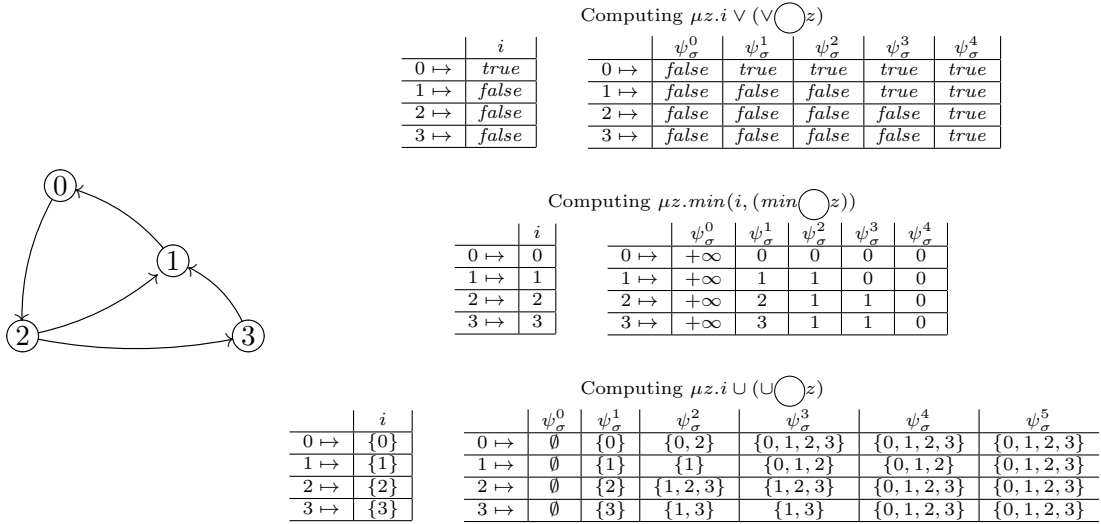
\begin{figure}
\begin{center}
\begin{minipage}{0.25\textwidth}
\begin{mdframed}[style=change]
\[
\begin{matrix}
\begin{xy}
(5,40)*{0}*\cir<6pt>{}="0";
(20,30)*{1}*\cir<6pt>{}="1";
(0,20)*{2}*\cir<6pt>{}="2";
(30,20)*{3}*\cir<6pt>{}="3";
\ar@{->}@/_0.4pc/ "0"; "2"
\ar@{->}@/_0.4pc/ "2"; "1"
\ar@{->}@/_0.4pc/ "2"; "3"
\ar@{->}@/_0.4pc/ "3"; "1"
\ar@{->}@/_0.4pc/ "1"; "0"
\end{xy}
\end{matrix}
\]
\end{mdframed}
\end{minipage}
\begin{minipage}{0.72\textwidth}
\begin{mdframed}[style=change]
{\tiny \centering

Computing $\mu z . i \vee (\aggregateOut{}{\mathbf{\vee}}{z})$

\begin{tabular}{c|c|}
                     & $i$ \\\hline
$0 \mapsto$ & $\mathit{true}$  \\\hline
$1 \mapsto$ & $\mathit{false}$  \\\hline
$2 \mapsto$ & $\mathit{false}$  \\\hline
$3 \mapsto$ & $\mathit{false}$  \\\hline
\end{tabular}
\hspace{0.5cm}
\begin{tabular}{c|c|c|c|c|c|c}
                     & $\psi_\sigma^0$ & $\psi_\sigma^1$  & $\psi_\sigma^2$  & $\psi_\sigma^3$  & $\psi_\sigma^4$   \\\hline
$0 \mapsto$ & $\mathit{false}$ & $\mathit{true}$ & $\mathit{true}$ & $\mathit{true}$ & $\mathit{true}$  \\\hline
$1 \mapsto$ & $\mathit{false}$ & $\mathit{false}$ & $\mathit{false}$ & $\mathit{true}$ & $\mathit{true}$ \\\hline
$2 \mapsto$ & $\mathit{false}$ & $\mathit{false}$ & $\mathit{false}$ & $\mathit{false}$ & $\mathit{true}$ \\\hline
$3 \mapsto$ & $\mathit{false}$ & $\mathit{false}$ & $\mathit{false}$ & $\mathit{false}$ & $\mathit{true}$ \\\hline
\end{tabular}

\vspace{0.5cm}
Computing $\mu z . \mathit{min}(i, (\aggregateOut{}{\mathit{min}}{z}))$

\begin{tabular}{c|c|}
                     & $i$ \\\hline
$0 \mapsto$ & $0$  \\\hline
$1 \mapsto$ & $1$  \\\hline
$2 \mapsto$ & $2$  \\\hline
$3 \mapsto$ & $3$  \\\hline
\end{tabular}
\hspace{0.5cm}
\begin{tabular}{c|c|c|c|c|c|c}
                     & $\psi_\sigma^0$ & $\psi_\sigma^1$  & $\psi_\sigma^2$  & $\psi_\sigma^3$ & $\psi_\sigma^4$  \\\hline
$0 \mapsto$ & $\mathit{+\infty}$ & $0$ & $0$ & $0$ & $0$  \\\hline
$1 \mapsto$ & $\mathit{+\infty}$ & $1$ & $1$ & $0$ & $0$  \\\hline
$2 \mapsto$ & $\mathit{+\infty}$ & $2$ & $1$ & $1$ & $0$  \\\hline
$3 \mapsto$ & $\mathit{+\infty}$ & $3$ & $1$ & $1$ & $0$  \\\hline
\end{tabular}

\vspace{0.5cm}
Computing $\mu z . i \cup (\aggregateOut{}{\mathbf{\cup}}{z})$

\begin{tabular}{c|c|}
                     & $i$ \\\hline
$0 \mapsto$ & $\{0\}$  \\\hline
$1 \mapsto$ & $\{1\}$  \\\hline
$2 \mapsto$ & $\{2\}$  \\\hline
$3 \mapsto$ & $\{3\}$  \\\hline
\end{tabular}
\hspace{0.5cm}
\begin{tabular}{c|c|c|c|c|c|c}
                     & $\psi_\sigma^0$ & $\psi_\sigma^1$  & $\psi_\sigma^2$  & $\psi_\sigma^3$ & $\psi_\sigma^4$ & $\psi_\sigma^5$  \\\hline
$0 \mapsto$ & $\emptyset$ & $\{0\}$ & $\{0,2\}$ & $\{0,1,2,3\}$ & $\{0,1,2,3\}$  & $\{0,1,2,3\}$\\\hline
$1 \mapsto$ & $\emptyset$ & $\{1\}$ & $\{1\}$ & $\{0,1,2\}$  & $\{0,1,2\}$ & $\{0,1,2,3\}$ \\\hline
$2 \mapsto$ & $\emptyset$ & $\{2\}$ & $\{1,2,3\}$ & $\{1,2,3\}$ & $\{0,1,2,3\}$ & $\{0,1,2,3\}$  \\\hline
$3 \mapsto$ & $\emptyset$ & $\{3\}$ & $\{1,3\}$ & $\{1,3\}$ & $\{0,1,2,3\}$ & $\{0,1,2,3\}$ \\\hline
\end{tabular}
}
\end{mdframed}
\end{minipage}
\end{center}
\caption{\change{Computation of the three \smuc{} formulas of Fig.~{\ref{fig:formulas3}} under a fair strategy.}}
\label{fig:formulas3}
\end{figure}

\begin{defi}[strategy]
Let $F$ be a field. A \emph{strategy} $\sigma$ is a possibly infinite sequence of patterns $\pi_1, \pi_2, \dots$ of $F$.
\end{defi}

As usual, we use $\epsilon$ for the empty sequence and, given a possibly infinite sequence $\sigma = \pi_1, \pi_2, \dots$, we use $\sigma_i$ for the $i$-th element (i.e. $\pi_i$), $\sigma^i$ for the suffix starting from the $i$-th element (i.e. $\pi_{i}, \pi_{i+1}, \dots$) and $\sigma[i..j]$ for the sub-sequence that starts from the $i$-th element and ends at the $j$-th element (i.e. $\pi_{i}, \dots , \pi_{j}$ if $i \leq j$ and $\epsilon$ otherwise).

\begin{defi}[strategy-restricted application] \label{def:chain-restricted-application}
Let $F$ be a field, $\sigma$ be a finite strategy and $\psi : (N \rightarrow A) \rightarrow N \rightarrow A$ be an update function. The $\sigma$-restricted application of $\psi$, denoted $\psi_\sigma$ is defined as:
\[
\psi_{\sigma} = \left\{
\begin{array}{ll}
\bot & \text{ if } \sigma = \epsilon \\
\psi_\pi \, \psi_{\sigma'} & \text{ if } \sigma = \sigma',\pi 
\end{array}
\right.
\]
\end{defi}
The intuition is that update function $\psi$ is applied to bottom $k=|\sigma|$ times, every time according to the $i$-th element $\sigma_i$ of the strategy $\sigma$, for $i$ ranging from $1$ to $k$. \change{As an example, Fig.~{\ref{fig:formulas3}} shows how the computations of Fig.~{\ref{fig:formulas1}} would be carried out under the strategy $\sigma$ where node 1 participates in odd rounds only, i.e. $\sigma$ is such that $\sigma_i = \{0,1,2,3\}$ if $i$ is odd and  $\sigma_i = \{0,2,3\}$ otherwise. One can see that the main effect of a strategy is to delay the achievement of the fixpoint.}

\change{Clearly, not all strategies correspond to realistic executions in practice. The situation under consideration here is that the underling middleware guarantees that nodes are able to participate infinitely often in the computations of formulas. We formalise this by considering a class of \emph{fair} strategies. The term \emph{fair} is inspired by classical notions of \emph{fairness} such as those used in concurrency theory, operating systems and formal verification.}

\begin{defi}[fair strategy]
Let $F$ be a field. A strategy $\sigma = \pi_1, \pi_2,\dots$ is \emph{fair} with respect to a field $F$ iff $\forall n \in N_F$ the set $\{ k \mid n \in \pi_k\}$ is infinite.
\end{defi}

Intuitively, a fair strategy allows every node to execute infinitely often. Clearly, only infinite strategies can be fair, \change{but when the fixpoint is reached in a finite number of steps, the strategy can be considered as finite/terminated. The example strategy discussed above is fair.}

\medskip

In the following we shall present lemmas and theorems related to \change{the robustness of} least fixpoints \change{under fair strategies}. Analogous results can be obtained for greatest fixpoints. 

\begin{restatable}[monotony of pattern-restricted application]{lemm}{asynchronylemma}
\label{th:asynchrony-lemma1}\label{th:asynchrony-lemma2}
Let $F$ be a field, $\psi : (N \rightarrow A) \rightarrow N \rightarrow A$ be a monotone update function, $\mathsf{f}$, $\mathsf{f1}$, $\mathsf{f2} : N \rightarrow A$ be node valuations, $\pi$, $\pi_1$, $\pi_2 $ be patterns, and $n \in N_F$ be a node. Then, the following holds:
\begin{enumerate}[label=(\roman*)]
\item[(i)] Function $\psi_\pi$ is monotone;
%
%
\item[(ii)]  $n \in \pi_1 \Leftrightarrow n \in \pi_2$ implies $\psi_{\pi_1}\mathsf{f}\, n = \psi_{\pi_2}\mathsf{f}\, n$;
\item[(iii)] $n \in \pi_1 \Leftrightarrow n \in \pi_2$ and $\mathsf{f}_1 \sqsubseteq \mathsf{f}_2$ implies $\psi_{\pi_1}\, \mathsf{f}_1\, n \sqsubseteq \psi_{\pi_2}\, \mathsf{f}_2\, n$.
\end{enumerate}
\end{restatable}

\noindent We recall that, for the sake of a lighter notation we drop  subscripts \changeR{when they are} clear from the context. This means that in the above definition and in what follows, for example, something like $\bot \sqsubseteq \psi$ abbreviates $\bot_{N \rightarrow A} \sqsubseteq_{N \rightarrow A} \psi$ in an unambigous manner (our notational convention for $\psi$ determines the field domain under consideration). 

\begin{restatable}[pattern-restricted application bounds]{lemm}{asynchronylemmabis}
\label{th:asynchrony-lemma3}
Let $F$ be a field, $\psi : (N \rightarrow A) \rightarrow N \rightarrow A$ be a monotone update function, $\sigma$ be a finite strategy and $\pi$ be a pattern. Then it holds $\psi_\sigma \sqsubseteq \psi_\pi \, \psi_\sigma \sqsubseteq \psi \, \psi_\sigma$
\end{restatable}

\change{A consequence of the lemma is that strategy-restricted applications yield partially ordered chains.}

\begin{restatable}[strategy-restricted applications yield chains]{corr}{asynchronycorollary}
\label{th:asynchrony-corollary1}\label{cor:asynchrony-corollary1}
Let $F$ be a field, $\psi : (N \rightarrow A) \rightarrow N \rightarrow A$ be a monotone update function and $\sigma$ be an infinite strategy. Then, the sequence $\bot \sqsubseteq \psi_{\sigma[1..1]} \sqsubseteq \psi_{\sigma[1..2]} \sqsubseteq  \dots $ is actually a partially ordered chain.
\end{restatable}

\change{The above results allow us to state now one of the main results of the paper, i.e. that fair strategies and the ideal situation (all nodes are always available) have the same bounds.}

\begin{restatable}[bounds under fair strategies]{thm}{asynchronytheorem}
\label{th:asynchrony-theorem1}
Let $F$ be a field, \change{with $N_F$ finite,} $\psi : (N \rightarrow A) \rightarrow N \rightarrow A$ be a monotone update function and $\sigma$ be a fair strategy. Then all elements of the partially ordered chains $\bot \sqsubseteq \psi_{\sigma[1..1]} \sqsubseteq \psi_{\sigma[1..2]} \sqsubseteq  \dots$ and $\bot \sqsubseteq \psi\, \bot \sqsubseteq \psi^2\, \bot \sqsubseteq  \dots$ have the same set of upper bounds and hence the same least upper bound, namely the least fixpoint of $\psi$. 
\end{restatable}

\change{The final result is the formalisation of robustness against node unavailability.}

\begin{restatable}[robustness against unavailability]{thm}{asynchronytheorembis}
\label{th:asynchrony-theorem2}
Let $F$ be a field with finite set of nodes $N_F$ and field domain $A$ with finite partially ordered chains only, $\sigma$ be a fair strategy and $\psi: (N \rightarrow A) \rightarrow N \rightarrow A$ be a monotone update function.
%
Then the partially ordered chain $\bot \sqsubseteq \psi_{\sigma[1..1]} \sqsubseteq \psi_{\sigma[1..2]} \sqsubseteq  \dots $
\begin{enumerate}[label=(\roman*)]
\item[(i)] stabilizes to its least upperbound $\mathsf{f}$; 
\item[(ii)]  its least upper bound $\mathsf{f}$ does not depend on the fair strategy: we always have $f = \mathit{lfp}\,  \psi$.
\end{enumerate}
\end{restatable}

\noindent This result is of utmost importance in practice since it guarantees that, under reasonable conditions, the computation of fixpoints can be performed asynchronously without the need of synchronising the agents, which may proceed at different relative speeds. 
%

%
The most significant restriction is the one that requires finite chains and stabilisation\change{, namely the recognition of a finite prefix $\sigma'$ of $\sigma$ which is enough to reach the fixpoint, i.e. such that
$\psi \psi_\sigma' = \psi_\sigma'$.} However, \change{we envisage in practice the use of libraries of function and formula patterns that already ensure those properties, so that the final user can just combine them at will. A typical example could be, for instance, to consider semirings as we did in~{\protect{\cite{DBLP:conf/coordination/Lluch-LafuenteL15}}} with finite discrete domains. Indeed, the semiring additive and multiplicative operations can both be used as aggregation functions (since both work on multisets), both are monotone (which ensures well-definedness of fixpoints), and the restriction to finite discrete domains ensures finite chains. 
}

\begin{mdframed}[style=change]

\subsection{Robustness against failures}

We now consider the possibility of agent failure. We restrict ourselves to the common case where not only agents can stay inactive for a (finite) period, but when they resume they enter a backup state they had in a previous iteration, possibly \changeR{the initial one} ($\bot$). Thus we exclude the erroneous behavior caused by an agent entering a completely unknown state, or occurring when the structure of the network is in any form damaged or modified. We prove that the stable, fixpoint state does not change, provided the system at some point enters a condition where no more failures occur.

\begin{defi}[failure sequence]
Let $\sigma$ be a finite strategy and $\psi : (N \rightarrow A) \rightarrow N \rightarrow A$ be an update function. A {\em $\sigma$ failure sequence of $\psi$}, denoted $\varsigma_\sigma$, is defined as:
\begin{equation}
\label{failure}
\varsigma_{\sigma}n = \left\{
\begin{array}{ll}
\bot_{A} & \text{ if } \sigma = \epsilon \\
\psi \varsigma_{\sigma'} n & \text{ if } \sigma = \sigma', \pi \mbox{ and } n \in \pi \\
\varsigma_{\sigma'}n & \text{ if } \sigma = \sigma', \pi  \mbox{ and } n \notin \pi \\
\varsigma_{\sigma'}n & \text{ if } \sigma = \sigma',\sigma'', \pi  \mbox{ with } \sigma'' \neq \epsilon \mbox{ and } n \notin \pi.
\end{array}
\right.
\end{equation}
\noindent Now let us extend $\varsigma_\sigma$ and $\psi_\sigma$ to infinite sequences $\tilde{\sigma}$: 
\[
\overline{\varsigma}_{\pi_1, \pi_2 \ldots } = \varsigma_\epsilon, \varsigma_{\pi_1}, \varsigma_{\pi_1, \pi_2}, \ldots  \;\;\;\;\; 
\overline{\psi}_{\pi_1, \pi_2 \ldots } = \psi_\epsilon, \psi_{\pi_1}, \psi_{\pi_1, \pi_2}, \ldots 
\]
\noindent An infinite sequence ${\overline{\varsigma}}_{\tilde{\sigma}}$ is called a {\em $\tilde{\sigma}$ failure sequence of $\psi$}. We call it {\em safe} if, for all finite $\sigma'$ larger than 
some finite $\hat{\sigma}$, the fourth option in equation~\ref{failure} has not been used for computing $\varsigma_{\sigma'}$. The idea is that, from some $\hat{\sigma}$ on failures never occur again and all nodes can progress, possibly skipping some rounds (but never returning to a backup/initial state).

\end{defi}

Notice that \changeR{in the above definition} $\varsigma_\sigma$ is not functional and models a non-deterministic presence of errors. Indeed if $n \notin \pi$ the value of node $n$ can be left unchanged (third option), can be initialized to $\bot_{A}$ (fourth option with $\sigma' = \epsilon$), or it can be assigned any previous value (fourth option). If $n \in \pi $ then it is updated using $\psi$. Also, if the fourth option is never used in a sequence, then $\varsigma_\sigma = \psi_\sigma$, namely a failure sequence is just an ordinary chain. Conversely, observe that a generic failure sequence is not a chain, since it is not necessarily increasing.

We can now prove our main result in this setting: a safe $\sigma$ failure sequence of $\psi$ has \changeR{a} least upper bound which is the fixpoint of $\psi$.

\begin{restatable}[least upper bound of a safe failure sequence]{thm}{failurethm}\label{theorem:failure}
Let $F$ be a field, with $N_F$ finite, $\psi : (N \rightarrow A) \rightarrow N \rightarrow A$ be a monotone update function, $\tilde{\sigma}$ be a fair strategy and ${\overline{\varsigma}}_{\tilde{\sigma}}$ be some infinite safe $\tilde{\sigma}$ failure sequence of $\psi$. Then ${\overline{\varsigma}}_{\tilde{\sigma}}$ has a least upper bound which is the least fix point of $\psi$.
\end{restatable}
\end{mdframed}


\changeR{
Similar results can be provided for additional failure situations. For example, the above results could be adapted to the situation in which errors can occur infinitely often, but sufficiently long progress is guaranteed between errors. This would be provided by a middleware that enforces a phase restore-progress-backup after each failure recovery.

More general kinds of failure, concerning unrecoverable failure of some node, or
possible changes in the structure of the network, cannot be recovered significantly. Specific recovery actions must be foreseen for maintaining networks with failures, which apply to our approach just as they concern similar coordination styles.}



\subsection{\smuc{} Programs}

The atomic computations specified by \smuc{} formulas can be embedded in any language. To ease the presentation we present a global calculus where atomic computations are embedded in a simple imperative language \changeR{similar to} the  \textsc{While}~\cite{DBLP:series/utcs/NielsonN07} \change{language, a core imperative language for imperative programming which has formal semantics.} 

\begin{defi}[\smuc{} syntax] 
The syntax of \smuc{} is given by the following grammar 
\[
\begin{array}{rcl}
P , Q & ::= & \mathsf{skip} \mid
    i \leftarrow \Psi \mid
    P\ \mathsf{;}\ Q \mid
    \mathsf{if}\ \Psi\ \mathsf{then}\ P\ \mathsf{else}\ Q 
   \mid 
    \mchange{\mathsf{until}}\ \Psi\ \mathsf{do}\ P
\end{array}
\]
\noindent 
where $i \in \mathcal{L}$, $\Psi$ is a \smuc{} formula (cf. Def~\ref{def:formulas}).
\end{defi}

The simple imperative language we used in~\cite{DBLP:conf/coordination/Lluch-LafuenteL15} featured $\mathsf{agree \cdot on}$ variants of the traditional control flow constructs 
 in order to remark the characteristics of the case study used there, where the global control flow depended on the existence of agreements among all agents in the field.
\change{This can be of course an expensive operation, which depends on the diameter of the graph.}

\begin{table}[t]
\begin{center}
\begin{tabular}{|rc|}
\hline
&\\
$( \mu\textsc{Step})$ 
&
$
\dfrac{\llbracket \Psi \rrbracket^{I_F}_{\emptyset}=f \quad 
I_F'=I_{F}[^{f}/_{i}]}{ 
\langle i \leftarrow \Psi , F \rangle \rightarrow \langle \mathsf{skip} , F [^{I_{F}'}/_{I_F}]  \rangle
}
 $
\\&\\
$(\textsc{Seq1})$ 
&
$
\dfrac{
\langle  P   ,   F  \rangle  \rightarrow \langle P' , F'\rangle
}{
\langle  P  \mathsf{;} Q ,  F \rangle \rightarrow \langle  P'\mathsf{;}Q ,  F' \rangle
}
$
\\&\\
$(\textsc{Seq2})$ 
&
$
\dfrac{
\langle  P   ,   F \rangle  \rightarrow \langle  P'   ,   F'  \rangle
}{
\langle  \mathsf{skip}  \mathsf{;} P ,  F \rangle \rightarrow \langle  P' ,  F' \rangle
}
$
\\&\\
$( \textsc{IfT} )$ 
&
$
\dfrac{
\llbracket \Psi \rrbracket_{\emptyset}^F\ =\ \lambda n . \emph{true}
}{
\langle  \mathsf{if }\ \Psi\  \mathsf{ then }\ P\ \mathsf{ else }\ Q   ,   F \rangle  
\rightarrow \langle  P , F \rangle 
}
$
\\&\\
$( \textsc{IfF} )$  
&
$
\dfrac{
\llbracket \Psi \rrbracket_{\emptyset}^F\ \neq \ \lambda n . \emph{true}
}{
\langle  \mathsf{if }\ \Psi\ \mathsf{ then }\ P\ \mathsf{ else }\ Q   ,   F \rangle  
\rightarrow \langle Q , F \rangle 
} 
$ 
\\&\\
$\ \ (\change{\textsc{UntilF}} )$  
& \ \ 
$
\dfrac{
\llbracket \Psi \rrbracket_{\emptyset}^F\ \neq \ \lambda n . \emph{true}
}{
\langle  \mchange{\mathsf{until }}\ \Psi\ \mathsf{ do }\ P ,   F \rangle  
\rightarrow \langle ( P\ \mathsf{ ; }\ \mchange{\mathsf{until }}\ \Psi\ \mathsf{ do }\ P ) , F \rangle 
} \ \ 
$
\\&\\
$\ \ (\change{\textsc{UntilT}} )$  
& \ \ 
$
\dfrac{
\llbracket \Psi \rrbracket_{\emptyset}^F\ = \ \lambda n . \emph{true}
}{
\langle  \mchange{\mathsf{until }}\ \Psi\ \mathsf{ do }\ P ,   F \rangle  
\rightarrow \langle \mathsf{skip}\  , F \rangle 
} \ \ 
$
\\&\\
\hline
\end{tabular}

\phantom{a}
\caption{Rules of the operational semantics}
\label{table:semantics-global1}
\end{center}
\end{table}

The use of traditional \change{control flow constructs} does not restrict the possibility to deal with agreements. Indeed, the existence of an agreement of all agents on an expression $\Psi$ can be easily verified by using the expression 
$
\mathit{eq}(\Psi,
\aggregateOut{\it{id}}{eq}{\Psi},
\aggregateIn{\it{id}}{eq}{\Psi}) 
\neq \emph{none}
$,
where $\mathit{id}$ is the identity function and $\mathit{eq}$ is a function equationally defined as follows:
\[
\begin{array}{c}
\mathit{eq}(\emptyset) = \mathit{any} \hspace{0.5cm}
\mathit{eq}(\{a\}) = \mathit{a} \hspace{0.5cm}
\mathit{eq}(\{a,\mathit{any}\} \cup B) = \mathit{eq}(\{a\} \cup B)
\\
\mathit{eq}(\{a,b\} \cup B) = \mathit{none} \hspace{0.5cm}
\mathit{eq}(\{a,a\} \cup B) = \mathit{eq}(\{a\} \cup B) 
\end{array}
\]

%
\noindent In words, the expression $\mathit{eq}(\Psi,
\aggregateOut{\it{id}}{eq}{\Psi},
\aggregateIn{\it{id}}{eq}{\Psi}) 
\neq \emph{none}
$ is true on all agents whenever $\Psi$ is evaluated to the same value on each node and its neighbours (both through in- and out-going edges).
Note that similar expressions can be used if one is interested in agreements that exclude certain values, say in a set $B$. The corresponding condition expression would be 
$
\mathit{eq}(\Psi,
\aggregateOut{\it{id}}{eq}{\Psi},
\aggregateIn{\it{id}}{eq}{\Psi}) 
\not\subseteq B \cup \{\emph{none}\}
$.

\medskip

The semantics of the calculus is straightforward, along the lines of \textsc{While}~\cite{DBLP:series/utcs/NielsonN07} with fields (and their interpretation functions) playing the role of memory stores. 
%
%

The semantics of our calculus is a transition system whose states are pairs of calculus terms and fields and whose transitions $\rightarrow \subseteq (P \times \mathcal{F})^2$ are defined by the rules of Table~\ref{table:semantics-global1}. 
Most rules are standard. Rule \textsc{IfT} and \textsc{IfF} are similar to the usual rules for conditional branching. It is worth to remark that the condition $\Psi$ must evaluate to $\emph{true}$ in each agent $n$ in the field $F$ for the $\mathsf{then}$ branch to be taken, otherwise the $\mathsf{else}$ branch is followed. Similarly for the $\mchange{\mathsf{until}}$ operator (cf. rules \change{\textsc{UntilF}} and $\mchange{\mathsf{until}}$). In particular, the \change{\textsf{until}} finishes when all agents agree on $\emph{true}$, \change{namely when the formula $\Psi$ is evaluated to $\lambda n.true$}. 
States of the form $\langle \mathsf{skip},I \rangle$ represent termination. 



\section{\smuc{} at Work: Rescuing Victims}\label{section:case-study}

\begin{figure}[tbp]
\begin{center}
\begin{tabular}{ccc}
\includegraphics[scale=0.4]{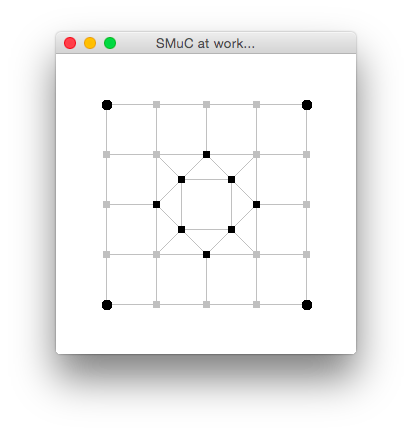} 
&\phantom{xx}&
\includegraphics[scale=0.4]{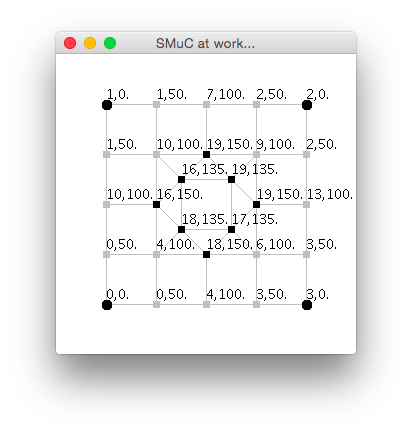} 
\end{tabular}
\end{center}
\caption{Execution of Robot Rescue \smuc{} Program (part 1)}
\label{figure:run_part1}
\end{figure}

%
The left side of Fig.~\ref{figure:run_part1}  depicts a simple instance of the considered scenario. There, victims are rendered as black circles while landmarks and rescuers are depicted via grey and black rectangles respectively. The length of an edge in the graph is proportional to the distance between the two connected nodes. The main goal is to assign rescuers to victims, where each victim may need more than one rescuer and we want to minimise the distance that rescuers need to cover to reach their assigned victims. 
We assume that all relevant information of the victim rescue scenario is suitably represented in field $F$. More details on this will follow, but for now it suffices to assume that nodes represent rescuers, victims or landmarks and edges represent some sort of direct proximity (e.g. based on visibility w.r.t. to some sensor).

\begin{figure}[t]
\scriptsize
\noindent\makebox[\linewidth]{\rule{\textwidth}{1pt}}\\
\begin{minipage}[t]{0.49\textwidth}
{\sf
{\color{blue} /* Initialisations */}
{\color{white}x}\\
$\mathsf{finish} \leftarrow \mathit{false}$;\\
$\mchange{\mathsf{until}}$ $\mathsf{finish}$ $\mathsf{do}$\\
\phantom{xxxx}{\color{blue} /* 1st Stage: Establishing the distance to victims */}\\
\phantom{xxxx} $\mchange{\mathsf{source} \leftarrow \mathsf{victim}\ ?\ (0,\mathsf{self}) : (+\infty,\mathsf{self})}$;\\
\phantom{xxxx} $\mathsf{D} \leftarrow \mu Z . \mathsf{min_1}(\mathsf{source}, 
\aggregateOut{\textsf{dst}}{\tiny\mathsf{min_1}}{Z}
%
)$;\\
\\
\phantom{xxxx} {\color{blue} /* 2nd Stage: Computing the rescuers paths */}\\
\phantom{xxxx} $\mathsf{rescuers} \leftarrow$ $\mu Z . \mathsf{init} \cup 
\aggregateIn{\tiny\textsf{grd}}{\bigcup}{Z}
$;\\
\\
\phantom{xxxx} {\color{blue} /* 3rd Stage: Engaging rescuers */}\\
\phantom{xxxx} {\color{blue} /* engaging the rescuers */}\\
\phantom{xxxx} $\mathsf{engaged} \leftarrow \mu Z . \mathsf{choose} \cup
\aggregateOut{\tiny\textsf{cgr}}{\bigcup}{Z}
$;\\
\phantom{xxxx} {\color{blue} /* updating victims and available rescuers */}\\
\phantom{xxxx} $\mathsf{victim'}  \leftarrow \mathsf{victim}$;\\
\phantom{xxxx} $\mathsf{victim}  \leftarrow \mathsf{victim} \wedge \neg \mathsf{saved}$;\\
\phantom{xxxx} $\mathsf{rescuer}  \leftarrow \mathsf{rescuer} \wedge 
\mchange{(}\mathsf{engaged} = \emptyset\mchange{)}$;\\
\phantom{xxxx} {\color{blue} /* determining termination */}\\
\phantom{xxxx} $\mathsf{finish} \leftarrow (\mathsf{victim}' == \mathsf{victim})$; \\
\\
{\color{blue} /* 4th Stage: Checking success */}\\
$\mathsf{if}$ $\neg\mathsf{victim}$\\
\phantom{xxxx} {\color{blue} /* ended with success */}\\
else\\
\phantom{xxxx} {\color{blue} /* ended with failure */}
}
\end{minipage}
\begin{minipage}[t]{0.49\textwidth}
\smallskip
\noindent
\fbox{
\parbox{0.95\textwidth}{
\[
\begin{array}{rcl}
\multicolumn{3}{l}{\textsf{\color{blue} /* Semiring types of node labels */}}  \\
\mathsf{source}, \mathsf{D}  & : & N \rightarrow T \times_1 \mchange{N_{\sqsubseteq}}\\
\mathsf{init},\mathsf{rescuers}  & : & N \rightarrow 2^{T \times N^*} \\
\mathsf{choose},\mathsf{engaged}  & : & N \rightarrow 2^{N^*}\\
\mchange{\mathsf{victim},\mathsf{victim'} } & : & \mchange{N \rightarrow \mathbb{B}}\\
\mchange{\mathsf{rescuer}, 
\mathsf{finish}} \\
\\
\multicolumn{3}{l}{\textsf{\color{blue} /* Semiring types of edge labels */}}  \\
\mathsf{dst}         & : & E \rightarrow T \times_1 \mchange{N_{\sqsubseteq}} \rightarrow T \times_1 \mchange{N_{\sqsubseteq}}\\
\mathsf{grd}         & : & E \rightarrow 2^{T \times N^*} \rightarrow 2^{T \times N^*}\\
\mathsf{cgr}         & : & E \rightarrow 2^{N^*}\rightarrow 2^{N^*}
\end{array}
\]
%
}
}
\end{minipage}
\noindent\makebox[\linewidth]{\rule{\textwidth}{1pt}}
\noindent
%
\caption{Robot Rescue \smuc{} Program}
\label{figure:program-global}
\end{figure}

We will use semirings as suitable structures for our operations. 
It is worth to remark that in practice it is convenient to define $A$ as a Cartesian product of semirings, e.g. for differently-valued node and edge labels. This is indeed the case of our case study. However, in order to avoid explicitly dealing with these situations (e.g. by resorting to projection functions, etc.) which would introduce a cumbersome notation, we assume that the corresponding semiring is implicit (e.g. by type/semiring inference) and that the interpretation of functions and labels are suitably specialised. For this purpose we decorate the specification in Fig.~\ref{figure:program-global} with the types of all labels. 

We now describe the coordination strategy specified in the algorithm of Fig.~\ref{figure:program-global}. 
\change{Many of the formulas that the algorithm uses are based on the formula patterns described in Examples~{\ref{ex:formulas1}}--{\ref{ex:mst}}.} 
The algorithm consists of a loop that is repeated until an iteration does not produce any additional matching of rescuers to victims. 
The body of the loop consist of different stages, each characterised by a fixpoint computation.

\paragraph*{1st Stage: Establishing the distance to victims.}
In the first stage of the algorithm the robots try to establish their closest victim.
Such information is saved in to $\mathsf{D}$, which is valued over the \change{lexicographical Cartesian product of domains $T = \langle \mathbb{N} \cup \{+\infty\}, \leq, +\infty, 0\rangle$ and $N_{\sqsubseteq} = \langle N , \sqsubseteq_{N},  n_{|N|} , n_1 \rangle $ given by some total ordering $n_1 \sqsubseteq_{N}  n_2 \sqsubseteq_{N} ..  \sqsubseteq_{N} n_{|N|}$ on nodes.}
%
%
In order to compute $\mathsf{D}$, some information is needed on nodes and arrows of the field. \change{For example, we assume that the boolean attribute $\mathsf{victim}$ initially records whether the node is a victim or not, while $\mathsf{source}$ is used to store that} victims initially point to themselves with no cost, while the rest of the nodes point to themselves with infinite cost. \change{The edge label $\mathsf{dst}$ is defined as} $I(\mathsf{dst})(n,n')$ = $\lambda(v,m).(distance(n,n')  + v,n')$
   where $distance(n,n')$  is the weight of $(n,n')$. Intuitively, $\mathsf{dst}$ provides a function to add the cost associated to the transition.  The second component of the value encodes the direction to go for the shortest path, while the total ordering on nodes is used for solving ties.

The desired information is then computed as $\mathsf{D} \leftarrow \mu Z . \mathsf{min_1}(\mathsf{source}, \aggregateOut{\textsf{dst}}{\tiny\mathsf{min_1}}{Z})$. \change{This formula is similar to the formulas  presented in Examples~{\ref{ex:shortestpathformulas1}}--{\ref{ex:mst}}.}
\change{Here} $\mathsf{min_1}$ 
is the \change{join of domain} $T \times_1 N_{\leq_N}$, specifically for a set $B \subseteq (\Real \cup \{+\infty\}) \times N$ the function $\mathsf{min_1}$ 
is defined as $\mathsf{min_1}(B) = (a,n) \in B$ such that $\forall (a',n') \in B : a \leq a'$ and if $a=a'$ then $n\leq n'$. 

At the end of this stage, $\mathsf{D}$ associates each element with the distance to its closest victim \change{and the identity of the neighbour on the next edge in the shortest path}. In the right side of Fig.~\ref{figure:run_part1} each node of our example is labeled with the computed distance. We do not include the second component of $\mathsf{D}$ (i.e. the identity of the closest neighbour) to provide a readable figure. In any case, the closest victim is easy to infer from the depicted graph: the closest victim of the rescuer in the top-left corner of the inner box formed by the rescuers is the victim at the top-left corner of the figure, and respectively for the top-right, bottom-left and bottom-right corners.

\begin{figure}[tbp]
\begin{center}
\begin{tabular}{ccc}
\includegraphics[scale=0.4]{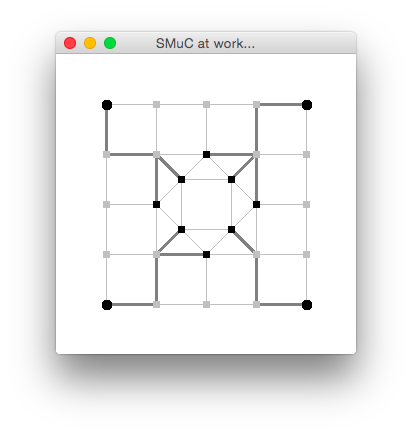} 
&\phantom{xx}&
\includegraphics[scale=0.4]{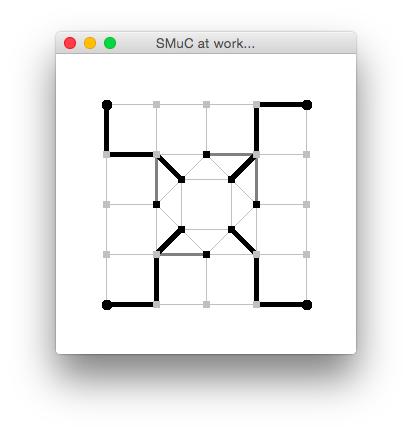} 
\end{tabular}

\end{center}
\caption{Execution of Robot Rescue \smuc{} Program (part 2)}
\label{figure:run_part2}
\end{figure}

\paragraph*{2nd Stage: Computing the rescuers paths to the victims.}
In this second stage of the algorithm, the robots try to compute, for every victim $v$, which are the paths from every rescuer $u$ to $v$ --- but only for those $u$ for which $v$ is the closest victim --- and the corresponding costs, as established by $\mathsf{D}$ in the previous stage. 
Here we use the semiring $2^{T \times N^*}$ with union as additive operator, i.e. $\langle 2^{T \times N^*}, \cup, \cap, T \times N^*, \emptyset \rangle$. We use here decorations $\mathsf{init}$ and $\mathsf{grd}$ whose interpretation is defined as 

\begin{itemize}
\item $I (\mathsf{init}) n$ = if $n \in$ rescuer then $\{(n',\epsilon) \mid \mathsf{D}(n) = (u,n')\}$ else $\emptyset$; 
\item $I(\mathsf{grd})(n,n')$ = $\lambda C .$ if 
\change{
$\mathsf{D}(n) = (u,n') \wedge 
\mathsf{D}(n') = (u',n'')$
}
%
%
then $n;C$ else $\emptyset$, where operation $;$ is defined as $n;C$ = $\{(cost,n;path ) \mid (cost,path) \in C\}$.
\end{itemize}

\noindent The idea of label \emph{rescuers} is to compute, for every node $n$, the set of rescuers whose path to their closest victim passes through $n$ (typically a landmark). However, the name of a rescuer is meaningless outside its neighbourhood, thus a path leading to it is constructed instead. In addition, each rescuer is decorated with its distance to its closest victim. Function $\mathsf{init}$ associates to a rescuer its name and its distance, the empty set to all the other nodes. Function $\mathsf{grd}$ checks if an arc $(n,n')$ is on the optimal path to the same victim $n''$ both in $n$ and $n'$. In the positive case, the rescuers in $n$ are considered as rescuers also for $n'$, but with an updated path; in the negative case they are discarded.
%
%

%
%
\mchange{On the} left side of Fig.~\ref{figure:run_part2} the result of this stage is presented. \change{Since all edges of the graph have a corresponding edge in the opposite direction we have depicted the graph as it would be undirected.} The edges that are part of a path from one rescuer to a victim are now marked (\change{where the actual direction is left implicit for simplicity}).  We can notice that some victims can be reached by more than one rescuer.  

\paragraph*{3rd Stage: Engaging the rescuers.} 
The idea of the third stage of the algorithm is 
 that each victim $n$, which needs $k$  rescuers,  will choose the $k$  closest rescuers, if there are enough, among those that have selected $n$  as target victim.
For this computation we use the decorations $\mathsf{choose}$ and $\mathsf{cgr}$.

\begin{itemize}
\item $I(\mathsf{choose})(n) =$ 
if $n \in victim$ and $\mathsf{saved}( n )$ then $\mathit{opt}(\mathsf{rescuers}(n), \mathit{howMany}(n))$ else $\emptyset$, where:
\begin{itemize}
\item $\mathsf{saved}(n) =  \big|\mathsf{rescuers}(n)\big| \leq \mathit{howMany}(n)$ and 
$\mathit{howMany}(n)$ 
returns the number of rescuers $n$ needs;
\item $\mathit{opt}(C,k)$ = 
$\{ \mathit{path} \mid (\mathit{cost},\mathit{path}) \in C\mbox{ and }$\\
$\phantom{\mathit{opt}(C,k)\ =\ xxxx}\big|\{(\mathit{cost}',\mathit{path}') \mid (\mathit{cost}',\mathit{path}') < (\mathit{cost},\mathit{path})\}\big| < k\}$
%
\\
where $(\mathit{cost},\mathit{path})<(\mathit{cost}',\mathit{path}')$ if $\mathit{cost} < \mathit{cost}'$ or $\mathit{cost}=\mathit{cost}'$ and $\mathit{path} < \mathit{path}'$, and paths are totally ordered lexicographically;
\end{itemize}
\item $I(\mathsf{cgr}(n,n' )= \lambda C .  \{\mathit{path} \mid n;\mathit{path} \in C\}$.
\end{itemize}

\begin{figure}[tbp]
\begin{center}
\includegraphics[width=0.66\textwidth]{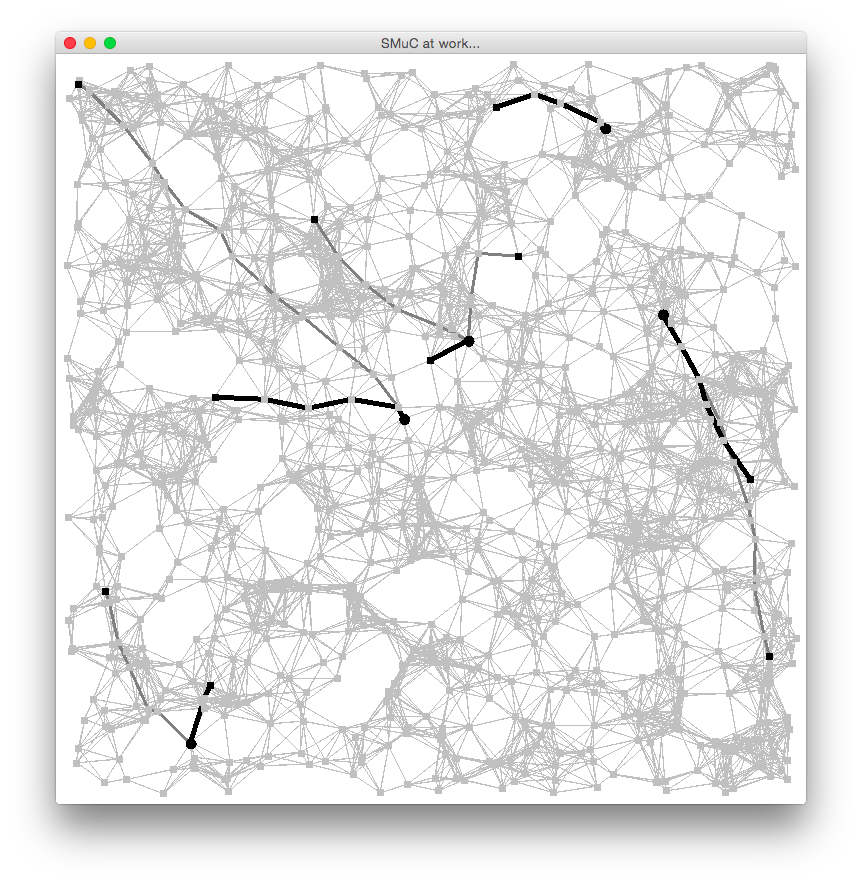} 
\end{center}
\caption{Execution of Robot Rescue \smuc{} Program on a random graph}
\label{figure:run_part3}
\end{figure}

Intuitively, $\mathsf{choose}$ allows a victim $n$ that has enough rescuers to choose and to record the paths leading to them.
The annotation $\mathsf{cgr}$ associates to each edge $(n,n')$ a function to select in a set $C$ of paths  those of the form $n;\mathit{path}$.

The computation in this step is $\mathsf{engaged} \leftarrow \mu Z . \mathsf{choose} \cup \aggregateOut{\tiny\textsf{cgr}}{\bigcup}{Z}$, which computes the desired information: in each node $n$ we will have the set of rescuer-to-victim paths that pass through $n$ and that have been chosen by a victim. 

The result of this stage is presented in the right side of Fig.~\ref{figure:run_part2}. Each rescuer has a route, that is presented in the figure with black edges, that can be followed to reach the assigned victim. Again, for simplicity we just depict some relevant information to provide an appealing and intuitive representation. 

Notice that this phase, and the algorithm, may fail even if there are enough rescuers to save some additional victims. For instance if there are two victims, each requiring two rescuers, and two rescuers, the algorithm fails if each rescuer is closer to a different victim.

\medskip
These three stages are repeated until there is agreement on whether to finish. The termination criteria is that an iteration did not update the set of victims. In that case the loop terminates and the algorithm proceeds to the last stage. 

\paragraph*{4th Stage: Checking succes.}
The algorithm terminates with success when $victim'=\emptyset$ and with failure when $victim'$ is not empty.
%
In Fig.~\ref{figure:run_part3} we present the result of the computation of program of Fig.~\ref{figure:program-global} on a randomly generated graph composed by $1,000$ landmarks, $5$ victims and $10$ rescuers, which actually need just one rescuer.  We can notice that, each victim can be reached by more than one rescuer and that the closer one is selected since.


\newcommand{\project}[1]{#1\!\!\upharpoonright}
\newcommand{\projectII}[2]{#2\!\!\downharpoonright_{#1}}
\newcommand{\sintll}[2]{#2\upharpoonright_{#1}}
\newcommand{\projectIII}[3]{#3\!\!\upharpoonright^{#1}_{#2}}
\newcommand{\frag}[5]{#1[~#2~|~#3:#4:#5]}

\section{On Distributing \smuc{} Computations}\label{section:calculus-local}

We discuss in this section the distributed implementation of \smuc{} computations. 
Needless to say, an obvious implementation would be based on a \emph{centralised} algorithm. In particular, the nodes could initially send all their information to a centralised coordinator that would construct the field, carry on the \smuc{} computations, and distribute the results back to the nodes. This solution is easy to realise and could be based on our prototype which indeed performs a centralised, global computation, as a sequential program acting on the field. 
However, such a solution has several obvious drawbacks: first, it creates a bottleneck in the coordinator. Second, there are many applications in which the idea of  constructing the whole field is not feasible and each agent needs to 
evolve independently. 
%

To provide a general framework for the distributed evaluation of \smuc{} computations we introduce some specific forms of programs that simplify their evaluation 
in a fully distributed environment.
We first define the class of \emph{elementary formulas} which are formulas that do not contain neither fixpoints nor formula variables, 
and that consist of at most one operator. 

\begin{defi}[elementary formulas]\label{def:elementary}
A \smuc{} formula $\Psi$ is \emph{elementary} if it has the following form:
\[
\Psi_{e} ::= j \mid 
%
f(j,\dots,j) 
\mid \aggregateOut{$\alpha$}{g}{j}
\mid \aggregateIn{$\alpha$}{g}{j}
\]
\noindent 
with $j \in \mathcal{L}$, $a \in \mathcal{L}'$, $f \in \mathcal{M}$. 
\end{defi}

Elementary formulas are used as  expressions in \emph{simple assignment} \smuc{} programs.  These are programs that only use \emph{elementary formulas}. Moreover, in this class of programs, boolean guards in $\mchange{\mathsf{until}}$ and $\mathsf{if-then-else}$ statements are always \emph{node labels}. 

\begin{defi}[simple assignment programs] A \smuc{} program $P$ is in \emph{simple assignment form} (SAF) if has the following syntax:
\[
\begin{array}{rcl}
S , R & ::= & \mathsf{skip} \mid
    i \leftarrow \Psi_e \mid
    S\ \mathsf{;}\ R \mid
    \mathsf{if}\ i\ \mathsf{then}\ S\ \mathsf{else}\ R 
   \mid 
    \mchange{\mathsf{until}}\ i\ \mathsf{do}\ S
    \mid
    \mathsf{free}( x,\ldots,x ) 
\end{array}
\]
\noindent 
where $i,j, x \in \mathcal{L}$, and $\Psi_e$ is a \smuc{} elementary formula (cf. Def~\ref{def:elementary}). 
\end{defi}

Above, we introduced a new costruct ($\mathsf{free}$) that is used to \emph{deallocate} labels during a program evaluation. The role of this construct will be clear later when
a distributed evaluation of \smuc{} programs is introduced. The operational semantics of Table~\ref{table:semantics-global1} is extended to consider the following rule:
\[
( \textsc{Free})
\qquad
\dfrac{
I_F'=I_{F}[^{\mathsf{undef}}/_{x_1},\ldots,^{\mathsf{undef}}/_{x_n}]}{ 
\langle \mathsf{free}(x_1,\ldots,x_n), F \rangle \rightarrow \langle \mathsf{skip} , F [^{I_{F}'}/_{I_F}]  \rangle
}
\]

\noindent
when $\mathsf{free}(x_1,\ldots,x_n)$ is executed all the labels $x_1$,\ldots, $x_n$, are removed from the interpretation in $F$. This is 
denoted with the value $\mathsf{undef}$.

One can also observe that a \smuc{} program $S$ in SAF does not contain any fixpoint formulas. However, if we consider only fields with a domain field with finite partially ordered chains only, like in Theorem~\ref{th:asynchrony-theorem2}, this does not limit the expressive power of our language. Indeed, fixpoints can be explicitly computed by using the other constructs of \smuc{}. In Table~\ref{def:transforminsaf} function $\mathcal{P}$ (and the auxiliary function $\mathcal{A}$)
is defined to transform a \smuc{} program $P$ into an equivalent program $S$ in SAF. This transformation introduces a set of 	\emph{auxiliary} node labels denoted by $x_{k}$ that
we assume to be distinct from all the other symbols and not occurring in $P$. 
From now on we will use $\mathcal{X}\subseteq \mathcal{L}$ to denote the set of this auxiliary \emph{node labels}. 
To guarantee the appropriate allocation of these symbols, function $\mathcal{P}$ (and the auxiliary function $\mathcal{A}$) is parametrised with a counter $c$ that indicates the number of \emph{auxiliary} symbols already introduced in the transformation. The result of $\mathcal{P}(P)_{c}$
is a pair $[ S,c']$: $S$ is a SAF \smuc{} program, while $c'$ indicates the number of symbols allocated in the transformation. 
In the following we will use $S_{P}$ to denote that, for some $c$ and $c'$, $\mathcal{P}(P)_{c}=[ S_{P} , c' ] $. Similary, we will use $S^{i}_{\Psi}$ to denote that, for some $c$ and $c'$, $\mathcal{A}(\Psi)_{c}^{i}=[ S_{\Psi}^{i} , c' ]$.

\begin{table}[tbp]
\begin{tabular}{c}
$\mathcal{P}( \mathsf{skip} )_{c} = [ \mathsf{skip}, c]$ \qquad
$
\dfrac{
\mathcal{A}( \Psi )^{i}_{c}=[ S , c' ]
}{
\mathcal{P}( i \leftarrow \Psi )_{c} = S;\mathsf{free}(x_{c},\ldots,x_{c'-1})
}
$\\[.5cm]
$\dfrac{
\mathcal{P}(P)_c=[ S, c'] \qquad
\mathcal{P}(Q)_{c'}=[ R, c'']
}{
\mathcal{P}(P\mathsf{;}Q)_{c} = [ S\mathsf{;}\mathsf{wait}(x_{c''})\mathsf{;}R,c''+1] 
}
$\\[.5cm]
$
\dfrac{
\mathcal{A}(\Psi)^{x_{c}}_{(c+1)}=[ R,c'] \quad
\mathcal{P}(P)_{c'}=[ S_1, c''] \qquad
\mathcal{P}(Q)_{c'''}=[ S_2, c''']
}{
\mathcal{P}(\mathsf{if}\ \Psi\ \mathsf{then}\ P\ \mathsf{else}\ Q)_c =
[ R\mathsf{;}\mathsf{if}\ x_{c}\ \mathsf{then}\ S_1\ \mathsf{else}\ S_2 , c''']
}
$\\[.5cm]
$
\dfrac{
\mathcal{A}(\Psi)^{x_{c}}_{(c+1)}=[ R,c'] \quad
\mathcal{P}(P)_{c'}=[ S, c''] \qquad
}{
\mathcal{P}(
\mchange{\mathsf{until}}\ \Psi\ \mathsf{do}\ P)_c =
[
R\mathsf{;}\mchange{\mathsf{until}}\ x_{c}\ \mathsf{do}\ S\mathsf{;}\mathsf{wait}(x_{c''})\mathsf{;}R , 
c''+1]
}$\\[1cm]
$\mathcal{A}( i )^{j}_{c} = [ j\leftarrow i , c ]$ 
\quad
$
\dfrac{
\mathcal{A}( \Psi_1 )^{x_{c}}_{c+1} = [ S_1,c_1] \quad \cdots\quad
\mathcal{A}( \Psi_n)^{x_{c_{n-1}}}_{c_{n-1}+1} = [ S_n,c_n]
}{
\mathcal{A}( f(\Psi_1,\dots,\Psi_n) )^{j}_{c} = [  S_1\ \mathsf{;}\cdots\mathsf{;}\ S_n\ \mathsf{;}\ j\leftarrow f(x_{c},\ldots,x_{c_{n-1}}),c_{n}]
}
$\\[.75cm]
$
\dfrac{
\mathcal{A}( \Psi )^{x_{c}}_{c+1} = [ S , c' ]
}{
\mathcal{A}( \aggregateOut{$\alpha$}{g}{\Psi} )^{j}_{c} = [ S\ \mathsf{;}\ i\leftarrow \aggregateOut{$\alpha$}{g}{x_{c}},c']
}
$\quad 
$
\dfrac{
\mathcal{A}( \Psi )^{x_{c}}_{c+1,\rho} = [ S , c' ]
}{
\mathcal{A}( \aggregateIn{$\alpha$}{g}{\Psi} )^{j}_{c} = [ S\ \mathsf{;}\ i\leftarrow \aggregateIn{$\alpha$}{g}{x_{c}},c']
}
$\\[.75cm]
$
\dfrac{
\mathcal{A}( \Psi[^{x_{c}}/_{z}] )^{x_{c+1}}_{c+2} = [ S , c' ]
}{
\mathcal{A}( \mu z.\Psi )^{j}_{c} = \left[ 
\begin{array}{l}
x_{c}\leftarrow \bot\ \mathsf{;}\\
x_{c+1}\leftarrow \bot\ \mathsf{;}\\
x_{c'}\leftarrow \mchange{\mathit{false}} \mathsf{;}\\
\mchange{\mathsf{until}}\ x_{c'}\ \mathsf{do}\\
\phantom{xxx} x_{c}\leftarrow x_{c+1}\mathsf{;}\\
\phantom{xxx} S\ \mathsf{;}\\
\phantom{xxx} x_{c'}\leftarrow x_{c}=x_{c+1}\\
j\leftarrow x_{c+1}
\end{array},c'+1\right]
}
$\\[4.0cm]
$
\dfrac{
\mathcal{A}( \Psi[^{x_{c}}/_{z}] )^{x_{c+1}}_{c+2} = [ S , c' ]
}{
\mathcal{A}( \nu z.\Psi )^{j}_{c} = \left[ 
\begin{array}{l}
x_{c}\leftarrow \top\ \mathsf{;}\\
x_{c+1}\leftarrow \top\ \mathsf{;}\\
x_{c'}\leftarrow \mchange{\mathit{false}} \mathsf{;}\\
\mchange{\mathsf{until}}\ x_{c'}\ \mathsf{do}\\
\phantom{xxx} x_{c}\leftarrow x_{c+1}\mathsf{;}\\
\phantom{xxx} S\ \mathsf{;}\\
\phantom{xxx} x_{c'}\leftarrow x_{c}=x_{c+1}\\
j\leftarrow x_{c+1}
\end{array},c'+1\right]
}
$\\[.75cm]

\end{tabular}
\caption{Function $\mathcal{P}$ that transforms a \smuc{} program $P$ in SAF.}
\label{def:transforminsaf}
\end{table}

Function $\mathcal{P}$ is inductively defined on the syntax of \smuc{} programs. The translation of $\mathsf{skip}$ and $P\mathsf{;}Q$ are straightforward. In the first case 
$\mathcal{P}$ does not change $\mathsf{skip}$ without allocating any auxiliary label while in the second case the translation of $P\mathsf{;}Q$ is obtained as \changeR{the} sequentialisation
of  $\mathcal{P}(P)_{c}$ and $\mathcal{P}(Q)_{c'}$, where $c'$ indicates the amount of symbols allocated in the translation of $P$.
\changeR{The macro $\mathsf{wait}$ is used between the two processes}. This is defined as follows:
\[
\mathsf{wait}(x) \equiv x\leftarrow \mchange{true}; \mchange{\mathsf{until}}~x~\mathsf{do}~\mathsf{skip}
\]
The use of this statement has no impact in the global evaluation of a \smuc{} program. However, when distributed executions will be considered, a $\mathsf{wait}$ \changeR{statement} 
can be used as a barrier for a global synchronisation in the field.

Each assignment $i\leftarrow \Psi$ is translated in\changeR{to} a program that first evaluates formula $\Psi$ and then assigns the result to $i$. After that all the auxiliary labels used in the computation of $\Psi$ are deallocated. The program computing $\Psi$ is obtained via 
function $\mathcal{A}(\Psi)^{i}_{c}$, that is also defined in Table~\ref{def:transforminsaf} and that is described below.
Function $\mathcal{P}$ translates a statement of the form $\mathsf{if}~\Psi~\mathsf{then}~P~\mathsf{else}~Q$ in a program that first evaluates formula $\Psi$ 
storing the result in the auxiliary label $x_{c}$ which is then used to select either the translation of $P$ or the translation of $Q$.  
Translation of $\mchange{\mathsf{until}}~\Psi~\mathsf{do}~P$ is similar to the previous one. At the end of the same body the construct $\mathsf{wait}(x)$
is used. Again, the role of this statement will be clear later when a distributed execution of \smuc{} programs is considered.

Function $\mathcal{A}$ is defined inductively on the syntax of formulas $\Psi$ and takes as parameter a counter $c$, that is used to allocate auxiliary labels.
Similarly to function $\mathcal{P}$, function $\mathcal{A}$ returns a pair consisting of a \smuc{} program and of a counter of allocated auxiliary node labels.
When $\Psi$ is a label $i$, $\mathcal{A}(\Psi)^{j}_{c}$ \change{(which arises from the translation of $j\leftarrow i$ in $\mathcal{P}( i \leftarrow \Psi )_{c}$)}  is just $j\leftarrow i$ and no auxiliary variable is created.  When $\Psi$ is 
$f(\Psi_1,\dots,\Psi_n)$  (resp. $\aggregateOut{$\alpha$}{g}{\Psi_1}$, $\aggregateIn{$\alpha$}{g}{\Psi_1}$), $\mathcal{A}(\Psi)^{j}_{c}$ consists of
the \smuc{} program that uses auxiliary variables $x_{c}$, \ldots, $x_{c_{n-1}}$ (resp. $x_{c}$) to store the evaluation to $\Psi_i$ and then
assings to $j$ the evaluation of the \emph{simple} formula $f(x_{c},\ldots,x_{c_{n-1}})$ (resp.   $\aggregateOut{$\alpha$}{g}{x_{c}}$, $\aggregateIn{$\alpha$}{g}{x_{c}}$).
The program that evaluates $\mu z.\Psi$ (resp. $\nu z.\Psi$) uses two auxiliary variables, namely $x_{c}$ and $x_{c+1}$. The former is initialised to $\bot$ (resp. $\top$),
the latter will contain the evaluation of $\Psi$ performed by  $\mathcal{A}(\Psi[^{x_{c}}/_{z}])^{x_{c+1}}_{c+2}$.This evaluation continues until $x_{c}$ will be
equal to $x_{c+1}$ (If not, $x_{c}$ is assigned to $x_{c+1}$).
This means that after $i>0$ iterations $x_{c}$ and $x_{c+1}$ contain the evaluation of approximants $(i-1)$ and $i$ of $\mu z.\Psi$ (resp. $\nu z.\Psi$).

The following Lemma guarantees that any formula $\Psi$, when interpreted over a field $F$ satisfying conditions of Theorem~\ref{th:asynchrony-theorem2},
can be evaluated by the SAF \smuc{} program obtained from the application of function $\mathcal{A}$.

\begin{restatable}{lemm}{exptosaf}
\label{lem:exptosaf}
Let $F$ be a field with field domain $A$ with finite chains only, $\Psi$ a formula, $c\in \mathbb{N}$, and label $i$. 
Let $\mathcal{A}(\Psi)_{c}^{i}=[ S , c' ]$, then:
\[
\llbracket \Psi\rrbracket^F_{\emptyset}=f 
\Leftrightarrow
 \langle S , F \rangle \rightarrow^{*} \langle \mathsf{skip} , F' \rangle \mbox{ and } I_{F'}(i)=f
 \]
%
%
\end{restatable}

\begin{restatable}{lemm}{eqone}
\label{lem:eq1}
Let $F$ be a field with field domain $A$ with finite chains only, for any $P$ the following holds:
\begin{itemize}
\item if $\langle  P , F \rangle  \rightarrow \langle P'  , F' \rangle$ then $\langle  S_{P} ,   F \rangle \rightarrow^{*} \langle S_{P'}  , F''\rangle$ and $F'=F''\backslash \mathcal{X}$;
\item if $\langle  S_{P} , F \rangle  \rightarrow \langle S' , F' \rangle$ then there exist $P'$ and $F''$ such that $\langle S'  , F' \rangle \rightarrow^{*} \langle S_{P'}  , F''\rangle$ and $\langle P,F\rangle \rightarrow \langle P',F''\backslash \mathcal{X}\rangle$.
\end{itemize}

\noindent
Above, we use $F'\backslash \mathcal{X}$ to refer to the field obtained from $F'$ by erasing all the labels in $\mathcal{X}$. 
\end{restatable}

\subsection{Asynchronous agreement.}

We describe now a technique that, by relying on a specific structure, can be used to perform \smuc{} computations in an fully distributed way. 
Here we assume that each node $n$ in the field has its own computational power and that it can interact with its neighbours to locally evaluate its part of the field.
However, to guarantee a correct execution of the program, a global coordination mechanism is needed. 
The corner stone of the proposed algorithm is a \emph{tree-based} infrastructure that spans the complete field. In this infrastructure each node/agent, that is referenced by a unique identifier, is responsible for the coordination of the computations occurring in its sub-tree. 
In the rest of this section we assume that this \emph{spanning tree} is computed in a \emph{set-up} phase executed when the system is deployed.

\begin{defi}[distributed field infrastructure]
A \emph{distributed field infrastructure} is a pair $\langle F,T\rangle$ where $F$ is a field
while $T\subseteq N_F\times N_F$ is a \emph{spanning tree} of the \change{underlying undirected} graph \change{$(N,E \cup E^{-1})$ of $F$, where} $(x,y)\in T$ if and only if $x$ is the parent of $y$ in the spanning tree.
Given a tree $T\subseteq N\times N$, let $children(T,n)=\{ n' | (n,n') \in T \}$, $rel(T,n)=\{ n' | (n,n')\in T\vee (n',n)\in T\}$, and $parent(T,n)=n'$ if and only if $(n',n)\in T$.
\end{defi}

Given a \emph{distributed field infrastructure} $\langle F,T\rangle$, we will use a variant of the \emph{Dijkstra-Scholten} algorithm~\cite{DS80} for termination detection to check if a global agreement 
on the evaluation of a given node label $x$ has been reached or not, where $x$ takes value on the standard boolean lattice $\{true,false\}$.
To check if an agreement has been reached or not each node uses two elements: an \emph{agreement store} 
$\chi:N\rightarrow L_N\rightarrow \mathbb{N}\rightarrow \{\mathsf{undef},?true,true,false\}$ and a \emph{counter} $\kappa: L_N\rightarrow \mathbb{N}$. 
Via the \emph{agreement store} $\chi$ each node stores the status of the agreement of labels collected from its relatives in the spanning tree $T$.
Since an agreement on the same label can be iterated in a \smuc{} program, and to avoid confusions among different iterations, a different value is stored for each
iteration. Counter $\kappa$ is then used to count the iterations associated with a label $x$.
If $\chi_{n}$ is the \emph{agreement store} used by node $n$, $\chi_{n}(n',k,x)$ is evaluated to: 
\begin{itemize}
\item $\mathsf{undef}$, when $n$ does not know the state of evaluation of 
label $x$ at $n'$ after $k$ iterations; 
\item $?true$ when $n'$ and all the nodes in its subtree have evaluated $x$ to $\mathit{true}$ at iteration $k$;
\item $\mathit{false}$ when at least one node in the field has evaluated $x$ to $\mathit{false}$ at iteration $k$;
\item $\mathit{true}$ when at iteration $k$ an agreement has been reached on $x=true$.
\end{itemize}
%
%
If a node $n$ at iteration $k$ evaluates $x$ to $\mathit{false}$, a message is sent to the relatives of $n$ in $T$. If at the same iteration $k$, the evaluation of $x$ at $n$ is $\mathit{true}$, 
for each child $n'$, $\chi_{n}(n',k,x)=?true$, $\chi_n$ is updated to let $\chi_n(n,k,x)=?true$ and a message is sent to its parent.
When this information is propagated in the spanning tree from leaves to the root, the latter is able to identify if at iteration $k$ an agreement on $x=true$ 
has been reached. After that, a notification message flows from the root  to the leaves of $T$ and each node $n$ will set $\chi_n(n,k,x)=true$.

\subsection{Distributed execution of \smuc{} programs}
A distributed execution of a \smuc{} program over a \emph{distributed field infrastructure} consists of a set of \emph{fragments} 
executed over each node in the field. In a \emph{fragment} each node computes its part of the field and interacts with its neighbours
to exchange the computed values. 

\begin{defi}[distributed execution]
Let $F$ be a \emph{field}, we let $\mathcal{D}$ be the set of  \emph{distributed fragments} $d$ of the form:
\[
d_{i}=\frag{n}{S}{ \iota}{\chi}{ \kappa}
\]

\noindent 
where $n\in N_{F}$, $S$ is SAF \smuc{} program, $\iota: N\rightarrow L_{N}\rightarrow N\rightarrow  A$ is a \emph{partial interpretation} of \emph{node labels} at $n$,
$\chi$ is an \emph{agreement store} and $\kappa$ is an \emph{agreement counter}.  
%
A \emph{distributed execution} $D$ for $F$ is a subset of $\mathcal{D}$ such that 
\change{$D$ consists of one fragment $d=\frag{n}{S}{ \iota}{\chi}{ \kappa}$ (for some $S$, $\iota$, $\chi$ and $\kappa$) for each node $n \in N_{F}$}.
\end{defi}

In a \emph{fragment} $d=\frag{n}{S}{ \iota}{\chi}{ \kappa}$, $S$ represents the portion of the program currently executed at $n$, $\iota$ is the portion of 
the field computed at $n$ together with the part of the field collected from the neighbour of $n$, $\chi$ and $\kappa$ are the structures described in the previous subsection to manage the agreement in \smuc{} computations.

\begin{table}[tbp]
{\begin{footnotesize}
\begin{tabular}{c}
(\textsc{D-Seq1} )
$
\dfrac{
\frag{n}{S}{ \iota}{\chi}{ \kappa} \xrightarrow{\lambda}_{\langle F,T\rangle} \frag{n}{S'}{ \iota'}{\chi'}{ \kappa'}
}{
\frag{n}{\mathsf{skip};S}{ \iota}{\chi}{ \kappa} \xrightarrow{\lambda}_{\langle F,T\rangle} \frag{n}{S'}{ \iota'}{\chi'}{ \kappa'}
}
$ 
\\[.75cm]
(\textsc{D-Seq2})
$
\dfrac{
\frag{n}{S}{ \iota}{\chi}{ \kappa} \xrightarrow{\lambda}_{\langle F,T\rangle} \frag{n}{S'}{ \iota'}{\chi'}{ \kappa'}
}{
\frag{n}{S;R}{ \iota}{\chi}{ \kappa} \xrightarrow{\lambda}_{\langle F,T\rangle} \frag{n}{S';R}{ \iota'}{\chi'}{ \kappa'}
}
$
\\[.75cm]
(\textsc{D-Step} )
$
\dfrac{
\Psi \mapsto_{F}^{n,\iota} v \qquad
X = \{ n' | (n,n') \in E_{F} \vee (n',n)\in E_{F} \}
}{
\frag{n}{i\leftarrow \Psi}{ \iota}{\chi}{ \kappa}
\xrightarrow{\overline{\langle n,i,v\rangle@X}}_{\langle F,T\rangle}
\frag{n}{\mathsf{skip}}{ \iota[^{v}/_{(i)(n)}]}{\chi}{ \kappa}
}
$
\\[.75cm]
(\textsc{D-Free} )
$
\dfrac{
\iota'=\iota[^{\lambda n.\mathsf{undef}}/_{x_1},\cdots,^{\lambda n.\mathsf{undef}}/_{x_n}]
}{
\frag{n}{\mathsf{free}(x_1,\ldots,x_n)}{ \iota}{\chi}{ \kappa}
\xrightarrow{\tau}_{\langle F,T\rangle}
\frag{n}{\mathsf{skip}}{ \iota'}{\chi}{ \kappa}
}
$
\\[.75cm]
(\textsc{D-IfT})
$
\dfrac{
\kappa(i) = c\qquad
\chi(n)(i)(k)=true 
}{
\frag{n}{\mathsf{if}~i~\mathsf{then}~S~\mathsf{else}~R}{ \iota}{\chi}{ \kappa}
\xrightarrow{\tau}_{\langle F,T\rangle}
\frag{n}{S}{ \iota}{\chi}{ \kappa[^{i}/_{c+1}]}
}
$
\\[.75cm]
(\textsc{D-IfF})
$
\dfrac{
\kappa(i) = c\qquad
\chi(n)(i)(k)=false 
}{
\frag{n}{\mathsf{if}~i~\mathsf{then}~S~\mathsf{else}~R}{ \iota}{\chi}{ \kappa}
\xrightarrow{\tau}_{\langle F,T\rangle}
\frag{n}{R}{ \iota}{\chi}{ \kappa[^{i}/_{c+1}]}
}
$
\\[.75cm]
(\change{\textsc{D-UntilF}})
$
\dfrac{
\kappa(i) = c\qquad
\chi(n)(i)(k)=\mchange{false} 
}{
\frag{n}{\mchange{\mathsf{until}}~i~\mathsf{do}~S}{ \iota}{\chi}{ \kappa}
\xrightarrow{\tau}_{\langle F,T\rangle}
\frag{n}{S;\mchange{\mathsf{until}}~i~\mathsf{do}~S}{ \iota}{\chi}{ \kappa[^{i}/_{c+1}]}
}
$\\[.75cm]
(\change{\textsc{D-UntilT}})
$
\dfrac{
\kappa(i) = c\qquad
\chi(n)(i)(k)=\mchange{true} 
}{
\frag{n}{\mchange{\mathsf{until}}~i~\mathsf{do}~S}{ \iota}{\chi}{ \kappa}
\xrightarrow{\tau}_{\langle F,T\rangle}
\frag{n}{\mathsf{skip}}{ \iota}{\chi}{ \kappa[^{i}/_{c+1}]}
}
$
\end{tabular}
\end{footnotesize}}
\caption{Distributed Semantics of  \smuc{} programs (\emph{fragments}).}
\label{tab:distopsem1}
\end{table}

The semantics of distributed \smuc{} programs is defined via the labelled transition relations defined in Tab.~\ref{tab:distopsem1}, Tab.~\ref{tab:distopsem2} and Tab.~\ref{tab:distopsem3}. 
\changeR{The} behaviour of the single fragment $d$ is described by the transition relation $\xrightarrow{\cdot}_{\langle F,T\rangle}\subseteq \mathcal{D}\times\Lambda\times\mathcal{D}$
defined in Tab.~\ref{tab:distopsem1} and Tab.~\ref{tab:distopsem2} where $\Lambda$ denotes the set of transition labels $\lambda$ having the following syntax:
\[
\lambda ::= \tau \mid m \mid \overline{m}\qquad m ::= \langle n , i , v \rangle@X \mid \langle n ,  i , c , v \rangle@X 
\]
where $n\in N_{F}$, $X\subseteq N_{F}$, $i\in \mathcal{L}_{N}$ and $v\in A_{F}$.
Following a standard notation in process algebras, transition label $\tau$ identifies internal operations. A transition is labelled with $\overline{m}$ when
a message $m$ \change{is sent}. Finally, transitions labelled with $m$ show how a fragment reacts when the message $m$ is received.

Rules in Tab.~\ref{tab:distopsem1} are similar to the corresponding ones in Tab.~\ref{table:semantics-global1}. 
However, while in the global semantics all elements of the field are synchronously evaluated, via the rules in Tab.~\ref{tab:distopsem1}  
each \emph{fragment} proceeds independently. 
Rules (\textsc{D-Seq1}) and (\textsc{D-Seq2}) are standard, while rule (\textsc{D-Step}) deserves more attention. It relies on relation $\mapsto_{F}^{n,\iota}$, defined in Tab.~\ref{tab:diseval} that is used 
to evaluate an \emph{elementary formula} $\Psi_e$ in a given node $n$ under a specific \emph{partial interpretation} $\iota$ and \emph{context} $\rho$.
A formula $\Psi_e$ can be directly evaluated when it is either a label $i$ or a function $f(i_1,\ldots,i_n)$. In this case, the evaluation simply relies on
the local evaluation $\iota$. However, when $\Psi_e$ is either $\aggregateOut{$\alpha$}{g}{i}$ or $\aggregateIn{$\alpha$}{g}{i}$, the evaluation is possible only when
for each node $n'$ in the poset (resp. preset) of $n$, $\iota(n')(i)$ is defined. When all these values are available, the evaluation of 
$\aggregateOut{$\alpha$}{g}{i}$ (resp. $\aggregateIn{$\alpha$}{g}{i}$) consists in the appropriate aggregation of values obtained from neighbours following outgoing 
($\aggregateOut{}{}{}$) or incoming ($\aggregateIn{}{}{}$) edges and using the edge capability $a$ with function $g$. 
When $\Psi_e$ can be evaluated to value $v$, $\frag{n}{i\leftarrow \Psi}{ \iota}{\chi}{ \kappa}$ can perform a step and $\iota$ is updated to consider
the new value for label $i$ ($\iota[^{v}/_{(i)(n)}]$) while the message $\langle n , i , v \rangle@X$ is sent to all the neighbours of 
$n$ ($X=\{ n' | (n,n')\in E_{F}\vee (n',n)\in E_F \}$) to notify them that the value of $i$ is changed at $n$. 

We want to remark that rule (\textsc{D-Step}) can be applied only when all the values needed to evaluate formula $\Psi_e$ are locally available. 
This means that an \emph{assignment} can be a barrier in a distributed computation. To guarantee that in the execution only updated values
are used, command $\mathsf{free}$ can be used. By applying rule (\textsc{D-Free}) all the labels passed as arguments are deallocated in $\iota$.  

Finally, rules (\textsc{D-IfT}), (\textsc{D-IfF}), (\change{\textsc{D-UntilF}}) and (\change{\textsc{D-UntilT}}) are as expected. We can notice that these rules are applied 
only when the label used as condition in the statement is evaluated in the agreement structure ($\chi(n)(c)(i)=v\in\{ true,false\}$). Moreover, when one of these rule is applied, the \emph{agreement counter} is updated ($\kappa[^{i}/_{c+1}]$) to avoid interferences with  subsequent evaluations of the same statement. 

%

\begin{table}[tbp]
\begin{tabular}{c}
$
\dfrac{
\iota(n)(i)=v
}{
i\mapsto^{n,\iota}_{F} v
}
$ \quad
$
\dfrac{
\iota(n)(i_1)=v_1\cdots \iota(n)(i_n)=v_n
}{
f(i_1,\ldots,i_n) \mapsto^{n,\iota}_{F} f(v_1,\ldots,v_n)
}
$ \\[.75cm]
$
\dfrac{
\forall n': (n,n')\in E_{F}: \iota(n')(i) \not= \mathsf{undef}
}{
\aggregateOut{$\alpha$}{g}{i}\mapsto^{n,\iota}_{F} \llbracket g \rrbracket_{A_F} ( \{ I_F (\alpha)(n,n')(\iota(n')(i)) \mid (n,n') \in  E_F\} )
}
$\\[.75cm]
$
\dfrac{
\forall n': (n',n)\in E_{F}: \iota(n')(i) \not= \mathsf{undef}
}{
\aggregateIn{$\alpha$}{g}{i}\mapsto^{n,\iota}_{F} \llbracket g \rrbracket_{A_F} ( \{ I_F (\alpha)(n',n)(\iota(n')(i)) \mid (n,n') \in  E_F\} )
}
$ 
\end{tabular}
\caption{Distributed evaluation of formula.}
\label{tab:diseval} 
\end{table}

Rules in Tab.~\ref{tab:distopsem2} show how a node interacts with the other nodes in the field to check if an agreement has been reached or not in the field
and implement the coordination mechanism discussed in the previous subsection. 
All these rules can be applied only when $isGuard( P , i )$ is true, namely when $P$ is either $\mathsf{if}~i~\mathsf{then}~P'~\mathsf{else}~Q'$
or $\mchange{\mathsf{until}}~i~\mathsf{do}~P'$. 

Rule (\textsc{D-AgreeF1}) is applied when in a node $n$ the label $i$ is evaluated to $\mathit{false}$ and no information about the agreement at the current iteration
$\kappa(i)=c$ is available ($\chi(i)(n)(c)=\mathsf{undef}$). In this case we can soon establish that the agreement is not reached at this iteration. Hence, this information
is locally stored in the \emph{agreement structure} ($\chi[^{false}/_{(n)(i)(c)}]$) while all the relatives in the spanning tree are notified with 
the message $\langle n,i,c,false\rangle$. 
Note that, after the application of rule (\textsc{D-AgreeF1}), either rule (\textsc{D-IfF}) or rule (\change{\textsc{D-UntilF}}) will be enabled.

When label $i$ is evaluated to $\mathit{true}$ ($\iota(i)(n)=true$) \changeR{at $n$}, data from the children of $n$ are needed to establish \changeR{whether} an agreement is possible.  
If one of the children of $n$ has notified $n$ that the agreement on $i$ has not been reached at iteration $c$ (i.e. $\exists n':(n,n')\in T: \chi(i)(n')(c)=false$)
rule (\textsc{D-AgreeF2}) is applied. Like for rule (\textsc{D-AgreeF1}), the \emph{agreement structure} is updated and all the relatives in the spanning tree are 
informed that an agreement has not \changeR{been} reached \changeR{yet}.
Otherwise, when $n$ has received information about a \emph{local agreement} \changeR{from all its children}, i.e. $\forall n':(n,n')\in T: \chi(i)(n')(c)=?true$, the local agreement 
structure is updated accordingly. If $n$ is not the root of $T$, rule (\textsc{D-AgreeT}) is applied and the parent of $n$ is notified about the possible agreement.
While, if $n$ is the root of $T$ an agreement is reached: rule (\textsc{D-AgreeN1}) is applied and all the children of $n$ are then notified.
At this point, rule (\textsc{D-AgreeP}) is used to propagate the status of the agreement from the root of the tree to its leaves.

\begin{table}[tbp]
{\begin{footnotesize}
\begin{tabular}{c}
(\textsc{D-AgreeF1})
$
\dfrac{
\begin{array}{c}
isGuard( S , i )\quad
\iota(i)(n)=false\quad
\kappa(i)=c\\[.1cm]
\chi(n)(i)(c)=\mathsf{undef}\quad 
X = \{ n' | (n',n) \in T \vee (n,n')\in T \}
\end{array}
}{
\frag{n}{S}{ \iota}{\chi}{\kappa}
\xrightarrow{\overline{\langle n,i,c,false\rangle@X}}_{\langle F,T\rangle}
\frag{n}{S}{ \iota}{\chi[^{false}/_{(n)(i)(c)}]}{\kappa}
}
$
\\[.75cm]
(\textsc{D-AgreeT})
$
\dfrac{
\begin{array}{c}
isGuard( S , i )\quad
\iota(i)(n)=true\quad
\kappa(i)=c\quad
\chi(n)(i)(c)=\mathsf{undef}\\[.1cm]
\forall n':(n,n')\in T: \chi(i)(n')(c)=?true\quad
(n',n)\in T
\end{array}
}{
\frag{n}{S}{ \iota}{\chi}{\kappa}
\xrightarrow{\overline{\langle n,i,c,?true\rangle@\{ n' \}}}_{\langle F,T\rangle}
\frag{n}{S}{ \iota}{\chi[^{?true}/_{(n)(i)(c)}]}{\kappa}
}
$
\\[.75cm]
(\textsc{D-AgreeF2})
$
\dfrac{
\begin{array}{c}
isGuard( S , i )\quad
\iota(i)(n)=true\quad
\kappa(i)=c\quad
\chi(n)(i)(c)=\mathsf{undef}\\[.1cm]
\exists n':(n,n')\in T: \chi(i)(n')(c)=false\quad 
X = \{ n' | (n',n) \in T \vee (n,n')\in T \}
\end{array}
}{
\frag{n}{S}{ \iota}{\chi}{\kappa}
\xrightarrow{\overline{\langle n,i,c,false\rangle@X}}_{\langle F,T\rangle}
\frag{n}{S}{ \iota}{\chi[^{\mchange{_\mathit{false}}}/_{(n)(i)(c)}]}{\kappa}
}
$
\\[.75cm]
(\textsc{D-AgreeN1})
$
\dfrac{
\begin{array}{c}
isGuard( S , i )\quad
\iota(i)(n)=true\quad
\kappa(i)=c\quad
\chi(n)(i)(c)=\mathsf{undef}\\[.1cm]
\forall n':(n,n')\in T: \chi(i)(n')(c)=?true\quad
isRoot(n,T)\quad
X = \{ n' | (n,n')\in T \}
\end{array}
}{
\frag{n}{S}{ \iota}{\chi}{\kappa}
\xrightarrow{\overline{\langle n,i,c,true\rangle@X}}_{\langle F,T\rangle}
\frag{n}{S}{ \iota}{\chi[^{\mchange{_\mathit{false}}}/_{(n)(i)(c)}]}{\kappa}
}
$
\\[.75cm]
(\textsc{D-AgreeP})
$
\dfrac{
\begin{array}{c}
isGuard( S , i )\quad
\kappa(i)=c\quad
\chi(n)(i)(c)=?v\\[.1cm]
(n',n)\in T\quad
\chi(n')(i)(c)=v'\in\{ true,false\}\quad
X = \{ n' | (n,n')\in T \}
\end{array}
}{
\frag{n}{S}{ \iota}{\chi}{\kappa}
\xrightarrow{\overline{\langle n,i,c,v\rangle@X}}_{\langle F,T\rangle}
\frag{n}{S}{ \iota}{\chi[^{v}/_{(n)(i)(c)}]}{\kappa}
}
$
\end{tabular}
\end{footnotesize}}
\caption{Distributed Semantics of  \smuc{} programs (\emph{agreement}).}
\label{tab:distopsem2}
\end{table}

The behaviour of a distribute execution $D$ is described via the transition relation 
$\xRightarrow{\cdot}_{\langle F,T\rangle}\subseteq 2^{\mathcal{D}}\times\Lambda\times2^{\mathcal{D}}$
defined in Tab.~\ref{tab:distopsem3}. 
These rules are almost standard and describe the interaction among the fragments of a distributed execution $D$. 
In Tab.~\ref{tab:distopsem3} we use $D=D_1\oplus D_2$ to denote that $D=D_1\cup D_2$ and $D_1\cap D_2=\emptyset$.
In the following we will write $D_1\xRightarrow{}_{\langle F,T\rangle} D_2$ to denote that $D_1\xRightarrow{\lambda}_{\langle F,T\rangle} D_2$,
with $\lambda=\tau$ of $\lambda=\overline{m}$; $D_1\xRightarrow{}^{*}_{\langle F,T\rangle} D_2$ is reflexive and transitive closure
of $D_1\xRightarrow{}_{\langle F,T\rangle} D_2$.

Rule (\textsc{D-Comp}) lifts  transitions from the level of fragments to the level of distributed executions. Rules (\textsc{R-Field}) and
(\textsc{R-Agree}) show how a fragment reacts when a new message is received, that is updating the partial field evaluation $\iota$ when 
a message of the form $\langle n', i, v\rangle$ is received, and updating the agreement structure $\chi$ when a message of the form
$\langle n', i, c, v\rangle$ is received. 
Rules (\textsc{I-Field}) and (\textsc{I-Agree}) are used when a node is not the recipient of a message, while (\textsc{D-Int}), (\textsc{D-Sync}) 
and (\textsc{D-Recv}) describe possible interactions at the level of systems.

\begin{table}[tbp]
{\begin{footnotesize}
\begin{tabular}{c}
(\textsc{D-Comp})
$
\dfrac{
\frag{n}{S}{ \iota}{\chi}{\kappa}
\xrightarrow{\lambda}
\frag{n}{S'}{ \iota'}{\chi'}{\kappa'}
}{
\{ \frag{n}{S}{ \iota}{\chi}{\kappa} \}
\xRightarrow{\lambda}
\{ \frag{n}{S'}{ \iota'}{\chi'}{\kappa'}\}
}
$
\\[.75cm]
(\textsc{R-Field})
$
\dfrac{
n\in X\quad
\iota(i)=\rho
}{
\{ \frag{n}{S}{ \iota}{\chi}{\kappa} \}
\xRightarrow{\langle n',i,v\rangle@X}
\{ \frag{n}{S}{ \iota[^{\rho[^{n'}/_{v}]}/_{i}}{\chi}{\kappa} \}
}
$
\\[.75cm]
(\textsc{R-Agree})
$
\dfrac{
n\in X
}{
\{ \frag{n}{S}{ \iota}{\chi}{\kappa} \}
\xRightarrow{\langle n',i,c,v\rangle@X}
\{ \frag{n}{S}{ \iota}{\chi[^{v}/_{(n)(i)(c)}]}{\kappa} \}
}
$
\\[.75cm]
(\textsc{I-Field})
$
\dfrac{
n\not\in X
}{
\{ \frag{n}{S}{ \iota}{\chi}{\kappa} \}
\xRightarrow{\langle n',i,v\rangle@X}
\{ \frag{n}{S}{ \iota}{\chi}{\kappa} \}
}
$
\\[.75cm]
(\textsc{I-Agree})
$
\dfrac{
n\not\in X
}{
\{ \frag{n}{S}{ \iota}{\chi}{\kappa} \}
\xRightarrow{\langle n',i,c,v\rangle@X}
\{ \frag{n}{S}{ \iota}{\chi}{\kappa} \}
}
$
\\[.75cm]
(\textsc{D-Int})
$
\dfrac{
D_1 \xRightarrow{\tau} D_1'
}{
D_1 \oplus D_2 \xRightarrow{\tau} D_1'\oplus D_2
}
$
\\[.75cm]
(\textsc{D-Sync})
$
\dfrac{
D_1 \xRightarrow{\overline{m}} D_1'\quad
D_2 \xRightarrow{m} D_2'
}{
D_1 \oplus D_2 \xRightarrow{\overline{m}} D_1'\oplus D_2'
}
$
\\[.75cm]
(\textsc{D-Recv})
$
\dfrac{
D_1 \xRightarrow{m} D_1'\quad
D_2 \xRightarrow{m} D_2'
}{
D_1 \oplus D_2 \xRightarrow{m} D_1'\oplus D_2'
}
$

\end{tabular}
\end{footnotesize}}
\caption{Distributed Semantics of  \smuc{} programs (\emph{interactions}).}
\label{tab:distopsem3}
\end{table}

We are now ready to introduce the key result of this section. Namely, that one is always able to switch from a centralised evaluation to a distributed execution. 
Indeed, we can define a \emph{projection operator} that given a \smuc{} program $S$ and a field $F$, distributes the execution of $S$ over the nodes in $N_{F}$.
To define this operator, we need first to introduce the projection of a an interpretation $I_{F}$ with respect to a set of nodes $X\subseteq N_{F}$.

\begin{defi}[interpretation projection]
Let $F$ a field, and $X\subseteq N_{F}$, the projection of $I_{F}$ to $X$ ($\projectII{X}{I_{F}}$) is the function $\iota$ such that:
\[
\iota(i)(n)= \left\{ \begin{array}{ll}
I_{F}(i)(n) \qquad & n\in X\\
\mathsf{undef} & \mbox{otherwise}
\end{array}\right.
\]
\end{defi}

\begin{defi}[program projection]
Let $F$ a field, and $S$ a program the projection of $S$ to $F$ ($\projectII{F}{S}$) is the function distributed execution $D$ such that:
\[
D =
\{
\frag{n}{S}{\projectII{{N(n)}}{I_{F}}}{\lambda n.\lambda i.\lambda c.\mathsf{undef}}{\lambda n.\lambda i.0}~|~n \in N_{F}
\}
\]
where for each $n\in N_{F}$, $N(n)=\{ n' | (n,n')\in E_{F} \vee (n',n)\in E_{F} \}$ is the set of neighbour of $n$ in $F$.
\end{defi}

Given a distributed execution $D$ we can reconstruct a global interpretation $I_F$ for a given field $F$. 

\begin{defi}[lifted interpretation] Let  be $F$ a field, and $D$ be a distributed execution, $\sintll{F}{D}$ denotes the interpretation $I$ such that:
\[
I(i)(n)=v \Leftrightarrow  \frag{n}{S}{\iota}{\chi}{\kappa}\in D\wedge \iota(i)(n)=v
\]
We say that a distributed execution $D$ \emph{agrees} with $F$ if and only if $I_{F}=\sintll{F}{D}$.
\end{defi}

Given a distributed execution $D$ it is sometime useful to check if all the nodes in $D$ are executing exactly the same piece of code. 

\begin{defi}[aligned execution] A distributed execution $D$ is \emph{aligned} at $S$ if and only if:
$\forall d\in D. d=\frag{n}{S}{\iota}{\chi}{\kappa}$, for some $n$, $\iota$, $\chi$ and $\kappa$. We will write $D_{S}$ to denote that the distributed
execution $D$ is aligned at $S$.
\end{defi}

We can notice that while the global operational semantics defined in Table~\ref{table:semantics-global1} is deterministic,
the distributed version considered in this section is not. This is due to the fact that each node can progress independently.
However, we will see below that in the case of our interest, we can guarantee the existence of a common \emph{flow}.
\change{We say that $D_1$ flows into $D_2$ if and only if any computation starting from $D_1$ eventually reaches $D_2$.}

\begin{defi}[execution flows]\label{def:confluence}
A distributed execution $D_1$ \emph{flows} into $D_2$ if and only if \change{$D_1\xRightarrow{}^{*}_{\langle F,T\rangle} D_2$} and
\begin{itemize}
\item \change{either $D_1=D_2$}
\item \change{or, for any $D'$ such that $D_1\xRightarrow{}_{\langle F,T\rangle} D'$, $D'$ flows into $D_2$;}
\end{itemize}
\end{defi}

\change{To reason about \emph{distributed computations} it is useful to introduce the appropriate notation that represents the execution of sequentially composed programs.}

\begin{defi}[concatenation]
Let $D$ be a distributed execution and $S$ a SAF \smuc{} program, we let $D;S$ denote:
\[
\{ \frag{n}{S';S}{\iota}{\chi}{\kappa} \mid \frag{n}{S'}{\iota}{\chi}{\kappa} \in D \}
\]
\end{defi}

\change{The following Lemma guarantees that sequential programs can be computed asynchronously, while preserving the final result, while each $\mathsf{wait}(x)$ represents a synchronization point.}

\begin{restatable}{lemm}{factsem}
\label{lem:factsem}
For any $S_1$ and $S_2$, and for any $D_{S_1}$ that flows into $D_{\mathsf{skip}}$, the following hold:
\begin{itemize}
\item if $D_{\mathsf{skip}};S_2$ flows into $D'$ then also $D_{S_1};S_2$ flows into $D'$;
\item $D_{S_1};\mathsf{wait}(x);S_2$ flows into $D_{\mathsf{skip}};\mathsf{wait}(x);S_2$.
\end{itemize}
\end{restatable}


The following theorem guarantees that, when we consider a field $F$ 
with field domain $A$ with finite chains only, any global computation of a \smuc{} program $P$ can be \changeR{realised} in terms of 
a distributed execution of its equivalent SAF program $S_{P}$. Moreover, any distributed execution will always converge to 
the same field computed by $P$. 

\begin{restatable}{lemm}{disexp}
\label{lem:disexp}
Let $F$ be a field with field domain $A$ with finite chains only, $\Psi$ a formula, and $S_{\Psi}^{x}$ the SAF \smuc{} program that evaluates $\Psi$, then
any distributed execution $D_{S_{\Psi}^{x}}$ that agrees with $F$,  $D_{S_{\Psi}^{x}}$ flows in a distributed execution $D_{\mathsf{skip}}$ such that
$\llbracket \Psi \rrbracket^{F}_{\emptyset}=\sintll{F}{D_{\mathsf{skip}}}(x)$.
\end{restatable}

\begin{restatable}{thm}{teoremaa}
Let $F$ be a field with field domain $A$ with finite chains only and $T$ be a spanning tree of $F$, for any \smuc{} program $P$,
for any $D_{S_{P}}$ that agrees with $F$:
\begin{enumerate}[label=(\roman*)]
\item[(i)] if $\langle P,F\rangle \rightarrow \langle P',F'\rangle$ then $D_{S_{P}}\xRightarrow{}_{\langle F,T\rangle} D_{S_{P'}}$ and 
$D_{S_{P'}}$ agrees with $F'$.
\item[(ii)] For any $D'$ such that $D_{S_{P}}\xRightarrow{}^{*}_{\langle F,T\rangle} D'$ there exists $P'$ such that $D'\xRightarrow{}^{*}_{\langle F,T\rangle} D_{S_{P'}}$,
$\langle P,F\rangle \rightarrow \langle P',F'\rangle$ and $D_{S_{P'}}$ agrees with $F'$.
\end{enumerate}
\end{restatable}

\section{Related Works}\label{section:related}

In recent years, spatial computing has emerged as a promising approach to model and control systems consisting of a large number of cooperating agents that are distributed over a physical or logical space~\cite{BDUVC12}.
This computational model starts from the assumption that, when the density of involved computational agents increases, the underlying network topology is strongly related to the geometry of the space through which computational agents are distributed. Goals are generally defined in terms of the system's spatial structure. A main advantage of these approaches is that their computations can be seen both as working on a single node, and as computations on the distributed data structures emerging in the network (the so-called ``computational fields"). 

\change{The first examples in this area} is Proto~\cite{BB06,BMS11}. This language aims at providing links between local and global computations and permits the specification of the individual behaviour of a node, typically in a sensor-like network, via specific space-time operators to situate computation in the physical world. In~\cite{VDB13,Damiani201617} a minimal core calculus, named \emph{field calculus}, has been introduced to capture the key ingredients of languages that make use of computational fields. 

The calculus proposed in this paper starts from a different perspective with respect to the ones mentioned above. In these calculi, computational fields result from (recursive) functional composition. These functions are used to compute a field, which may consists of a tuple of different values. 
\change{Each function in the \emph{field calculus} or in Proto plays a role similar to a \smuc{} formula.
In our approach, each step of a \smuc{} program \changeR{computes} a different field, \changeR{which is} then used in the rest of the computation.  
This step is \emph{completed} only when a \emph{fixpoint} is reached. 
This is possible because in \smuc{} only specific functions over the appropriate domains are considered. This guarantees the existence of fixpoints and the possibility to identify a global stability in the field computation.} 
%

\change{In \smuc{} a formula is evaluated when a \emph{fixpoint} is reached. The resulting value is then used in the continuation of the program. 
The key feature of our approach is that \changeR{our} formulas have a declarative, global meaning, which restates in this setting the well understood interpretation of temporal logic and $\mu$-calculus formulas, originally defined already on graphical structures, namely on labelled transition systems or on Kripke frames. In addition, the chains approximating the fixpoints can be understood as propagation processes, thus giving also a pertinent operational interpretation to the formulas.}

\change{Our approach is also reminiscent of the bulk synchronous model of computation, as adopted for instace by} Pregel~\cite{DBLP:conf/sigmod/MalewiczABDHLC10} and Giraph~\cite{DBLP:journals/pvldb/ChingEKLM15}. 
\change{\changeR{In this model} computations consist of a sequence of iterations, called \emph{supersteps}. 
During a \emph{superstep} the framework evaluates a user-defined function at each vertex. 
Such evaluations are  conceptually executed in parallel. After each superstep the computed value is propagated to the neighbours via the outgoing edges and used in the next supersteps.} \changeR{Approaches similar to ours can be found in the field of distributed and parallel programming, in particular in early works on distributed fixpoint computations. For example~{\cite{DBLP:journals/mp/Bertsekas83}} presents a general distributed algorithm schema for 
the computation of fixpoints where iterations are not synchronised. The spirit of the work is similar to our results on robustness, but is focused on functions on domains of real numbers, while we consider the more general case of field domains. Another related example is~{\cite{DBLP:journals/pc/SavariB96}}, which presents solutions for distributed termination in asynchronous iterative algorithms for fixpoint computations.}

\change{
In the \emph{field calculus} the evaluation of a program yields a continuous stream of data (the field) that does not stop even if a fixpoint is reached. 
Under this perspective, one interesting propert\changeR{y} is the \emph{self-stabilisation}, \changeR{i.e. the ability} of a \emph{field} to reach a \emph{stable} value after a perturbation or starting from some initial conditions.  
}
In~\cite{VD14,DamianiV15} \change{\emph{self-stabilisation} for the \emph{field calculus} is studied. In these papers sufficient conditions for self-stabilisation are presented as the ability to react to changes in the environment finding a new stable state in finite time. A type-based approach is used to provide a correct checking procedure for self-stabilisation.}
%

\change{
%
%
A deep comparison of the field calculus and the \smuc{} calculus would require an extensive discussion. We focus here on the two main aspects in which these calculi differ.
First, even if  the two calculi share the starting motivations, the \emph{field-calculus} and \smuc{} operate at two different levels of abstraction. Indeed, while the former aims at defining a general and universal framework for \emph{computational fields}, \smuc{} operates at a higher level where devices (i.e. the nodes in the graph) achieve results (the fixpoints) definable in expressive declarative ways, but rely on an underlying framework that can be used to support communications and to check termination of formula evaluation\changeR{s} (correct computation of fixpoints). Moreover, the \emph{field-calculus} is mainly functional while \smuc{}, as already mentioned, is based on a declarative definition of \emph{fields} computed by two kinds of recursions (least and greatest fixpoint).
Second, the underlying graph \changeR{in \smuc{}} is explicitly considered in the operational semantics while it is abstract in the \emph{field-calculus}. Moreover, links are also equipped with network capabilities that can be used to transform communicated values. 
}

Different middleware/platforms have been proposed to support coordination of distributed agents via computational fields~\cite{MZ09,VCMZ11,MVFDZ13,PVB15}.
Protelis~\cite{PVB15}\footnote{http://protelis.github.io/}\change{ is a language that, inspired by Proto and integrating the Field Calculus features, aims at providing a Java framework for simplifying development of networked systems. }
In~\cite{MZ09} the framework TOTA (\emph{Tuples On The Air}), is introduced to provide spatial abstractions for a novel approach to distributed systems development and management, and is suitable to tackle the complexity of modern distributed computing scenarios, and promotes self-organisation and self-adaptation. In~\cite{VCMZ11} a similar approach has been extended to obtain a chemical-inspired model. This extends tuple spaces with the ability of evolving tuples mimicking chemical systems and provides the machinery enabling agents coordination via spatial computing patterns of competition and gradient-based interaction.
\change{In~{\cite{DBLP:conf/icse/SebastioAL14}} computational fields and ant colony optimisation techniques are combined in a cloud computing scenario. The idea in that approach is to populate the network with mobile agents that explore and build a computational field of pheromones to be exploited when looking for computational resources in the cloud system. The approach is validated using a simulator of cloud systems~{\cite{DBLP:journals/spe/SebastioAL16}}.} 
In~\cite{MVFDZ13} a framework for distributed agent coordination via \emph{eco-laws} has been proposed. This kind of laws generalise the chemical-inspired ones~\cite{VCMZ11} in a framework where self-organisation can be injected in pervasive service ecosystems in terms of spatial structures and algorithms for supporting the design of context-aware applications.
The proposed calculus considers computational fields at a more higher level of abstraction with respect to the above mentioned frameworks. However, these frameworks could provide the means for developing a distributed implementation of \smuc{}. 

\changeR{
Finally, we can observe that \smuc{}, like many of the languages and frameworks referenced above, is remindful of \emph{gossip protocols}~}\cite{Jelasity2011,Birman:2007:PLG:1317379.1317382}.
\changeR{These are a class of communication protocol that, inspired by the form of gossip experienced in social networks, try to solve coordination/computational problems in distributed systems: each node in the network spreads/collets relevant information to/from its neighbours until a global equilibrium is reached. 
\smuc{} somehow generalizes some classes of gossip protocols. Functions associated with edges and nodes via labels can be used to control and aggregate the data exchanged among nodes while providing a general framework that can be used to model/program many of the existing protocols. 
However, many gossip protocols are probabilistic in nature, while \smuc{} computations are deterministic. 
Further investigations are definitively needed to assess the exact relation between gossip protocols and \smuc{}.
}
	

\section{Conclusion}\label{section:conclusion}

We have presented a simple calculus, named \smuc{}, that can be used to program and coordinate the activities of distributed agents via computational fields. In \smuc{} a computation consists of a sequence of fixpoints computed in a \changeR{fixed} graph-shaped field that represents the space topology modelling the underlying network. 
Our graph-based fields have attributes on both nodes and arcs, where the latters represent interaction capabilities between nodes. 
Under reasonable conditions, fixpoints can be computed via asynchronous iterations. At each iteration the attributes of some nodes are updated according to the values of neighbors in the previous iteration. The fixpoint \change{computation is robust against certain forms of unavailability and failure situations}. 
\smuc{} is also equipped with a set of control-flow constructs which allow one to conveniently structure the fixpoint computations. 
We have also developed a prototype tool for our language, equipped with a graphical interface that provides useful visual feedback. Indeed we employ those visual features to illustrate the application of our approach to a robot rescue case study, for which we provide a novel rescue coordination strategy, programmed in \smuc{}.

Finally, we have presented a distributed implementation of our calculus. The translation is done in two phases, from \smuc{} programs into normal form \smuc{} programs and then into distributed programs. The correctness of translations exploits the above mentioned results on asynchronous computations.

As future work we plan to deploy the implementation specified in the paper on a suitable distributed architecture, and to carry \changeR{out} experiments about case studies of aggregate programming \changeR{and gossip protocols}. \change{Specific domains of applications are the Internet-of-Things and Big (Graph) Data analytics. The former has been subject of focus by seminal works on aggregate programming~{\cite{DBLP:journals/computer/BealPV15}}, while the latter seems particularly attractive given the similarities between the model of computation of \smuc{} and the BSP model of computation~{\cite{DBLP:journals/cacm/Valiant90}} on which parallel graph analysis frameworks like Google's Pregel~{\cite{DBLP:conf/sigmod/MalewiczABDHLC10}} and Apache's Giraph~{\cite{DBLP:journals/pvldb/ChingEKLM15}} are based on.} 
\changeR{We will also consider gossip based protocols for aggregate computations in large dynamic and p2p networks (see for instance} \cite{Jelasity:2005:GAL:1082469.1082470,Makhloufi2009}%
\changeR{).  
}
\changeR{Another possible field of application could be distributed and parallel model checking, given that \smuc{} formulas generalise some well-known temporal logics used in the field of model checking.}

\change{Furthermore, we plan to compare the \emph{expressivity aspects} of \smuc{} with respect to the languages and calculi previously proposed in literature, and to the \emph{field calculus}}~\cite{Damiani201617} in particular. 
\change{This comparison
 is not only interesting from a theoretical point of view, but could also provide a deeper understanding of possible alternative \emph{paradigms} for \emph{aggregate programming}.} 

\change{We also plan to study mechanisms that allow dynamic deployment of new \smuc{} code fragments. From this point of view a source of inspiration could be the works presented} in~\cite{Damiani2015} 
\change{where a higher-order version of the \emph{field calculus} is presented} \changeR{and~{\cite{DBLP:conf/coordination/PianiniBV16}} where overlapping fields are used to adapt to network changes.}

%

\section*{Acknowledgement}
The authors wish to thank Carlo Pinciroli for interesting discussions in preliminary stages of the work and the anonymous referees from the conference version of this work for their insightful and encouraging comments. 
This work has been supported by the European projects IP 257414 ASCENS and STReP 600708 QUANTICOL, and the Italian PRIN 2010LHT4KM CINA. 
%

\appendix

\section{Proofs} \label{section:proofs}

\semiringmonotony*

\proof
The proof is easily obtained from the monotony requirements and the properties of the semiring operators.
\qed

\asynchronylemma*

\proof
We prove the above statements {(i-iii)} separately.

\noindent
\begin{enumerate}[label=(\roman*)]
\item
We have to show that for all node valuations $\mathsf{f}_1,\mathsf{f}_2$ such that $\mathsf{f}_1 \sqsubseteq \mathsf{f}_2$ we have that $\psi_{\pi}\, \mathsf{f}_1  \sqsubseteq \psi_{\pi}\, \mathsf{f}_2$, i.e. that for all nodes $n \in N_F$ we have that $\psi_{\pi}\, \mathsf{f}_1 \, n   \sqsubseteq \psi_{\pi}\, \mathsf{f}_2 \, n$.
%
We assume $\mathsf{f}_1 \sqsubseteq \mathsf{f}_2$ and we consider two cases for $n$ depending on whether it belongs to $\pi$ or not.
If $n \in \pi$ then, according to Definition~\ref{def:pattern-restricted-application}, $\psi_{\pi}\mathsf{f}_1\, n = \psi\, \mathsf{f}_1\, n$ and $\psi_{\pi}\mathsf{f}_2\, n = \psi\, \mathsf{f}_2\, n$. Since $\psi$ is monotone it follows that $\psi\, \mathsf{f}_1\, n \sqsubseteq \psi\, \mathsf{f}_2\, n$ and hence $\psi_{\pi}\, \mathsf{f}_1\, n \sqsubseteq \psi_{\pi}\, \mathsf{f}_2\, n$. 
Otherwise, if $n \not\in \pi$ then, according to Definition~\ref{def:pattern-restricted-application}, $\psi_{\pi}\mathsf{f}_1\, n = \mathsf{f}_1\, n$ and $\psi_{\pi}\mathsf{f}_2\, n = \mathsf{f}_2\, n$. Since $\mathsf{f}_1 \sqsubseteq \mathsf{f}_2$ it clearly follows that $\mathsf{f}_1\, n \sqsubseteq \mathsf{f}_2\, n$ and hence $\psi_{\pi}\, \mathsf{f}_1\, n \sqsubseteq \psi_{\pi}\, \mathsf{f}_2\, n$. 

\item
  This can be easily derived from Definition~\ref{def:pattern-restricted-application}. Indeed, if both $n \in \pi_1$ and $n \in \pi_2$ then $\psi_{\pi_1}\mathsf{f}\, n = \psi\, \mathsf{f}\, n$ and $\psi_{\pi_2}\mathsf{f}\, n = \psi\, \mathsf{f}\, n$. Otherwise, if $n \not\in \pi_1$ and $n \not\in \pi_2$ then $\psi_{\pi_1}\mathsf{f}\, n = \mathsf{f}\, n$ and $\psi_{\pi_2}\mathsf{f}\, n = \mathsf{f}\, n$. 
\item
  This follows immediately from {(ii)} and {(iii)}. Indeed, from {(i)} we have that $\psi_{\pi_1}\, \mathsf{f}_1\, n \sqsubseteq \psi_{\pi_1}\, \mathsf{f}_2\, n$, and from {(ii)} we have that  $\psi_{\pi_1}\, \mathsf{f}_2\, n = \psi_{\pi_2}\, \mathsf{f}_2\, n$. \qedhere
\end{enumerate}

\asynchronylemmabis*

\proof
The proof is by induction on the length of $\sigma$. 
\noindent
\begin{description}
\item[Case $|\sigma| = 0$, {\rm i.e.} $\sigma = \epsilon\,$]
We have to prove that $\psi_\epsilon \sqsubseteq \psi_\pi \, \psi_\epsilon \sqsubseteq \psi \, \psi_\epsilon$. Since $\psi_\epsilon = \bot$ this amounts to proving $\bot \sqsubseteq \psi_\pi \, \bot \sqsubseteq \psi \, \bot$. 
First, we have $\bot \sqsubseteq \psi_\pi \, \bot$ by definition. And, second, $\psi_\pi \, \bot \sqsubseteq \psi \, \bot$ 
can be shown pointwise on nodes $n$ distiguishing whether $n$ belongs to $\pi$ or not, as in the above proofs. Indeed, for $n \in \pi$ we have $\psi_\pi \, \bot \, n = \psi \, \bot \, n$ by definition. If instead,  $n \not\in \pi$ we have $\psi_\pi \, \bot \, n = \bot \, n = \bot_A$ by definition and, clearly, $\bot_A \sqsubseteq \psi \, \bot \, n$. 

\item[Case $|\sigma| > 0\,$]
Assume as induction hypothesis that for all prefixes $\sigma[1..i]$ of $\sigma[1..|\sigma|-1]$ the lemma holds, i.e. for all patterns $\pi'$ we have $\psi_{\sigma[1..i]} \sqsubseteq \psi_{\pi'} \, \psi_{\sigma[1..i]} \sqsubseteq \psi \, \psi_{\sigma[1..i]}$.
%
%
We prove that the theorem follows pointwise for every node $n \in N_F$, i.e. that $\forall n \in N_F : \psi_\sigma\, n\sqsubseteq \psi_\pi \, \psi_\sigma \, n \sqsubseteq \psi \, \psi_\sigma \, n$.

We start proving the first part of the inequality in the theorem, i.e. $\psi_\sigma \sqsubseteq \psi_\pi \, \psi_\sigma$. 
We consider two cases for $n$ depending on whether it belongs to $\pi$ or not. 
The easiest case is when $n \not\in \pi$. Indeed, in this case we obtain $\psi_\pi \psi_\sigma \, n = \psi_\sigma\, n$ from Definition~\ref{def:pattern-restricted-application}.  
If instead $n \in \pi$ the proof is more elaborated. Let $\sigma[1..j]$ be the longest prefix of $\sigma$, if any, where $n$ has been updated. We have $\sigma = \sigma[1..j] , \sigma_{j+1} , \sigma[j+1..|\sigma|]$ and $n \not\in (\sigma_{j+1} \cup .. \cup \sigma_{|\sigma|})$.   
We have $\psi_{\sigma[1..j]} \sqsubseteq \psi_\sigma$ by the inductive hypothesis, thus $\psi_{\sigma_{j+1}}\, \psi_{\sigma[1..j]} \, n \sqsubseteq \psi_\pi\, \psi_\sigma \, n$ \change{by Lemma~{\ref{th:asynchrony-lemma1}}}. 
But $\psi_{\sigma_{j+1}} \, \psi_{\sigma[1..j]} \, n = \psi_{\sigma[1..j+1]} \, n = .. = \psi_\sigma\, n$ by Definition~\ref{def:pattern-restricted-application} since $n$ does not belong to any pattern in $\sigma[j+1..|\sigma|]$ (otherwise $\sigma[1..j]$ would not be the longest prefix of $\sigma$ where $n$ has been updated) 
%
Therefore, $\psi_\sigma\, n \sqsubseteq \psi_\pi \, \psi_\sigma \, n$. 

We now prove the second part of the inequality in the theorem, i.e. $\psi_\pi \, \psi_\sigma \, n \sqsubseteq \psi \, \psi_\sigma \, n$. 
Again, we consider two cases for $n$ depending on whether it belongs to $\pi$ or not. 
If $n \in \pi$ then, from Definition~\ref{def:pattern-restricted-application}, we have that $\psi_\pi \,  \psi_\sigma \, n = \psi \, \psi_\sigma \, n$. 
If instead $n \not \in \pi$, we have that, according to Definition~\ref{def:pattern-restricted-application}, $\psi_\pi \,  \psi_\sigma \, n = \psi_\sigma\, n$ and $\psi_\sigma\, n  \sqsubseteq \psi \, \psi_\sigma \, n$ by letting $\pi = N_F$ in the previous result. \qed 
\end{description}

\asynchronycorollary*

\proof
The corollary follows immediately from Lemma~\ref{th:asynchrony-lemma3}. 
\qed

\asynchronytheorem*

\proof

We prove that for any element of a chain there is an element in the other chain which is larger or equal.

We start first proving that this holds for chain $\bot \sqsubseteq \psi_{\sigma[1..1]} \sqsubseteq \psi_{\sigma[1..2]} \sqsubseteq  \dots$ with respect to $\bot \sqsubseteq \psi\, \bot \sqsubseteq \psi^2\, \bot \sqsubseteq  \dots$. In particular, for every chain element $\psi_{\sigma[1..k]}$ there is a chain element $\psi^{l}\, \bot$ such that $\psi_{\sigma[1..k]} \sqsubseteq \psi^{l}\, \bot$. Indeed, this holds for $k = l$. We hence prove that $\psi_{\sigma[1..k]} \sqsubseteq \psi^k\, \bot$ for every $k \in \Nat$. The proof is by induction on $k$. For $k=0$ we have, $\psi_{\epsilon} = \bot = \bot = \psi^k\, \bot$. Assuming $\psi_{\sigma[1..k]} \sqsubseteq \psi^k\, \bot$ as induction hypothesis we can easily prove that $\psi_{\pi_{k+1}}\, \psi_{\sigma[1..k]}  =  \psi_{\sigma[1..k+1]} \sqsubseteq \psi^{k+1}\, \bot = \psi\, \psi^k\, \bot$.
In fact,
$\psi_{\pi_{k+1}}\, \psi_{\sigma[1..k]}  \sqsubseteq  \psi\, \psi_{\sigma[1..k]}$ holds by Lemma~\ref{th:asynchrony-lemma3} and $\psi\, \psi_{\sigma[1..k]}  \sqsubseteq  \psi\, \psi^k\, \bot $ follows from the inductive hypothesis ($\psi_{\sigma[1..k]}  \sqsubseteq  \psi^k\, \bot$) and the monotony of $\psi$. 

We now prove the other direction, i.e. that for every element $\psi^k\, \bot$ in $\bot \sqsubseteq \psi\, \bot \sqsubseteq \psi^2\, \bot \sqsubseteq  \dots$ there an element $\psi_{\sigma[1..h]}$ in $\bot \sqsubseteq \psi_{\sigma[1..1]} \sqsubseteq \psi_{\sigma[1..2]} \sqsubseteq  \dots$ that is larger or equal (i.e. $\psi^k\, \bot  \sqsubseteq  \psi_{\sigma[1..h]}$). 
We prove this by induction on $k$. 
For $k = 0$ we can choose $h$ to be 0 as well, so that we trivially have $\psi^0\, \bot = \bot = \psi_{\epsilon}$.
Assume by induction that there is an $h \in \Nat$, such that  $\psi^k\, \bot  \sqsubseteq  \psi_{\sigma[1..h]}$. We will show that there is an element $\psi_{\sigma[1..h+i]}$ that can bound $\psi^{k+1}\, \bot$. In particular, we choose such an element (determined by $i$) that satisfies $ \forall n . n \in \sigma_{h+1} \cup ..  \cup \sigma_{h+i}$. In words, every node will be updated within the next $i$ steps.   
Such an $i$ does necessarily exist since the strategy $\sigma$ is fair. 
We prove $\psi^{k+1}\, \bot  \sqsubseteq  \psi_{\sigma[1..h+i]}$ pointwise.
Let $h+j$ be the last index such that $n \in \sigma_{h+j}$. Thus we have $\psi \, \psi_{\sigma[1..h+j]}\, n = \psi_{\sigma[1..h+j+1]}\, n = .. = \psi_{\sigma[1..h+i]}\, n$.
Finally we conclude: 
\[
\begin{array}{rcll}
\psi^k\, \bot  & \sqsubseteq  &\psi_{\sigma[1..h]} & \text{by the inductive hypothesis} \\
\psi_{\sigma[1..h]} & \sqsubseteq & \psi_{\sigma[1..h+j]} & \text{by Corollary~\ref{cor:asynchrony-corollary1}} \\
\psi^k\, \bot  &\sqsubseteq & \psi_{\sigma[1..h+j]} & \text{by transitivity} \\ 
\psi^{k+1}\, \bot\, n  &\sqsubseteq& \psi \, \psi_{\sigma[1..h+j]} \, n & \text{by monotonicity and pointwise ordering} \\
\psi^{k+1}\, \bot\, n  &\sqsubseteq&  \psi_{\sigma[1..h+i]}\, n & \text{by the above equality}
\end{array}\vspace{-16 pt}
\]
\qed

\asynchronytheorembis*

\proof
It is immediate to see that domain $N \rightarrow A$ has only finite chains. Thus given any fair strategy $\sigma$, the chain $\bot  \sqsubseteq  \psi_{\sigma[1..1]}  \sqsubseteq \psi_{\sigma[1..2]} \sqsubseteq \dots$ reaches in, say, $k'$ steps its least upper bound (which proves {(i)}). 
Property {(ii)} immediately follows from Theorem~\ref{th:asynchrony-theorem1}.
\qed

\begin{mdframed}[style=change]
\failurethm*

\proof

We prove that: (i) for any element of the failure sequence 
${\overline{\varsigma}}_{\tilde{\sigma}}$ 
there is an element in the chain ${\overline{\psi}}_{\tilde{\sigma}}$ which is larger or equal of it; and (ii) viceversa. Thus both the failure sequence and the chain have the same set of upper bounds: but the chain is guaranteed to have a least upper bound, which, by Theorem~\ref{th:asynchrony-theorem2}, is the minimal fix point of $\psi$.

For (i), given $\varsigma_\sigma$, with $\sigma$ finite, we choose just $\psi_\sigma$. We have to prove $\varsigma_\sigma \sqsubseteq \psi_\sigma$ for all finite $\sigma$.
First, we have $\varsigma_\epsilon = \bot_{N \rightarrow A}= \psi_\epsilon$.
Also, assuming $\varsigma_\sigma \sqsubseteq \psi_\sigma$ we have $\psi_\pi \varsigma_\sigma \sqsubseteq \psi_\pi \psi_\sigma$. But $\forall n. \varsigma_{\sigma, \pi}n \sqsubseteq \psi_\pi \varsigma_\sigma n$, and thus $\varsigma_{\sigma, \pi} \sqsubseteq \psi_{\sigma, \pi}$, as required. In fact, if $n \in \pi$, or $n \notin \pi$ and in equation~\ref{failure} the third option is taken, then $\psi_\pi$ has the same effect as the recursive call in equation~\ref{failure}, while if the fourth option is taken  we have $\varsigma_{\sigma, \pi}n = \varsigma_{\sigma'} n \sqsubseteq \psi_{\sigma'} n \sqsubseteq \psi_\sigma n$, where the former inclusion holds for the inductive hypothesis, while the latter inclusion holds since ${\overline{\psi}}_{\tilde{\sigma}}$ is a chain.

For (ii), given a $\tilde{\sigma}$ safe failure sequence ${\overline{\varsigma}}_{\tilde{\sigma}}$, let $\tilde{\sigma} = \hat{\sigma}, \tilde{\sigma}'$, where $\hat{\sigma}$, finite, corresponds to the only part of ${\overline{\varsigma}}_{\tilde{\sigma}}$ where the fourth option of equation~\ref{failure} has been employed. 
Such $\hat{\sigma}$ do exists since $\overline{\sigma}$ is safe.
Thus ${\overline{\varsigma}}_{\tilde{\sigma}'}$ and ${\overline{\psi}}_{\tilde{\sigma}'}$ coincide, while for every $\psi_{\sigma''}$ in ${\overline{\psi}}_{\tilde{\sigma}'}$ we have   $\psi_{\sigma''} \sqsubseteq \varsigma_{\hat{\sigma},\sigma''}$ for all finite $\sigma''$. 
In fact, 
$\psi_\epsilon = \bot_{N \rightarrow A} \sqsubseteq \varsigma_{\hat{\sigma}}$ 
and then the same update functions have been applied on both sides according to $\sigma''$.
Furthermore, chains ${\overline{\psi}}_{\tilde{\sigma}'}$ and ${\overline{\psi}}_{\tilde{\sigma}}$, according to Theorem~\ref{th:asynchrony-theorem1} have the same least upper bound, the fix point of $\psi$. Thus given any element $\psi_\sigma$ in ${\overline{\psi}}_{\tilde{\sigma}}$ there exists a $\psi_{\sigma'}$ in ${\overline{\psi}}_{\tilde{\sigma}'}$ with 
$\psi_\sigma \sqsubseteq \psi_{\sigma'}$. But in ${\overline{\varsigma}}_{\tilde{\sigma}}$ we have an even larger element: $\psi_{\sigma'} \sqsubseteq \varsigma_{\hat{\sigma},\sigma'}$.
Thus we conclude $\psi_\sigma \sqsubseteq \varsigma_{\hat{\sigma},\sigma'}$.
\qed
\end{mdframed}

\exptosaf*

\begin{proof}
By induction on the syntax of $\Psi$. \medskip

\noindent\textit{Base of Induction:} If $\Psi=j$ then statement follows directly from the fact that $\llbracket j\rrbracket^F_{\rho} = I_F (j)$, $\mathcal{A}_{c}^{i}= i\leftarrow j$ and from rule ($\mu$\textsc{Step}) in Table~\ref{table:semantics-global1}. \medskip

\noindent\textit{Inductive Hypothesis:} Let \change{$\Psi_1$,\ldots, $\Psi_k$} be such that, for any $j$:
\[
\llbracket \Psi_{j}\rrbracket^F_{\emptyset}=f 
\Leftrightarrow
 \langle S_{\Psi_{j}}^{i_j} , F \rangle \rightarrow^{*} \langle \mathsf{skip} , F' \rangle \mbox{ and } I_{F'}(i_j)=f
 \]

\noindent\textit{Inductive Hypothesis:} Many cases are standard and we provide the details here only for the cases $\Psi=f(\Psi_1,\ldots,\Psi_k)$ and
$\Psi = \mu z.\Psi_i$ while we omit $\Psi=\aggregateOut{$\alpha$}{g}{\Psi_i}$, $\Psi=\aggregateIn{$\alpha$}{g}{\Psi_i}$ and $\Psi=\nu z.\Psi_i$ that are 
similar to the ones considered in the proof.
In this case we have that \change{$\llbracket f(\Psi_1,\ldots,\Psi_j)\rrbracket^F_{\rho} = f( 
\llbracket \Psi_1\rrbracket^F_{\rho},\ldots \llbracket \Psi_n\rrbracket^F_{\rho})$}. Moreover:
 \[
 S_{\Psi}^{i}=S_{\Psi_1}^{x_{c_1}}\ \mathsf{;}\cdots\mathsf{;}\ S_{\Psi_k}^{x_{c_k}}\ \mathsf{;}\ i\leftarrow f(x_{c_1},\ldots,x_{c_{k}})
 \]
 for the appropriate $x_{c_1}$,\ldots, $x_{c_n}$. By \emph{Inductive Hypothesis} we have that for each $k$:
\[
\llbracket \Psi_{k}\rrbracket^{F_k}_{\emptyset}=f 
\Leftrightarrow
 \langle S_{\Psi_{k}}^{x_k} , F_k \rangle \rightarrow^{*} \langle \mathsf{skip} , F_{k+1} \rangle \mbox{ and } I_{F_{k+1}}(x_k)=f
 \]
 where $F_1=F$. Moreover, since each $S_{\Psi_{k}}^{x_{c_k}}$ uses different auxiliary labels, we also have that: 
 \begin{itemize}
 \item $\langle S_{\Psi_1}^{x_{c_1}}\ \mathsf{;}\cdots\mathsf{;}\ S_{\Psi_n}^{x_{c_n}}\ \mathsf{;}\ i\leftarrow f(x_{c_1},\ldots,x_{c_{n}}) , F\rangle
 \rightarrow^{*} \langle j\leftarrow f(x_{c_1},\ldots,x_{c_{n}}) , F_{n+1}\rangle$;
 \item for any $k$,  $I_{F_{n+1}}(x_{k})=I_{F_{k+1}}(x_{k})$ and \change{$\llbracket \Psi_k\rrbracket^F_{\rho}=
 \llbracket \Psi_k\rrbracket^{F_k}_{\rho}=\llbracket \Psi_k\rrbracket^{F_{n+1}}_{\rho})$}.
 \end{itemize}
 
\noindent
Hence, we have that $\langle i\leftarrow f(x_{c_1},\ldots,x_{c_{n}}) , F_{n+1}\rangle\rightarrow \langle \mathsf{skip} , F'\rangle$,
where:
\[
\begin{array}{rcl}
 I_{F'}(i) & = & \lambda n . \llbracket f \rrbracket_{A_F}  ( I_{F_k+1}(x_{c_1}) n , \ldots , I_{F_k+1}(x_{c_{k}}) n)\\
 & = & \lambda n . \llbracket f \rrbracket_{A_F}  ( \llbracket \Psi_1\rrbracket^F_{\rho}\, n , .. , \llbracket \Psi_k\rrbracket^F_{\rho}\, n)\\
 & = & \llbracket \Psi \rrbracket^F_{\rho}
\end{array}
 \] 
Notice that similar considerations apply when $\Psi=\aggregateOut{$\alpha$}{g}{\Psi_i}$, $\Psi=\aggregateIn{$\alpha$}{g}{\Psi_i}$.  \\

Let us now consider the case $\Psi=\mu \kappa.\Psi_i$ (Similarly considerations can be used when $\Psi=\nu \kappa.\Psi_i$). We can first of all notice that since $A_F$ has only only finite partially ordered chains, 
we have that there exists a value $k$ such that $\llbracket \mu z.\Psi\rrbracket^{F}_{\emptyset} = \llbracket \tilde{\Psi}_{k}\rrbracket^{F}_{\emptyset}$  where $\tilde{\Psi}_{0}=\bot$ while $\tilde{\Psi}_{k+1}=\Psi[^{\tilde{\Psi}_{k}}/_{z}]$. Moreover, for any $k'\geq k$ we also have that:
$\llbracket \tilde{\Psi}_{k'}\rrbracket^{F}_{\emptyset}=\llbracket \tilde{\Psi}_{k'+1}\rrbracket^{F}_{\emptyset}$.
We can also notice that $S_{\Psi}$ has the form:
\[
\begin{array}{l}
x_{c}\leftarrow \bot\ \mathsf{;}\\
x_{c+1}\leftarrow \bot\ \mathsf{;}\\
x_{c'}\leftarrow \bot \mathsf{;}\\
\mathsf{while}\ x_{c'}\ \mathsf{do}\\
\phantom{xxx} x_{c}\leftarrow x_{c+1}\mathsf{;}\\
\phantom{xxx} S_{\Psi[^{x_c}/_{z}]}\ \mathsf{;}\\
\phantom{xxx} x_{c'}\leftarrow x_{c}=x_{c+1}\\
i\leftarrow x_{c+1}
\end{array}
\]

Let $F_{k}$ be the field obtained after $k$ iterations, of the program above. By using inductive hypothesis we have that 
$I_{F_{k}}(x_{c+1})= \llbracket \tilde{\Psi}_{k}\rrbracket^{F}_{\emptyset}$ while, if $k>1$, $I_{F_{k}}(x_{c})= \llbracket \tilde{\Psi}_{k-1}\rrbracket^{F}_{\emptyset}$. The program above terminates only when $x_{c}$ and $x_{c+1}$ are equal, i.e. when  $\llbracket \tilde{\Psi}_{k}\rrbracket^{F}_{\emptyset}=\llbracket \tilde{\Psi}_{k+1}\rrbracket^{F}_{\emptyset}=\llbracket \mu z.\Psi\rrbracket^{F}_{\emptyset}$, and $i$ is assigned to $x_{c+1}$. 
\end{proof}

\eqone*

\begin{proof} We prove this lemma by induction on the syntax of $P$.

  \medskip\noindent\textit{Base of Induction:} When $P=\mathsf{skip}$
  the thesis follows directly from the fact that $S_{P}=P$, while when
  $P=i\leftarrow \Psi$ we can directly use Lemma~\ref{lem:exptosaf}.

\medskip\noindent\textit{Inductive Hypothesis:} For any $P_1$ and $P_2$:
\begin{itemize}
\item if $\langle  P_i , F \rangle  \rightarrow \langle P'  , F' \rangle$ then $\langle  S_{P_1} ,   F \rangle \rightarrow^{*} \langle S_{P'}  , F''\rangle$ and $F'=F''\backslash \mathcal{X}$;
\item if $\langle  S_{P_1} , F \rangle  \rightarrow \langle S' , F' \rangle$ then exist $P'$ and $F''$ such that $\langle S'  , F' \rangle \rightarrow^{*} \langle S_{P'}  , F''\rangle$ and $\langle P,F\rangle \rightarrow \langle P',F''\backslash \mathcal{X}\rangle$.
\end{itemize}

\medskip\noindent\textit{Inductive Step:} We have to consider the following cases:
\begin{itemize}
\item $P=P_1;P_2$. If $P_1=\mathsf{skip}$ we have that $\mathsf{skip};P_2$ has exactly the same computations of $P_2$. Moreover,
$S_{P}=\mathsf{skip};\mathsf{wait}(x_{c});S_{P_2}$ and we can directly derive our thesis from the inductive hypothesis and from the fact that:
$\langle \mathsf{skip};\mathsf{wait}(x_{c});S_{P_2} , F \rangle \rightarrow \langle \mathsf{skip}; S_{P_2} , F \rangle$. 

Let $P_1\not=\mathsf{skip}$.
If $\langle  P_1;P_2 , F \rangle  \rightarrow \langle P'  , F' \rangle$ then $P'=P_1';P_2$ (for some $P_1'$) and 
$\langle  P_1 , F \rangle  \rightarrow \langle P_1'  , F' \rangle$. By inductive hypothesis we have that 
$\langle  S_{P_1} , F \rangle  \rightarrow^{*} \langle S_{P_1'}  , F'' \rangle$ and $F'=F''\backslash \mathcal{X}$. However, $S_{P}=S_{P_1};\mathsf{wait}(x_{c});S_{P_2}$, $\langle  S_{P_1};\mathsf{wait}(x_{c});S_{P_2} , F \rangle  \rightarrow^{*} \langle S_{P_1'};\mathsf{wait}(x_{c});S_{P_2}  , F'' \rangle$ and $S_{P'}=S_{P_1'};\mathsf{wait}(x_{c});S_{P_2}$. 

Similarly, if $\langle S_{P_1};\mathsf{wait}(x_{c});S_{P_2} , F\rangle \rightarrow \langle S' , F'\rangle$ then there exists $S_1'$ such that 
$\langle S_{P_1} , F\rangle \rightarrow \langle S_1' , F'\rangle$. By inductive hypothesis, there exist $P_1'$ and $F''$ such that
$\langle S_1'  , F' \rangle \rightarrow^{*} \langle S_{P_1'}  , F''\rangle$ and $\langle P_1,F\rangle \rightarrow \langle P_1',F''\backslash \mathcal{X}\rangle$. However, by using rule (\textsc{Seq1}) in Table~\ref{table:semantics-global1}, we have that
$\langle S_1';\mathsf{wait}(x_{c});S_{P_2}  , F' \rangle \rightarrow^{*} \langle S_{P_1'};\mathsf{wait}(x_{c});S_{P_2} , F''\rangle$ and
$\langle P_1;P_2,F\rangle$ $\rightarrow$ $\langle P_1';P_2,F''\backslash \mathcal{X}\rangle$.

\item $P=\mathsf{if}\ \Psi\ \mathsf{then}\ P_1\ \mathsf{else}\ P_2$. We have that $S_{P}=S_{\Psi}^{x_{c}};\mathsf{if}\ x_c\ \mathsf{then}\ S_{P_1}\ \mathsf{else}\ S_{P_2}$.
If $\langle P , F\rangle \rightarrow \langle P', F'\rangle$ then  $F=F'$ and either 
$\llbracket \Psi \rrbracket_{\emptyset}^F\ =\ \lambda n . \emph{true}$ and $P'=P_2$ or $\llbracket \Psi \rrbracket_{\emptyset}^F\ \not=\ \lambda n . \emph{true}$
and $P'=P_2$. 
We can consider only the first case, since the other one follows with similar considerations. By using Lemma~\ref{lem:exptosaf}, we have that if $\llbracket \Psi \rrbracket_{\emptyset}^F\ =\ \lambda n . \emph{true}$  then
$\langle S_{\Psi}^{x_{c}} , F\rangle \rightarrow^{*} \langle \mathsf{skip} , F'\rangle$ and 
$I_{F'}(x_{c})=\lambda n.true$. Hence:
\[
 \langle S_{P}, F \rangle \rightarrow^{*} \langle \mathsf{skip}; \mathsf{if}\ x_c\ \mathsf{then}\ S_{P_1}\ \mathsf{else}\ S_{P_2} , F''\rangle\rightarrow 
 \langle S_{P_1}, F''\rangle
 \] 
where $F=F''\backslash \mathcal{X}$ since in $S_{\Psi}^{x_{c}}$ only auxiliary labels can be assigned.  

If $\langle S_{P} , F\rangle \rightarrow \langle S', F'\rangle$ then $S' = S'';\mathsf{if}\ x_c\ \mathsf{then}\ S_{P_1}\ \mathsf{else}\ S_{P_2}$ and
$\langle S_{\Psi}^{x_{c}} , F\rangle \rightarrow \langle S'' , F'\rangle$. Moreover, thanks to Lemma~\ref{lem:exptosaf}, we have that 
$\langle S'',F'\rangle \rightarrow^{*} \langle \mathsf{skip} , F'' \rangle$ and $I_{F''}(x_{c})=\llbracket \Psi \rrbracket_{\emptyset}^F$.
This implies that either $\langle S' , F' \rangle \rightarrow \langle S_{P_{1}} , F''\rangle$ (when $I_{F''}(x_{c})=\llbracket \Psi \rrbracket_{\emptyset}^F=\lambda n.true$)
or $\langle S' , F' \rangle \rightarrow \langle S_{P_{2}} , F''\rangle$ (when $I_{F''}(x_{c})=\llbracket \Psi \rrbracket_{\emptyset}^F\not=\lambda n.true$).
In the first case $\langle P, F \rangle \rightarrow \langle P_1 , F \rangle$ while in the second case $\langle P, F\rangle \rightarrow \langle P_2 , F \rangle$. 
In both the cases $F=F''\backslash \mathcal{X}$.

\item $P=\mathsf{while}\ \Psi\ \mathsf{do}\ P_1$. We have that 
\[
S_{P}=R_{\Psi}^{x_{c}};\mathsf{while}~x_{c}~\mathsf{do}~S_{P_1}\mathsf{;}\mathsf{wait}(x_{c''})\mathsf{;}R_{\Psi}^{x_{c}}
\]
If $\langle P,F\rangle \rightarrow \langle P',F'\rangle$ then $F=F'$ and either 
$\llbracket \Psi \rrbracket_{\emptyset}^F\ =\ \lambda n . \emph{true}$ and $P'=\mathsf{skip}$, or 
$\llbracket \Psi \rrbracket_{\emptyset}^F\ \not=\ \lambda n . \emph{true}$ and $P'=P_1;\mathsf{while}\ \Psi\ \mathsf{do}\ P_1$.
In both the cases the statement follows like in the previous item by using Lemma~\ref{lem:exptosaf}.\qedhere
\end{itemize}
\end{proof}

\factsem*

\begin{proof}
Both the cases follow directly from Def.~\ref{def:confluence} and from the rules in Tab.~\ref{tab:distopsem1},Tab.~\ref{tab:distopsem2} and Tab.~\ref{tab:distopsem3}
\end{proof}

\disexp*

\begin{proof}
We proceed by induction on the syntax of $\Psi$.

\medskip\noindent\textit{Base of Induction:} If $\Psi=j$ then statement follows directly from the fact that $\llbracket j\rrbracket^F_{\rho} = I_F (j)$ and from
the fact that $S_{\Psi}^{x}=x\leftarrow j$ via the appropriate application of rule (\textsc{D-Step}) in Table~\ref{tab:distopsem1} and of 
rules in Table~\ref{tab:distopsem3}.

\medskip\noindent\textit{Inductive Hypothesis:} Let $\Psi_1$,\ldots, $\Psi_k$ be such that for any distributed execution $D_{S_{\Psi_i}^{x_i}}$ that agrees with $F$, it flows in a distributed execution $D_{\mathsf{skip}}$ such that
$\llbracket \Psi_{k}\rrbracket^{F}_{\emptyset}=\sintll{F}{D_{\mathsf{skip}}}(x)$.

\medskip\noindent\textit{Inductive Step:} 
%
\begin{itemize}
\item $\Psi = f(\Psi_1,\ldots,\Psi_k)$. In this case we have that 
$\llbracket f(\Psi_1,\ldots,\Psi_k)\rrbracket^F_{\rho} = f(\llbracket \Psi_1\rrbracket^F_{\rho},\ldots \llbracket \Psi_k\rrbracket^F_{\rho})$. Moreover:
 \[
 S_{\Psi}^{i}=S_{\Psi_1}^{x_{c_1}}\ \mathsf{;}\cdots\mathsf{;}\ S_{\Psi_k}^{x_{c_k}}\ \mathsf{;}\ i\leftarrow f(x_{c_1},\ldots,x_{c_{k}})
 \]
 for the appropriate $x_{c_1}$,\ldots, $x_{c_k}$. By \emph{Inductive Hypothesis} we have there exists $F_1$,\ldots, $F_{k+1}$, such
 that for each $j$, $D_{S_{\Psi_j}^{x_j}}$ agrees with $F_j$ and it flows in a distributed execution $D^{j}_{\mathsf{skip}}$ such that
\[
\llbracket \Psi_{j}\rrbracket^{F_{j}}_{\emptyset}=\sintll{{F_j}}{{D^{j}_{\mathsf{skip}}}}(x_j)
\]
where $F_1=F$ while $F_{j+1}=F_{j}[^{\sintll{{F_{j}}}{D^{j}_{\mathsf{skip}}}}/_{I_{F_{j}}}]$. 
Let $S=S_{\Psi_1}^{x_{c_1}}\ \mathsf{;}\cdots\mathsf{;}\ S_{\Psi_k}^{x_{c_k}}$, by Lemma~\ref{lem:factsem}, we have that if $D_{S}$ agrees with $F$
then $D_{S}$ flows in $D_{\mathsf{skip}}$ and for any $j$,  $\sintll{{F_k+1}}{{D_{\mathsf{skip}}}}(x_j)=\llbracket \Psi_{j}\rrbracket^{F_{j}}_{\emptyset}$.
However, $\Psi_{j}$ refers only to labels in $F$. Hence, $\llbracket \Psi_{j}\rrbracket^{F_{j}}_{\emptyset}=\llbracket \Psi_{j}\rrbracket^{F}_{\emptyset}$.
Moreover, it is easy to see that via the appropriate application of rule of rule (\textsc{D-Step}) in Table~\ref{tab:distopsem1} and of 
rules in Table~\ref{tab:distopsem3}, $D_{\mathsf{skip}};i\leftarrow f(x_{c_1},\ldots,x_{c_{k}})$ flows in $D'_{\mathsf{skip}}$ and
for any $n\in N_{F}$,  $\sintll{F}{D'_{\mathsf{skip}}}(x)=\llbracket \Psi \rrbracket^{F}_{\emptyset}$.


%

\item $\Psi=\aggregateOut{$\alpha$}{g}{\Psi_i}$ or $\Psi=\aggregateIn{$\alpha$}{g}{\Psi_i}$, in both the cases we can proceed as in the previous case. 
Indeed, the SAF program that evaluates $\Psi$ is $S_{\Psi}^{x}=S_{\Psi_i}^{x_i};x \leftarrow \aggregateOut{$\alpha$}{g}{x_i}$
or $S_{\Psi}^{x}=S_{\Psi_i}^{x_i};x \leftarrow \aggregateIn{$\alpha$}{g}{x_i}$. In both the cases we have that
and via notice that $D_{S_{\Psi}^{x}}$ flows to  $D_{\mathsf{skip}}$ while $D_{\mathsf{skip}};x \leftarrow \aggregateOut{$\alpha$}{g}{x_i}$ (resp. 
$D_{\mathsf{skip}};x \leftarrow \aggregateIn{$\alpha$}{g}{x_i}$) flows to the appropriate $D'_{\mathsf{skip}}$ such that 
$\sintll{F}{D'_{\mathsf{skip}}}(x)=\llbracket \Psi \rrbracket^{F}_{\emptyset}$.
 

\item $\Psi=\mu \kappa.\Psi_i$ or $\Psi=\mu \kappa.\Psi_i$. We consider here $\Psi=\mu \kappa.\Psi$ since the proof for the other case is similar.
First of all we can observe that $S_{\Psi}$ has the form:
\[
\begin{array}{l}
x_{c}\leftarrow \bot\ \mathsf{;}\\
x_{c+1}\leftarrow \bot\ \mathsf{;}\\
x_{c'}\leftarrow \bot \mathsf{;}\\
\mathsf{while}\ x_{c'}\ \mathsf{do}\\
\phantom{xxx} x_{c}\leftarrow x_{c+1}\mathsf{;}\\
\phantom{xxx} S_{\Psi[^{x_c}/_{z}]}\ \mathsf{;}\\
\phantom{xxx} x_{c'}\leftarrow x_{c}=x_{c+1}\\
x\leftarrow x_{c+1}
\end{array}
\]

Let us consider the sequence $D_0$,\ldots,$D_i$,\ldots of distributed programs starting from $D_{S_{\Psi}}$, where $D_0=D_{S_{\Psi}}$ and
for any $i$, $D_i\xRightarrow{}_{\langle F,T\rangle} D_{i+1}$. Let $\Psi_{i} $ be the interpretation $x_{c+1}$ at $D_i$ ($\Psi_{i}=\sintll{F}{D_i}(x_{c+1})$), we can consider
the sequence of elements $\Psi_{j}$ such that $\Psi_{j}\not=\Psi_{j-1}$. 
We can note that this sequence $\Psi_{0},\Psi_{j_1},\ldots$ defines a monotonic sequence that is the result of the application of a \emph{fair strategy} (the
one induced by the application of the distributed operational semantics). Since $A_F$ has only only finite partially ordered chains we are under the hypotheses of Theorem~\ref{th:asynchrony-theorem2}. Hence a fixpoint is eventually reached. Hence, there exists an index $k$ such that: $\sintll{F}{D_k}(x_{c})=\sintll{F}{D_i}(x_{c+1})=
\llbracket \mu \kappa.\Psi_i\rrbracket^{F}_{\emptyset}$. 
Moreover, $\sintll{F}{D_k}(x_{c'})=\lambda n. true$. 
This implies that, after a number of reductions, distributed execution $D_{\mathsf{skip}}$ is reached where 
$\sintll{F}{D_{\mathsf{skip}}}(x)= \llbracket \mu \kappa.\Psi_i\rrbracket^{F}_{\emptyset}$.\qedhere
\end{itemize}
\end{proof}

\teoremaa*

\begin{proof} $\ $
\begin{enumerate}[label=(\roman*)]
\item[(i)] 
We proceed by induction on the syntax of $P$. 

\medskip\noindent%
\textit{Base of Induction:} Let $P=\mathsf{skip}$ or $P=i\leftarrow \Psi;$. In the first case the thesis is trivially verified while in the second case 
our statement follows directly from Lemma~\ref{lem:disexp} and from Lemma~\ref{lem:factsem}.

\medskip\noindent%
\textit{Inductive Hypothesis:} Let $P_1$, $P_2$, $D_{P_1}$ and $D_{P_2}$ be such that $D_{S_{P_i}}$ that agrees with $F$,  if $\langle P_i,F\rangle \rightarrow \langle P_i',F'\rangle$ then $D_{S_{P_i}}\xRightarrow{}_{\langle F,T\rangle} D_{S_{P_i'}}$ and 
$D_{S_{P_i'}}$ agrees with $F'$.

\medskip\noindent%
\textit{Inductive Step:} The following cases can be considered:

\begin{itemize}
\item $P=P_1;P_2$. We have that $S_{P}=S_{P_1};\mathsf{wait}(x);S_{P_2}$. We have that $\langle P, F\rangle \rightarrow \langle P',F'\rangle$ if and only if 
either $P_1=\mathsf{skip}$ and $\langle P_2,F\rangle \rightarrow \langle P',F'\rangle$ or $\langle P_1,F\rangle \rightarrow \langle P_1',F'\rangle$
and $P'=P_1';P_2$. In both the cases the statement follows by inductive hypothesis and by Lemma~\ref{lem:factsem}.

\item $P=\mathsf{if}\ \Psi\ \mathsf{then}\ P_1\ \mathsf{else}\ P_2$. We have that $S_{P}=R_{\Psi}^{x};\mathsf{if}\ x\ \mathsf{then}\ S_{P_1}\ \mathsf{else}\ S_{P_2}$.
Moreover,  if $\langle P,F\rangle\rightarrow \langle P', F'\rangle$ then $F=F'$ and either $\llbracket \Psi \rrbracket^{F}_{\emptyset}=\lambda n.true$, $P'=P_1$
or $\llbracket \Psi \rrbracket^{F}_{\emptyset}=\lambda n.true$, $P'=P_2$. Let us assume that $\llbracket \Psi \rrbracket^{F}_{\emptyset}=\lambda n.true$.
From Lemma~\ref{lem:disexp} we have that $D_{R_{\Psi}^{x}}\xRightarrow{}^{*}_{\langle F,T\rangle} D_{\mathsf{skip}}$ and by applying rules in 
Table~\ref{tab:distopsem1}, Table~\ref{tab:distopsem2}  
and Table~\ref{tab:distopsem3} we have that 
\[
D_{S}\xRightarrow{}_{\langle F,T\rangle}^{*} D_{\mathsf{skip}};\mathsf{if}\ x\ \mathsf{then}\ S_{P_1}\ \mathsf{else}\ S_{P_2}
\]
Moreover, $\sintll{F}{D_{\mathsf{skip}}}(x)=\lambda n.true$ (resp.$\sintll{F}{D_{\mathsf{skip}}}(x)\not=\lambda n.true$), hence
\[
D_{\mathsf{skip}};\mathsf{if}\ x\ \mathsf{then}\ S_{P_1}\ \mathsf{else}\ S_{P_2}\xRightarrow{}_{\langle F,T\rangle}^{*} D_{S_{P_1}}
\]
and $D_{S_{P_1}}$ agrees with $F'$.

\item $P=\mathsf{while}\ \Psi\ \mathsf{do}\ P_1$. This case follows similarly to previous one.
\end{itemize}\medskip

\item[(ii)]
We proceed by induction on the syntax of $P$.

\medskip\noindent%
\textit{Base of Induction:} If $P=\mathsf{skip}$ or $P=i\leftarrow \Psi$, like for the previous point, the statement follows directly from Lemma~\ref{lem:disexp} and from Lemma~\ref{lem:factsem}.

\medskip\noindent%
\textit{Inductive Hypothesis:}  Let $P_1$, $P_2$, $D_{P_1}$ and $D_{P_2}$ be such that 
if $D_{S_{P_i}}\xRightarrow{}^{*}_{\langle F,T\rangle} D'$ there exists $P'$ such that $D'\xRightarrow{}^{*}_{\langle F,T\rangle} D_{S_{P'}}$,
$\langle P_i,F\rangle \rightarrow \langle P',F'\rangle$ and $D_{S_{P'}}$ agrees with $F'$.

\medskip\noindent%
\textit{Inductive Step:} The following cases can be considered:

\begin{itemize}
\item $P=P_1;P_2$. We have that $S_{P}=S_{P_1};\mathsf{wait}(x);S_{P_2}$. Let $D_{S_{P}} \xRightarrow{}^{*}_{\langle F,T\rangle} D'$. 
By Lemma~\ref{lem:factsem} we have that $D_{S}$ flows into $D_{\mathsf{wait}(x);S_{P_2}}$. For this reason we can distinguish three cases:
(1) there exists $D''$ such that $D_{S_{P_1}}\xRightarrow{}_{\langle F,T\rangle}^{*} D''$ and $D'=D''; \mathsf{wait}(x);S_{P_2}$; 
(2) $D' \xRightarrow{}_{\langle F,T\rangle}^{*} D_{\mathsf{wait}(x);S_{P_2}}$; 
(3) $D' \xRightarrow{}_{\langle F,T\rangle}^{*} D_{\mathsf{wait}(x);S_{P_2}}$. In all the cases the statement follows directly from the inductive hypothesis.

\item $P=\mathsf{if}\ \Psi\ \mathsf{then}\ P_1\ \mathsf{else}\ P_2$. We have that $S_{P}=R_{\Psi}^{x};\mathsf{if}\ x\ \mathsf{then}\ S_{P_1}\ \mathsf{else}\ S_{P_2}$.
By Lemma~\ref{lem:factsem} we have that we can distinguish three cases: (1) there exists $D''$ such that $D_{R_{\Psi}^{x}} \xRightarrow{}^{*}_{\langle F,T\rangle} D''$ and $D'=D'';\mathsf{if}\ x\ \mathsf{then}\ S_{P_1}\ \mathsf{else}\ S_{P_2}$; (2) $\llbracket \Psi \rrbracket^{F}_{\emptyset}=\lambda n.true$ and 
$D_{S_{P_1}}\xRightarrow{}^{*}_{\langle F,T\rangle} D'$; (3) $\llbracket \Psi \rrbracket^{F}_{\emptyset}\not=\lambda n.true$ and $D_{S_{P_2}}\xRightarrow{}^{*}_{\langle F,T\rangle} D'$. In all the cases the statement follows directly from the inductive hypothesis and by simple applications of rules in Table~\ref{tab:distopsem1}, Table~\ref{tab:distopsem2}   and Table~\ref{tab:distopsem3}.

\item $P=\mathsf{while}\ \Psi\ \mathsf{do}\ P_1$. This case follows similarly to previous one by noticing that $S_{P}=R_{\Psi}^{x_{c}}\mathsf{;}\mathsf{while}\ x_{c}\ \mathsf{do}\ S\mathsf{;}\mathsf{wait}(x_{c''})\mathsf{;}R_{\Psi}^{x_{c}}$ and that after $k$ iterations $D_{S_{P}}$ flows into $D_{\mathsf{wait}(x_{c''});S_{P}}^{k}$.\qedhere
\end{itemize}
\end{enumerate}
\end{proof}


\clearpage
\section{Symbols} \label{section:proofs}

\begin{table}[h]
\centering \scriptsize
\begin{tabular}{|r|l|}\hline
$A$ & field domain carrier \\\hline 
$a$ & element of $A$ \\\hline
$B$ & subset of $A$ \\\hline
$\sqsubseteq$ & field domain ordering relation \\\hline 
$\top$ & field domain top element \\\hline 
$\bot$ & field domain bottom element \\\hline 
$N$ & set of field nodes  \\\hline 
$\mathcal{N}$ & universe of nodes  \\\hline 
$n,n',\dots$ & node  \\\hline 
$E$ & set of field edges  \\\hline 
$\mathcal{E}$ & universe of edges  \\\hline 
$e,e',\dots$ & edge  \\\hline 
$L$ & set of field labels \\\hline 
$\mathcal{X}$ & set of auxiliary field labels \\\hline
$\mathcal{L}$ & set of all labels \\\hline 
$i,j,x,\dots$ & node labels \\\hline 
$\alpha$ & edge label \\\hline 
$I$ & interpretation of field labels \\\hline 
$F$ & field \\\hline 
$\mathcal{F}$ & set of all fields \\\hline 
$f$ & combination operations \\\hline 
$g$ & aggregation operations \\\hline 
$\mathcal{M}$ & set of all function symbols \\\hline 
$\mathsf{f}$ & node valuation \\\hline 
$z$ & recursion variable \\\hline
$\mathcal{Z}$ & set of all recursion variables \\\hline  
$\rho$ & recursion variable envionment \\\hline  
$\Psi$ & \smuc{} formula \\\hline 
$\psi$ & update function \\\hline 
$\mu$ & least fixpoint operator \\\hline 
$\nu$ & greatest fixpoint operator \\\hline 
$$\aggregateOut{}{}{}$ $ & modal operator (out)  \\\hline 
$$\aggregateIn{}{}{}$$ & modal operator (in) \\\hline 
$\pi$ & execution pattern \\\hline 
$\sigma$ & execution strategy \\\hline 
$\varsigma$ & failure sequence \\\hline 
$P$ & \smuc{} program  \\\hline 
$S,R$ & \smuc{} program in SAF  \\\hline 
$c$ & counter \\\hline 
$\chi$ & agreement store \\\hline 
$\kappa$ & agreement counter \\\hline 
$\iota$ & interpretation of node labels \\\hline 
$d$ & fragment \\\hline 
$\lambda$ & transition label \\\hline 
$\tau$ & silent action  \\\hline 
$m$ & message \\\hline 
$D$ & distributed execution  \\\hline 
$X$ & set of nodes  \\\hline 
\end{tabular}
\caption{Symbol notation}
\label{symbols}
\end{table}
\clearpage

\bibliographystyle{plain}
\bibliography{bibliography}

\end{document}